\documentclass[usegraphicx, longauth]{aa}

\usepackage{natbib}
\bibpunct{(}{)}{;}{a}{}{,} 
\usepackage{subfigure}
\usepackage{longtable}
\usepackage{graphicx}
\usepackage{graphics}
\usepackage{keyval}
\usepackage{trig}
\usepackage[american]{babel}
\usepackage{url}
\usepackage{color}
\usepackage{amsmath}
\usepackage{bm}
\usepackage{longtable}

\def\beq{\begin{equation}}
\def\eeq{\end{equation}}
\def\bey{\begin{eqnarray}}
\def\eey{\end{eqnarray}}

\def\lsim{\mathrel{\raise.3ex\hbox{$<$\kern-.75em\lower1ex\hbox{$\sim$}}}}
\def\gsim{\mathrel{\raise.3ex\hbox{$  $\kern-.75em\lower1ex\hbox{$\sim$}}}}

\def\a0{A_{\mathrm{0}}}

\def\teff{T_\mathrm{eff}}
\def\fehzw{\rm [Fe/H]_{ZW}}
\def\fehuves{\rm [Fe/H]_{UVES}}

\def\apjl{ApJ}

\def\aap{A\&A}      
\def\apjs{ApJS}       

\newcommand\Eq[1]{Eq.~(\ref{#1})}
\newcommand\Fig[1]{Fig.~\ref{#1}}
\newcommand\Tab[1]{Table~\ref{#1}}
\newcommand\Sec[1]{Sec.~\ref{#1}}

\begin{document}
   \title{Estimating the parameters of globular cluster M 30 (NGC 7099) from time-series photometry\thanks{The full light curves presented in this paper (\Tab{tab:onlinedata}) are only available in electronic form at the Centre de donn\'{e}es astronomiques de Strasbourg (CDS) via anonymous ftp to \tt{cdsarc.u-strasbg.fr (130.79.128.5)}
or via \tt{http://cdsweb.u-strasbg.fr/cgi-bin/qcat?J/A+A/}}\fnmsep\thanks{This work is based on data collected by MiNDSTEp with the Danish 1.54m telescope at the ESO La Silla Observatory.}}

   \author{N. Kains
          \inst{\ref{eso}}
          \and
          D. M. Bramich\inst{\ref{eso}}
          \and
          A. Arellano Ferro\inst{\ref{unam}}
           \and
          R. Figuera Jaimes\inst{\ref{eso}, \ref{standrews}}
          \and
          U. G. J{\o}rgensen\inst{\ref{nbi}, \ref{copenhagen}}
          \and
          S. Giridhar\inst{\ref{iia}}\\
          and\\
          K.A. Alsubai \inst{\ref{qnrf}} \and
          J. M. Andersen \inst{\ref{bostonu}, \ref{copenhagen}} \and
	V. Bozza \inst{\ref{salerno}, \ref{fisicanucleare}} \and
	P. Browne \inst{\ref{standrews}} \and
	M. Burgdorf \inst{\ref{hespace}} \and 
	S. Calchi Novati \inst{\ref{salerno}, \ref{iiass}} \and
	Y. Damerdji \inst{\ref{liege}} \and
	C. Diehl \inst{\ref{ari}, \ref{hamburg}} \and
	P. Dodds \inst{\ref{standrews}} \and
	M. Dominik \inst{\ref{standrews}}\fnmsep\thanks{Royal Society University Research Fellow} \and
	A. Elyiv \inst{\ref{liege}, \ref{ukraine}} \and 
	X.-S. Fang \inst{\ref{yunan}, \ref{chinasci}} \and
	E. Giannini \inst{\ref{ari}} \and
	S.-H. Gu \inst{\ref{yunan}, \ref{chinasci}} \and
	S. Hardis \inst{\ref{nbi}} \and
	K. Harps{\o}e \inst{\ref{nbi}, \ref{copenhagen}} \and
	T.C. Hinse \inst{\ref{kasi},\ref{nbi}} \and
	A. Hornstrup \inst{\ref{dantek}} \and
	M. Hundertmark \inst{\ref{standrews}} \and
	J. Jessen-Hansen \inst{\ref{aarhus}} \and
	D. Juncher  \inst{\ref{nbi}, \ref{copenhagen}} \and
	E. Kerins \inst{\ref{manchester}} \and
	H. Kjeldsen \inst{\ref{aarhus}} \and
	H. Korhonen \inst{\ref{nbi}, \ref{copenhagen}} \and
          C. Liebig \inst{\ref{standrews}} \and
          M. N. Lund \inst{\ref{aarhus}} \and
	M. Lundkvist \inst{\ref{aarhus}} \and
	L. Mancini \inst{\ref{mpia}} \and
	R. Martin \inst{\ref{perthobs}} \and
	M. Mathiasen \inst{\ref{nbi}} \and
	M.T. Penny \inst{\ref{osu}} \and
	M. Rabus \inst{\ref{puc}} \and
	S. Rahvar \inst{\ref{sharif}, \ref{perimeter}} \and
	D. Ricci \inst{\ref{unam_ens}, \ref{liege}} \and
	K. Sahu \inst{\ref{stsci}}\and
	G. Scarpetta \inst{\ref{salerno},\ref{infn}} \and
	J. Skottfelt \inst{\ref{nbi}, \ref{copenhagen}} \and
	C. Snodgrass \inst{\ref{mps}} \and
	J. Southworth \inst{\ref{keele}} \and
	J. Surdej \inst{\ref{liege}} \and
	J. Tregloan-Reed \inst{\ref{keele}} \and
	C. Vilela \inst{\ref{keele}} \and
	O. Wertz \inst{\ref{liege}} \and
	A. Williams \inst{\ref{perthobs}} \\
          (The MiNDSTEp consortium)\\
          }

   \institute{European Southern Observatory, Karl-Schwarzschild Stra\ss e 2, 85748 Garching bei M\"{u}nchen, Germany\label{eso}\\
              \email{nkains@eso.org}
         \and Instituto de Astronom\'{i}a, Universidad Nacional Aut\'{o}noma de Mexico\label{unam}\\
         \and SUPA School of Physics \& Astronomy, University of St Andrews, North Haugh, St Andrews, KY16 9SS, United Kingdom \label{standrews}\\
	\and Niels Bohr Institute, University of Copenhagen, Juliane Maries vej 30, 2100 Copenhagen, Denmark \label{nbi}\\	
	\and Centre for Star and Planet Formation, Geological Museum, {\O}ster Voldgade 5, 1350 Copenhagen, Denmark \label{copenhagen}\\	
      	\and Indian Institute of Astrophysics, Koramangala 560034, Bangalore, India\label{iia}\\	
          \and Qatar Foundation, P.O. Box 5825, Doha, Qatar \label{qnrf}\\	
          \and Department of Astronomy, Boston University, 725 Commonwealth Ave, Boston, MA 02215, United States of America \label{bostonu}\\	
          \and Dipartimento di Fisica ``E.R Caianiello", Universitˆ di Salerno, Via Ponte Don Melillo, 84084 Fisciano, Italy \label{salerno}\\	
	\and Istituto Nazionale di Fisica Nucleare, Sezione di Napoli, Italy \label{fisicanucleare}\\
	\and HE Space Operations GmbH, Flughafenallee 24, 28199 Bremen, Germany \label{hespace}\\
	\and Istituto Internazionale per gli Alti Studi Scientifici (IIASS), Vietri Sul Mare (SA), Italy \label{iiass}\\		
	\and Institut d'Astrophysique et de G\'{e}ophysique, Universit\'{e} de Li\`{e}ge, All\'{e}e du 6 Ao\^{u}t 17, Sart Tilman, B\^{a}t.\ B5c, 4000 Li\`{e}ge, Belgium \label{liege}\\	
	\and Astronomisches Rechen-Institut, Zentrum f\"{u}r Astronomie der Universit\"{a}t Heidelberg (ZAH),  M\"{o}nchhofstr.\ 12-14, 69120 Heidelberg, Germany \label{ari}\\
	\and Hamburger Sternwarte, Universit\"{a}t Hamburg, Gojenbergsweg 112, 21029 Hamburg \label{hamburg}\\
	\and Main Astronomical Observatory, Academy of Sciences of Ukraine, vul. Akademika Zabolotnoho 27, 03680 Kyiv, Ukraine \label{ukraine}\\
	\and National Astronomical Observatories/Yunnan Observatory, Chinese Academy of Sciences, Kunming 650011, China \label{yunan}\\
	\and Key Laboratory for the Structure and Evolution of Celestial Objects, Chinese Academy of Sciences, Kunming 650011, China \label{chinasci}\\
	\and Korea Astronomy and Space Science Institute, Daejeon 305-348, Korea \label{kasi}\\
	\and Institut for Rumforskning og -teknologi, Danmarks Tekniske Universitet, K{\o}benhavn, Denmark \label{dantek}\\
	\and Stellar Astrophysics Centre, Department of Physics and Astronomy, Aarhus University, Ny Munkegade 120, DK-8000 Aarhus C, Denmark \label{aarhus}\\	
	\and Jodrell Bank Centre for Astrophysics, University of Manchester, Oxford Road,Manchester, M13 9PL, UK \label{manchester}\\	
	\and Max Planck Institute for Astronomy, K\"{o}nigstuhl 17, 69117 Heidelberg, Germany \label{mpia}\\	
	\and Perth Observatory, Walnut Road, Bickley, Perth 6076, Australia \label{perthobs}\\ 
	\and Department of Astronomy, Ohio State University, 140 West 18th Avenue, Columbus, OH 43210, United States of America \label{osu}\\	
	\and Departamento de Astronom\'ia y Astrof\'isica, Pontificia Universidad Cat\'olica de Chile, Av. Vicu\~na Mackenna 4860, 7820436 Macul, Santiago, Chile \label{puc}\\
	\and Department of Physics, Sharif University of Technology, P.~O.\ Box 11155--9161, Tehran, Iran \label{sharif}\\	
	\and Perimeter Institute for Theoretical Physics, 31 Caroline St. N., Waterloo ON, N2L 2Y5, Canada \label{perimeter}\\	
	\and Instituto de Astronom\'ia - UNAM, Km 103 Carretera Tijuana Ensenada, 422860, Ensenada (Baja Cfa), Mexico \label{unam_ens}\\
	\and Space Telescope Science Institute, 3700 San Martin Drive, Baltimore, MD 21218, United States of America \label{stsci}\\
	\and INFN, Gruppo Collegato di Salerno, Sezione di Napoli, Italy \label{infn}\\	
	\and Max Planck Institute for Solar System Research, Max-Planck-Str. 2, 37191 Katlenburg-Lindau, Germany \label{mps}\\	
	\and Astrophysics Group, Keele University, Staffordshire, ST5 5BG, United Kingdom \label{keele}\\	
	}

   \date{Received ... ; accepted ...}


  \abstract
   {}
   {We present the analysis of 26 nights of $V$ and $I$ time-series observations from 2011 and 2012 of the globular cluster M 30 (NGC 7099). We used our data to search for variable stars in this cluster and refine the periods of known variables; we then used our variable star light curves to derive values for the cluster's parameters.}
   {We used difference image analysis to reduce our data to obtain high-precision light curves of variable stars. We then estimated the cluster parameters by performing a Fourier decomposition of the light curves of RR Lyrae stars for which a good period estimate was possible. We also derive an estimate for the age of the cluster by fitting theoretical isochrones to our colour-magnitude diagram (CMD).
   }
   {Out of 13 stars previously catalogued as variables, we find that only 4 are bona fide variables. We detect two new RR Lyrae variables, and confirm two additional RR Lyrae candidates from the literature. We also detect four other new variables, including an eclipsing blue straggler system, and an SX Phoenicis star. This amounts to a total number of confirmed variable stars in M 30 of 12. We perform Fourier decomposition of the light curves of the RR Lyrae stars to derive cluster parameters using empirical relations. We find a cluster metallicity $\fehzw=-2.01 \pm 0.04$, or $\fehuves=-2.11 \pm 0.06$, and a distance of $8.32 \pm 0.20$ kpc (using RR0 variables), 8.10 kpc (using one RR1 variable), and $8.35 \pm 0.42$ kpc (using our SX Phoenicis star detection in M 30). Fitting isochrones to the CMD, we estimate an age of $13.0\pm 1.0$ Gyr for M 30.
}
   {}


   \maketitle
%

\section{Introduction}\label{sec:intro}

In this paper we analyse time-series observations of M 30 (NGC 7099, or C2137-234 in the IAU nomenclature; $\alpha = 21^h40^m22^s, \delta = -23^{\circ}10'47.5''$ at J2000.0), one of the most metal-poor globular clusters known, with [Fe/H] $\sim$ -2.1, located at a distance of $\sim$ 8 kpc. This cluster is thought to be of extra-Galactic origin, due to its retrograde orbit \citep{allen06}, pointing to its possible accretion by the Milky Way following an encounter with a satellite, as well as its position on the age-metallicity diagram \citep{forbes10}. M 30 is also thought to have undergone core collapse \citep[e.g.][]{djorgovski86}. Here we detect and classify already reported, as well as new variables, and use Fourier decomposition to derive properties of the RR Lyrae stars in this cluster. We then estimate the cluster parameters of M 30, providing us with an independent estimate of the metallicity, distance and age of this cluster, and the first using this method.

Our observations are detailed in \Sec{sec:observations}; we discuss the variables in M 30 in \Sec{sec:variables}, and in \Sec{sec:fourdec} we derive properties of the RR Lyrae stars in this cluster using Fourier decomposition and empirical relations. We use this to derive properties of the cluster in \Sec{sec:clusterprop} and summarise our findings.

\section{Observations and reductions}\label{sec:observations}
\subsection{Observations}

We obtained Johnson $V$- and $I$-band data with the Danish Faint Object Spectrograph and Camera (DFOSC) imaging camera at the Danish 1.54m telescope at La Silla in Chile, as part of a programme with the MiNDSTEp consortium. In this programme we are making use of parts of the night when the MiNDSTEp primary-science microlensing observations towards the Galactic Bulge are not possible. The data were obtained in 2011 and 2012, with the 2011 data consisting mostly of $I$-band observations, while the 2012 observations are evenly split between $V$ and $I$-band images. The observations are summarised in \Tab{tab:observations}. 

The imaging CCD is 2147 $\times$ 2101 pixel, with a scale of 0.396 arcsec per pixel, giving a field of view of 14.2 $\times$ 13.9 arcmin$^2$.

\begin{table}
\begin{center}
  \begin{tabular}{ccccc}

     \hline
    Date		 &$N_V$ 	&$t_V$(s)	&$N_I$	&$t_I$(s) \\
  \hline  

    20110728		&$-$		&$-$		&48		&10	\\
    20110730		&$-$		&$-$		&27		&15-20\\
    20110803		&$-$		&$-$		&12		&30	\\
    20110804		&$-$		&$-$		&4		&30	\\
    20110806		&6		&30		&14		&30	\\
    20110809		&$-$		&$-$		&16		&30-40	\\
    20110810		&1		&30		&12		&30	\\
    20110811		&$-$		&$-$		&12		&30	\\
    20110813		&6		&40		&6		&30	\\
    20110814		&$-$		&$-$		&10		&27	\\
    20110815		&$-$		&$-$		&6		&27	\\
    20110817		&$-$		&$-$		&6		&30	\\
    20110820		&$-$		&$-$		&16		&30	\\
    20110821		&$-$		&$-$		&16		&30	\\
    20110822		&$-$		&$-$		&7		&30	\\
    20110828		&$-$		&$-$		&8		&30	\\
    20110830		&6		&20		&11		&30	\\
    20110831		&$-$		&$-$		&16		&30	\\
    20120807		&37		&10-20	&36		&10-20	\\
    20120810		&38		&20		&38		&20	\\
    20120811		&34		&20		&41		&20	\\
    20120819		&6		&100	&4		&100 \\
    20120820		&24		&100	&22		&100 \\
    20120821		&38		&70-100	&38		&70-100	\\
    20120828		&65		&50-100	&65		&40-50	\\
    20120919		&40		&50		&39		&40	\\
\hline
   Total		&301	&	&530	& \\
\hline \hline
  \end{tabular}
  \caption{Numbers of images and exposure times for the $V$ and $I$ band observations of M 30. When varying exposure times were used, a range is given. \label{tab:observations}}
  \end{center}
\end{table}

\subsection{Difference image analysis}

As in our previous studies of variables in globular clusters (Figuera et al. 2013, \citealt{kains12b, arellano11}), we used the {\tt DanDIA}\footnote{{\tt DanDIA} is built from the {\tt DanIDL} library of {\tt IDL} routines available at {\tt http://www.danidl.co.uk}} pipeline \citep{bramich13, bramich08} to reduce our observations. Using difference image analysis (DIA) enabled us to obtain high-precision photometry for sources within our field of view. The reader is referred to our previous papers \citep[e.g.][]{bramich11} for a detailed description of the software used; we provide a short summary of the main steps here.

After preprocessing (bias level correction and flat-fielding), our images were blurred with a Gaussian of  full-width half-maximum (FWHM) of 3 pixels to avoid undersampling, which is detrimental for determining the kernel in DIA. We then produce a reference image for each filter, stacking images within 10\% of the best seeing ($\sim 1.2''$), also taking care to avoid including images with many saturated stars. Our reference image consists of 5 stacked images, with a combined exposure time of 100 s, and a FWHM of the point-spread function (PSF) of 3.88 pixels ($\sim 1.5''$) in $V$, while in $I$ the combined exposure time is 80 s from 4 images, and the PSF FWHM is 3.31 pixels ($\sim 1.3''$). For each filter, the source positions and reference fluxes were then extracted from the reference image. Images were then registered with the reference, and the reference was convolved with the kernel solution and subtracted from each image, producing a set of difference images. Finally, difference fluxes were measured from each difference image for each source, which we used to construct a set of light curves. In \Tab{tab:onlinedata}, we outline the format of the data as it is provided in electronic format with the online version of this paper. 

As was noted in previous papers, blending due to the presence of other objects near a source may lead to the reference flux being overestimated. However, non-variable sources are still fully subtracted in the difference images, while variable objects have the amplitude of their variability underestimated if their reference flux is affected by blending, although the shape of their light curve, crucially, will remain unaffected.

\begin{table*}
\begin{center}
  \begin{tabular}{ccccccccccc}

     \hline
    \#		&Filter	&HJD 	&$M_{\rm std}$	  & $m_{\rm ins}$ 	&$\sigma_m$		&$f_{\rm ref}$ 	&$\sigma_{\rm ref}$	& $f_{\rm diff}$ &$\sigma_{\rm diff}$ &$p$  \\
	&	&($d$)	&(mag)	&(mag)	&(mag)	&(ADU s$^{-1}$)	&(ADU s$^{-1}$)	&(ADU s$^{-1}$)	&(ADU s$^{-1}$)	&	\\
  \hline  
   
V1 &V & 2455779.84508 &    15.360 &    14.475 &     0.004 &  4619.581 &     1.991 &  -500.380 &     4.612 &    0.3633 \\
V1 &V & 2455779.84637 &    15.378 &    14.493 &     0.005 &  4619.581 &     1.991 &  -455.515 &     4.326 &    0.3185 \\
\vdots &\vdots&\vdots&\vdots&\vdots&\vdots&\vdots&\vdots&\vdots&\vdots&\vdots \\
V1 &I & 2455770.83621 &    14.628 &    14.797 &     0.007 &  2843.148 &     3.237 &   -68.927 &     2.472 &    0.1593 \\
V1 &I & 2455770.83726 &    14.639 &    14.809 &     0.007 &  2843.148 &     3.237 &   -72.977 &     2.461 &    0.1593 \\
\vdots &\vdots&\vdots&\vdots&\vdots&\vdots&\vdots&\vdots&\vdots&\vdots&\vdots \\

\hline \hline
  \end{tabular}
  \caption{Format for the time-series photometry of all confirmed variables in our field of view. The standard $M_{\rm std}$ and instrumental $m_{\rm ins}$ magnitudes listed in columns 4 and 5 respectively correspond to the variable star, filter and epoch of mid-exposure listed in columns 1-3, respectively. The uncertainty on $m_{\rm ins}$ is listed in column 6, which also corresponds to the uncertainty on $M_{\rm std}$. For completeness, we also list the reference flux $f_{\rm ref}$ and the differential flux $f_{\rm diff}$ (columns 7 and 9 respectively), along with their uncertainties (columns 8 and 10), as well as the photometric scale factor $p$. Definitions of these quantities can be found in e.g. \cite{bramich11}, Eq. 2-3. This is a representative extract from the full table, which is available with the electronic version of the article. \label{tab:onlinedata}}
  \end{center}
\end{table*}

\subsection{Photometric calibration}

\subsubsection{Self-calibration}

Although all photometry is affected by systematic errors, steps can be taken to correct for this, in particular with time-series photometry, for which we can carry out a \textit{self-calibration} of the data \citep[e.g.][]{padmanabhan08}. To do this, we use the method of \cite{bramich12b} to derive magnitude offsets to be applied to each photometric measurement. In practice, this equates to correcting for errors in the fitted value of the photometric scale factor $p$. We find magnitude offsets of the order of 2-3 mmag for most data points, which leads to small difference in the resulting light curves.

\subsubsection{Photometric standards}

\begin{figure}
  \centering
  \includegraphics[width=8cm, angle=0]{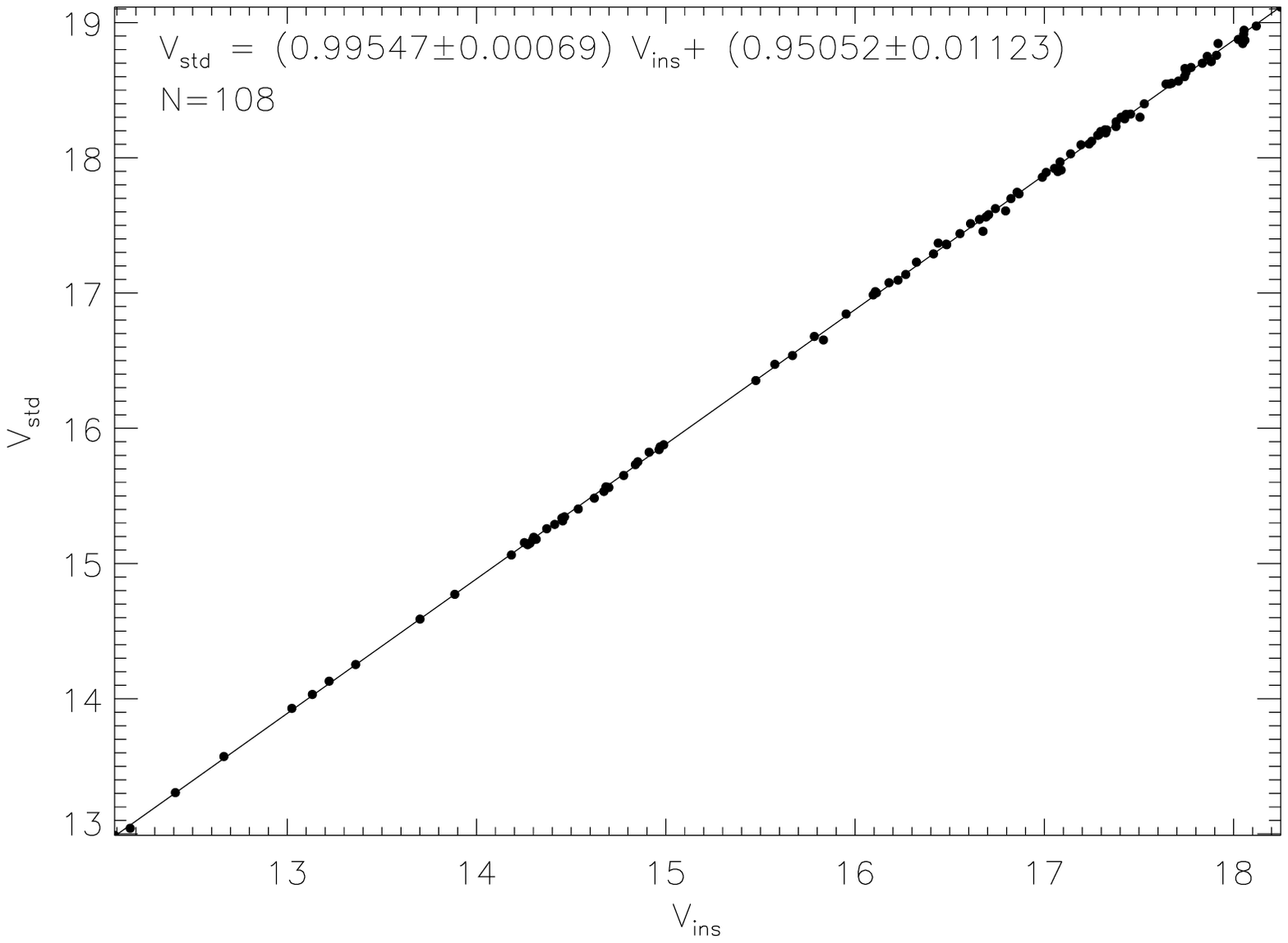}
  \includegraphics[width=8cm, angle=0]{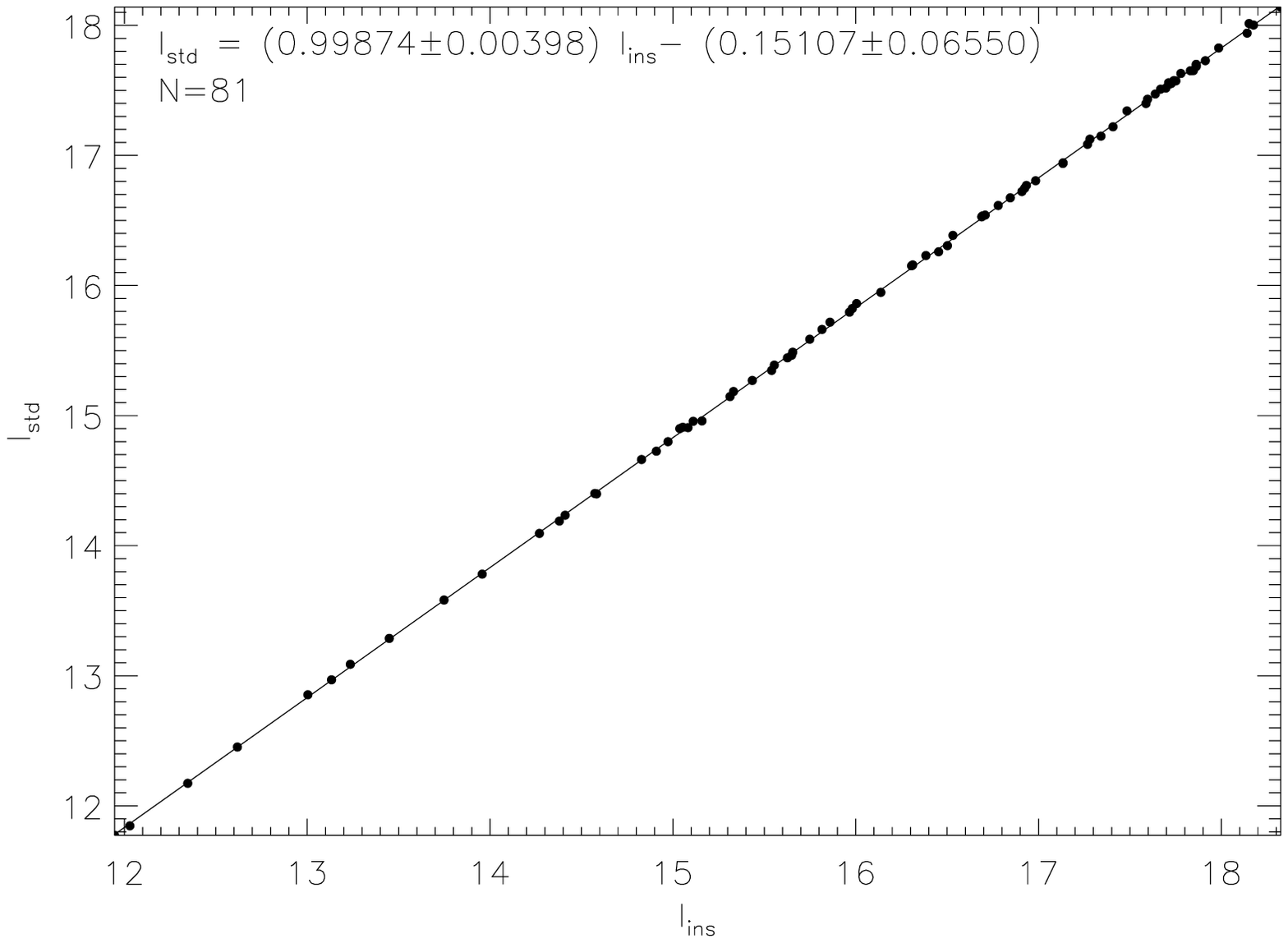}
  \caption{Linear regression fits used to convert from instrumental to standard Johnson-Kron-Cousins magnitudes for the $V$ (top) and $I$ (bottom) bands. \label{fig:transf}}

\end{figure}

We converted the instrumental magnitudes from our data reduction pipeline to standard Johnson-Kron-Cousins magnitudes by carrying out a linear fit to the relation between our light curve magnitudes and those of the photometric standards of \cite{stetson00} in the field of this cluster. The relations, shown in \Fig{fig:transf}, were used to obtain light curves in standard magnitudes. The standard stars we used cover the full range of colour and magnitude spanned by our CMD. No significant colour term was found and the correlation coefficients are $\sim 1$ for both filters.

\subsection{Astrometry}

We used GAIA to perform the astrometry by matching $\sim$400 manually picked stars from our reference images with the USNO-B1.0 catalogue of stars \citep{monet03} to derive a linear transformation between our image positions and equatorial coordinates. This means that the coordinates in \Tab{tab:coordinates} are given at the effective epoch of our reference image, HJD$\sim$2456151.84. The root mean square (rms) of the residuals from our astrometric fit is 0.23 arcsec, or 0.59 pixels.

\begin{table}
\begin{center}
  \begin{tabular}{ccc}

     \hline
    \#		 	&RA 		&DEC \\
  \hline  

    V1			&21:40:24.25	&-23:11:46.4	\\
    V2			&21:40:26.44	&-23:12:51.2	\\
    V3			&21:40:15.04	&-23:11:27.9	\\
    V4			&21:39:58.46	&-23:11:43.3	\\
    V14		&21:40:21.10	&-23:11:30.8	\\
    V15		&21:40:21.79	&-23:10:50.1	\\
    V16		&21:40:24.81	&-23:11:48.8	\\
    V17		&21:40:23.66	&-23:11:01.1	\\
    V18		&21:40:21.88	&-23:09:32.4	\\
    V19		&21:40:22.09	&-23:10:43.3	\\
    V20		&21:40:29.83	&-23:11:05.9	\\
    V21		&21:40:47.80	&-23:10:23.9	\\

\hline \hline
  \end{tabular}
  \caption{Equatorial celestial coordinates of confirmed variables in M 30 at the epoch of the reference image, HJD$\sim$2456151.84. \label{tab:coordinates}}
  \end{center}
\end{table}

\section{Variables in M 30}\label{sec:variables}

The first three variables (V1-3) in this cluster were detected by \cite{bailey1902} using photographic observations made at the Harvard College Station in Arequipa, Peru. V4 was then reported by \cite{rosino49}, who discovered it using observations of M 30 at 175 epochs, taken between 1946 and 1948 with the 60cm reflecting telescope at the Lojano Astronomical Station, near Bologna in Italy. Rosino also derived periods for V1-3, and refined his period for V1 in a later paper \citep{rosino61}. He also noted that V4 did not present the characteristics of an RR Lyrae type variable. He described it as an unusual object, reaching a magnitude of 16.4, remaining at that level for several days, before dropping below the limits of detectability of his photographic plates, i.e. below $\sim$ 18th magnitude. Other light curve features led him to conclude that V4 is a variable of type U Geminorum. This was confirmed in several subsequent publications by \cite{margon83}, who obtained spectroscopic evidence that V4 is a cataclysmic variable, and \cite{machin91}, who also concluded that V4 is most likely a foreground object rather than a cluster member. \cite{pietrukowicz08} found a relatively high $V$-band brightness consistent with that assessment, and also noted that V4 has a likely X-ray counterpart detected by the $ROSAT$ satellite. 

Finally, V5-12 were all reported as variable by \cite{terzan68}, and V13 was reported by \cite{terzan75}, based on photographic observations taken at the 1.52m telescope at ESO between 1972 and 1974.; however they did not publish light curves or periods for any of those variables.
	
\cite{pietrukowicz04} analysed HST archival data of M 30 and identified two previously unknown RR Lyrae variables, as well as four W UMa-type contact binaries, and an eclipsing close-binary variable with ellipsoidal variations. However, they did not assign those variables catalogue V numbers. Finally, although \cite{smitka07} also studied this cluster, their photometric accuracy is poor due to the cluster's low altitude for Northern hemisphere observations. They reported three new RR Lyrae star candidates, but did not provide coordinates or a finding chart, so we are not able to cross-correlate our findings with theirs; they also did not assign their variables V numbers.

This amounts to a total of 13 catalogued variables in the cluster, although V4 is now thought to be a foreground object. There are also 7 additional variables reported by \cite{pietrukowicz04}, including 2 new RR Lyrae. The present study of M 30 is the first from the Southern hemisphere using CCD photometry, allowing us to carry out a significant update of the variable star population in this cluster.

\subsection{Stars that do not show variability}

We find that all of the stars V5-V13, reported by \cite{terzan68} and \cite{terzan75} as variables, do not show signs of variability in our data, to within the limits of the rms scatter in our light curves; those limits are given in \Tab{tab:rms_nonvar}. We are also unable to detect variability of the five contact binaries listed by \cite{pietrukowicz04}; for most of these, we attribute this to the fact that the variations have amplitudes of $\sim 0.3$ mag, for objects with $V\sim20$ mag, which is very challenging to detect within the rms of our data (Fig. \ref{fig:rms}). This is especially difficult in the crowded central core, where blending leads to photon noise dominating any intrinsic variability signal. This is also true for the two brightest eclisping variables, which have $V$ magnitudes of $\sim$17.3 and $\sim$17.9 and variation amplitudes of $\sim0.3$ mag. The advent of electron-multiplying CCD (EMCCD) cameras, coupled with DIA, will allow us in the future to obtain high-quality photometry even for stars in the crowded cores of clusters \cite{skottfelt13}, and to verify the variability status of these objects using ground-based photometry.

\begin{table}
\begin{center}
  \begin{tabular}{ccccc}

     \hline
    \#		 	&$<V>$	&rms ($V$)	&$<I>$ 	&rms ($I$) \\
  \hline  

    V5			&17.42	&0.035	&16.75	&0.047\\
    V6			&15.34	&0.013	&15.13	&0.037\\
    V7			&16.63	&0.021	&15.84	&0.018\\
    V8			&15.00	&0.010	&14.31	&0.011\\
    V9			&15.40	&0.012	&15.08	&0.021\\
    V10		&15.44	&0.010	&15.22	&0.017\\
    V11		&16.11	&0.013	&15.25	&0.012\\
    V12		&15.49	&0.011	&15.38	&0.015\\
    V13		&15.69	&0.040	&14.78	&0.030\\

\hline \hline
  \end{tabular}
  \caption{Mean magnitudes and rms limits for the stars previously catalogued as variable, but for which we do not find variability. \label{tab:rms_nonvar}}
  \end{center}
\end{table}

\subsection{Detection of known variables}

We recover the first three known RR Lyrae in this cluster (V1-V3) in our data, and calculate periods for each of them, using phase dispersion minimisation (PDM; \citealt{stellingwerf78}) and the ``string length" method \citep{lafler65}. In order to use the longest possible baseline to derive precise periods for these three variables, we used the data from \cite{rosino49}; these data are published in a table in that paper, and we provide them as an additional resource with the electronic version of this paper\footnote{The light curves for V1-V3 from \cite{rosino49}, as well as the additional data for V1 from \cite{rosino61}, are available in electronic form at the CDS via anonymous ftp to \tt{cdsarc.u-strasbg.fr (130.79.128.5)}
or via \tt{http://cdsweb.u-strasbg.fr/cgi-bin/qcat?J/A+A/}}. We also used these data to refine our periods by optimising the alignment in phase of that data set with ours; this is highly sensitive to the period, thanks to the baseline of $\sim$63 years. We also recovered the two additional RR Lyrae stars reported by \cite{pietrukowicz04} and assign them catalogue numbers V15 and V19, and calculated refined periods for these two stars. We note that V19 is highly blended due to its location in the very centre of the cluster, which explains its peculiar position on the CMD.

We also detect the U Gem variable V4, including some data taken during an outburst period. This star is discussed in more detail in \Sec{sec:individual}.

\subsection{Detection of new variables}

As in our previous studies, we searched for new variables using three methods. Firstly, we constructed a stacked image $S$ consisting of the sum of the absolute values of the deviations $D$ of each image from the convolved reference image, divided by the pixel uncertainty $\sigma$, so that

\begin{equation}
S_{ij} = \sum_{k}\frac{|D_{kij}|}{\sigma_{kij}} \, .
\end{equation}

\begin{figure}
  \centering
  \includegraphics[width=8cm, angle=0]{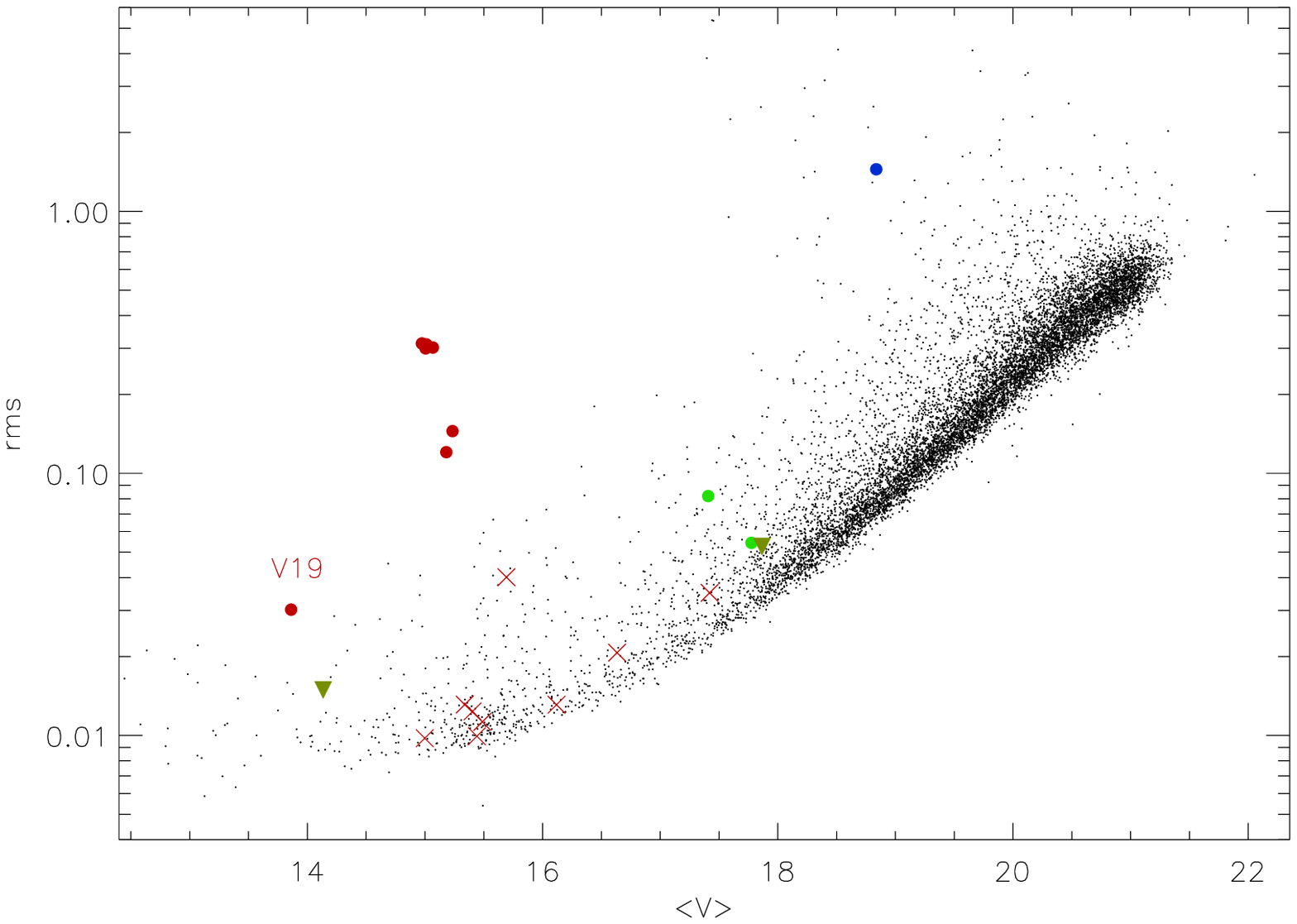}
  \includegraphics[width=8cm, angle=0]{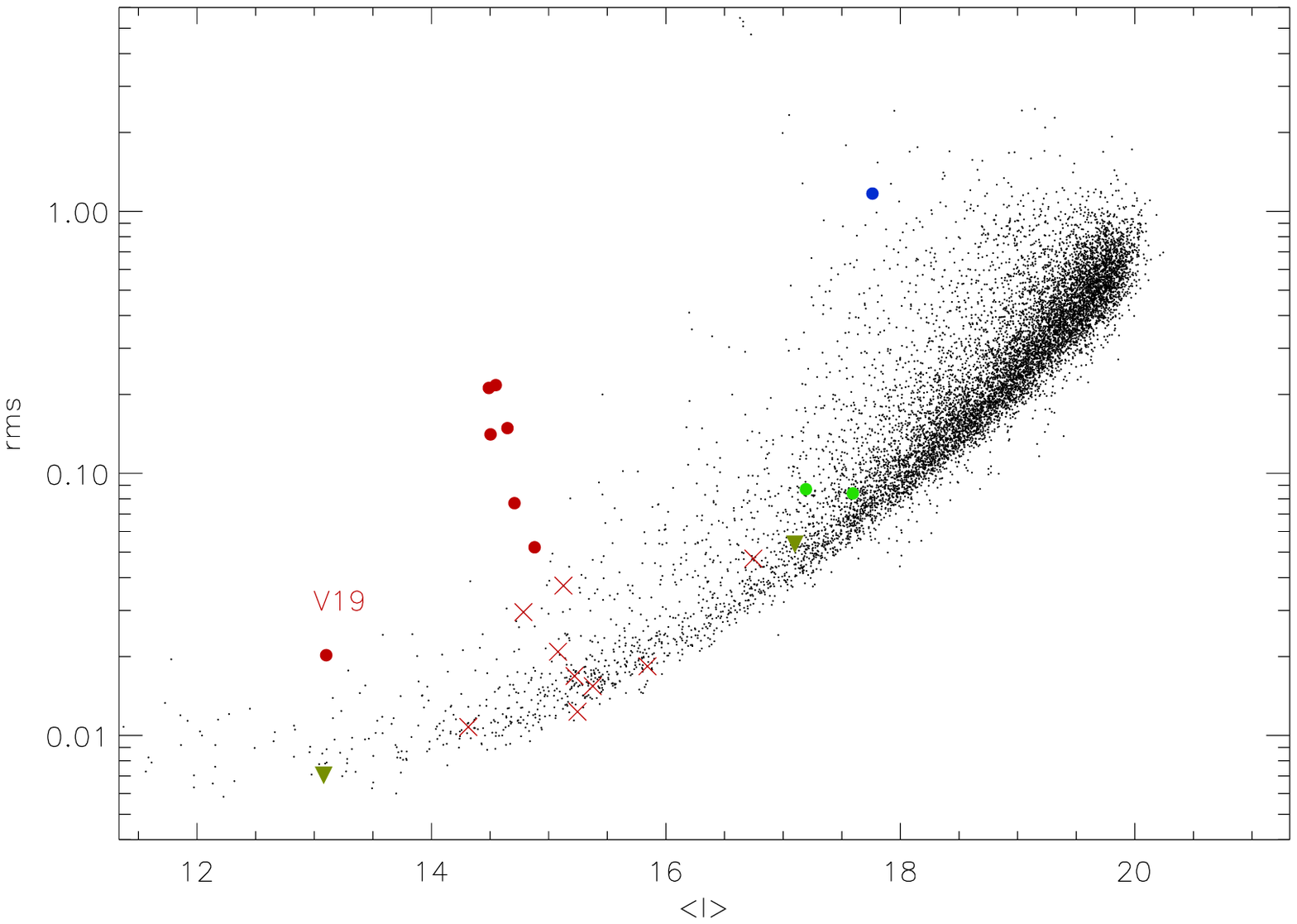}
  \caption{Plot of root mean square magnitude deviation versus mean magnitude for all stars for which photometry was obtained. Plots are for the $V-$band (top) and $I-$band (bottom). Classified variables are marked as filled circles, with RR Lyrae in red, SX Phoenicis and blue stragglers in light green, U Geminorum in blue, and variables of unknown type as dark green triangles. Non-variable objects previously catalogued as variable in the literature are marked with red crosses. \label{fig:rms}}

\end{figure}

Stars that deviate consistently from the reference image then stand out in this stacked image. Using this method, we discovered 2 new RR Lyrae stars, V14 and V16, both of RR1 type, with V16 also showing signs of amplitude and period modulation due to Blazhko effects or non-radial pulsation (see Sec. \ref{sec:individual}). Secondly, we inspected the light curves of objects which stand out on a plot of root mean square magnitude deviation versus mean magnitude, shown in \Fig{fig:rms}.

\begin{figure}
  \centering
  \includegraphics[width=8cm, angle=0]{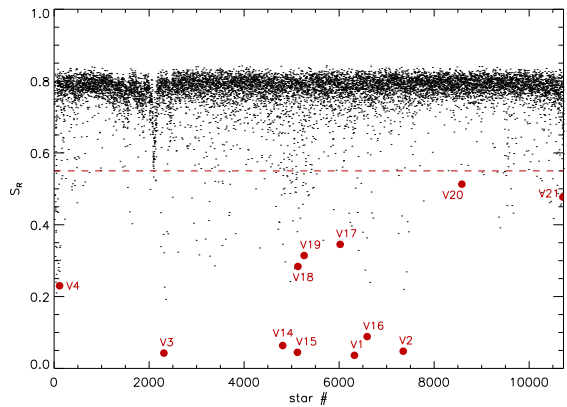}
  \caption{Plot of the distribution of the $S_R$ statistic as defined in the text, for our $V$-band light curves. \label{fig:sq}}

\end{figure}

Finally, we also searched for variables by conducting a period search for all our light curves using the string length method, and computing the ratio $S_R$ of the string length for the best- to that of the worst-fit periods, with the smallest ratios expected for true periodic variations. We inspected visually all of the light curves of stars with $S_R < 0.55$, where the threshold value of 0.55 was chosen by inspecting the distribution of $S_R$ (see \Fig{fig:sq}). Using this method, we discovered V17, V18, V20 and V21. 

\begin{figure*}
  \centering
  \includegraphics[width=14cm, angle=0]{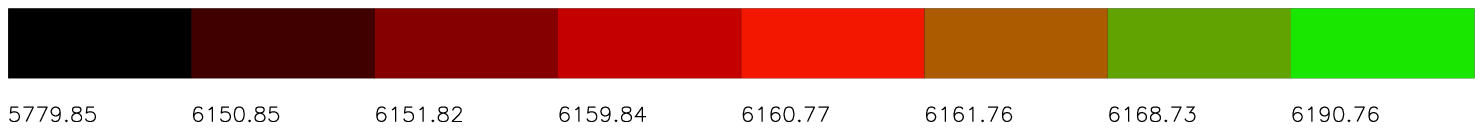}   \\
  \includegraphics[width=6cm, angle=0]{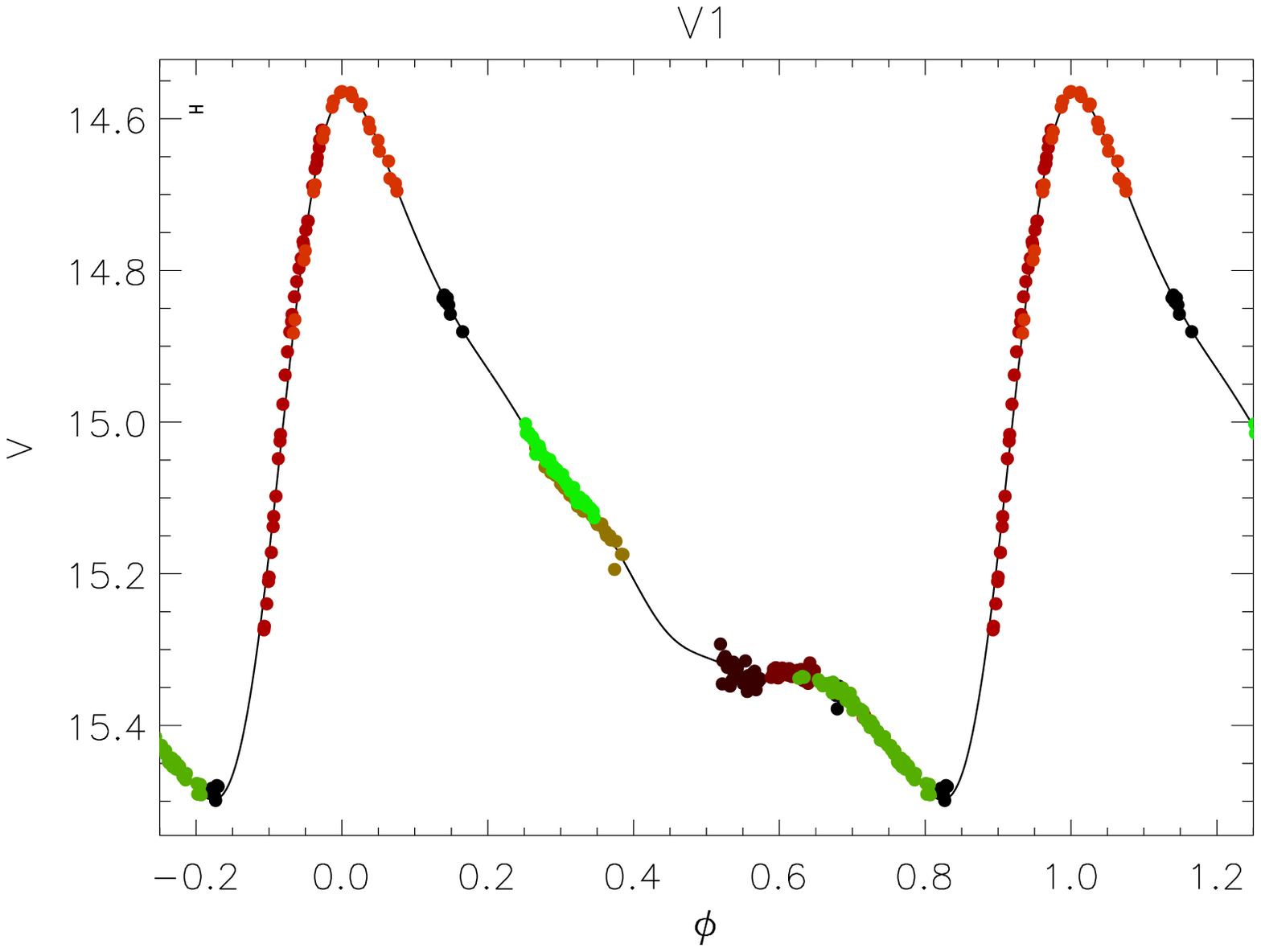}
  \includegraphics[width=6cm, angle=0]{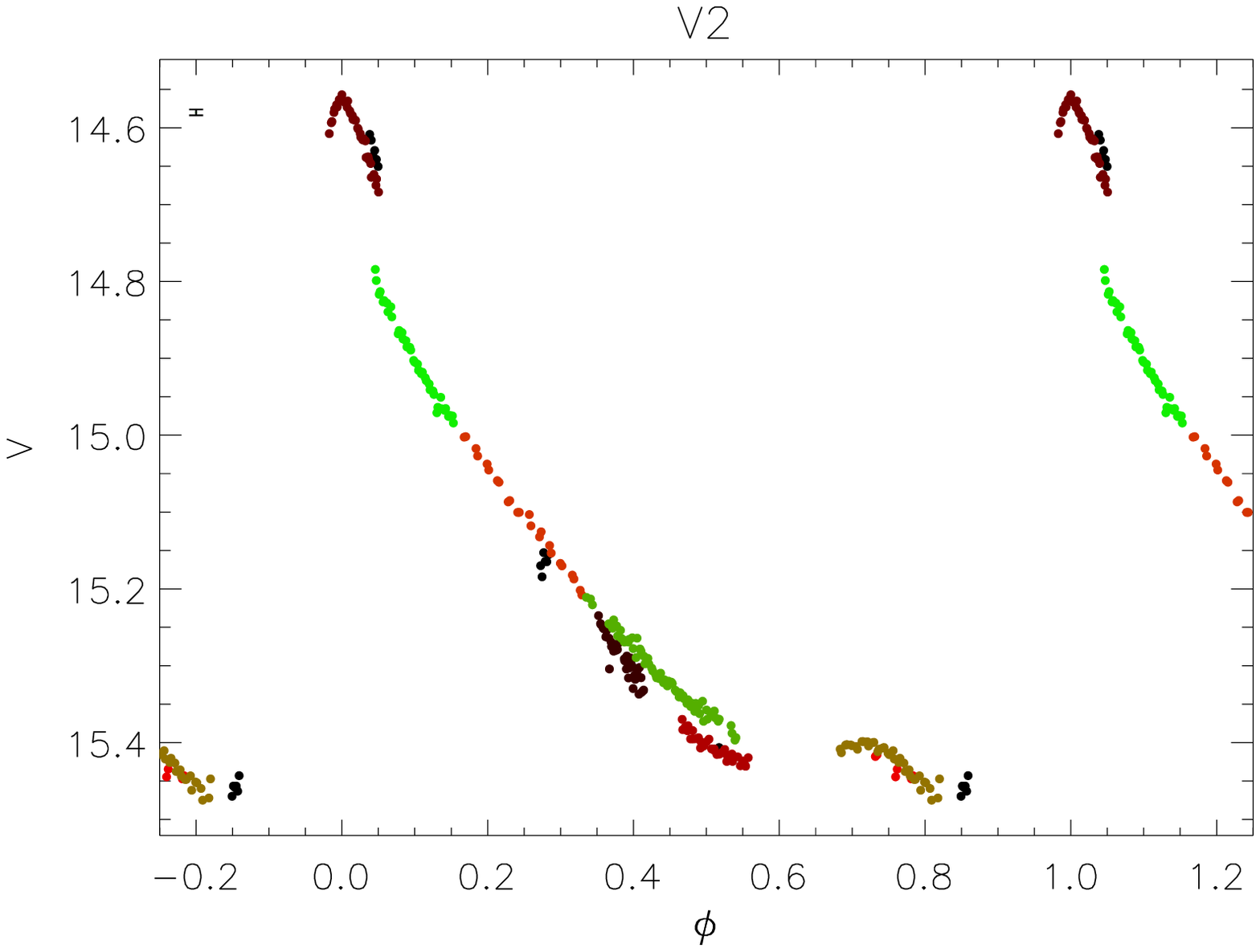}
  \includegraphics[width=6cm, angle=0]{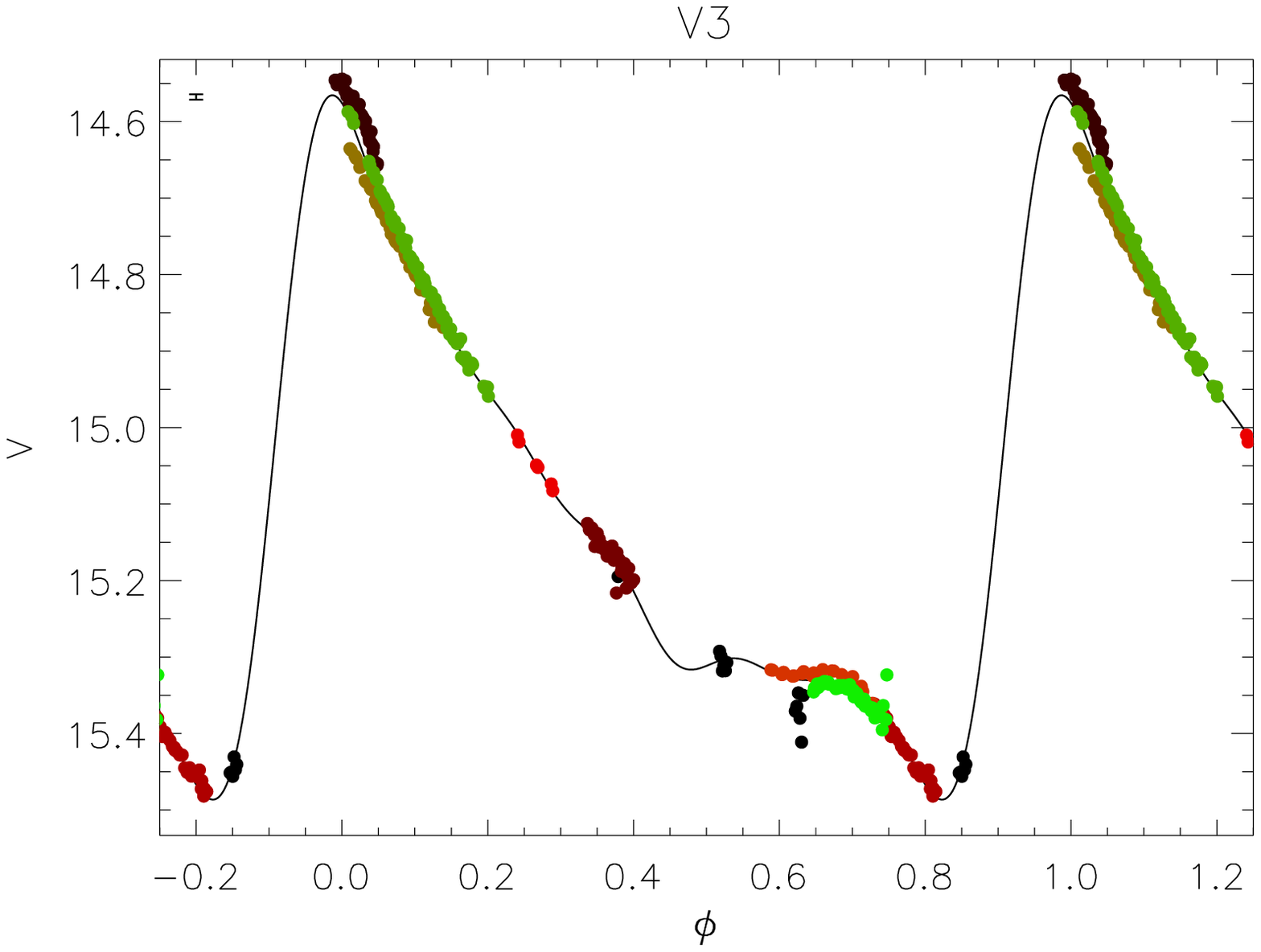}
  \includegraphics[width=6cm, angle=0]{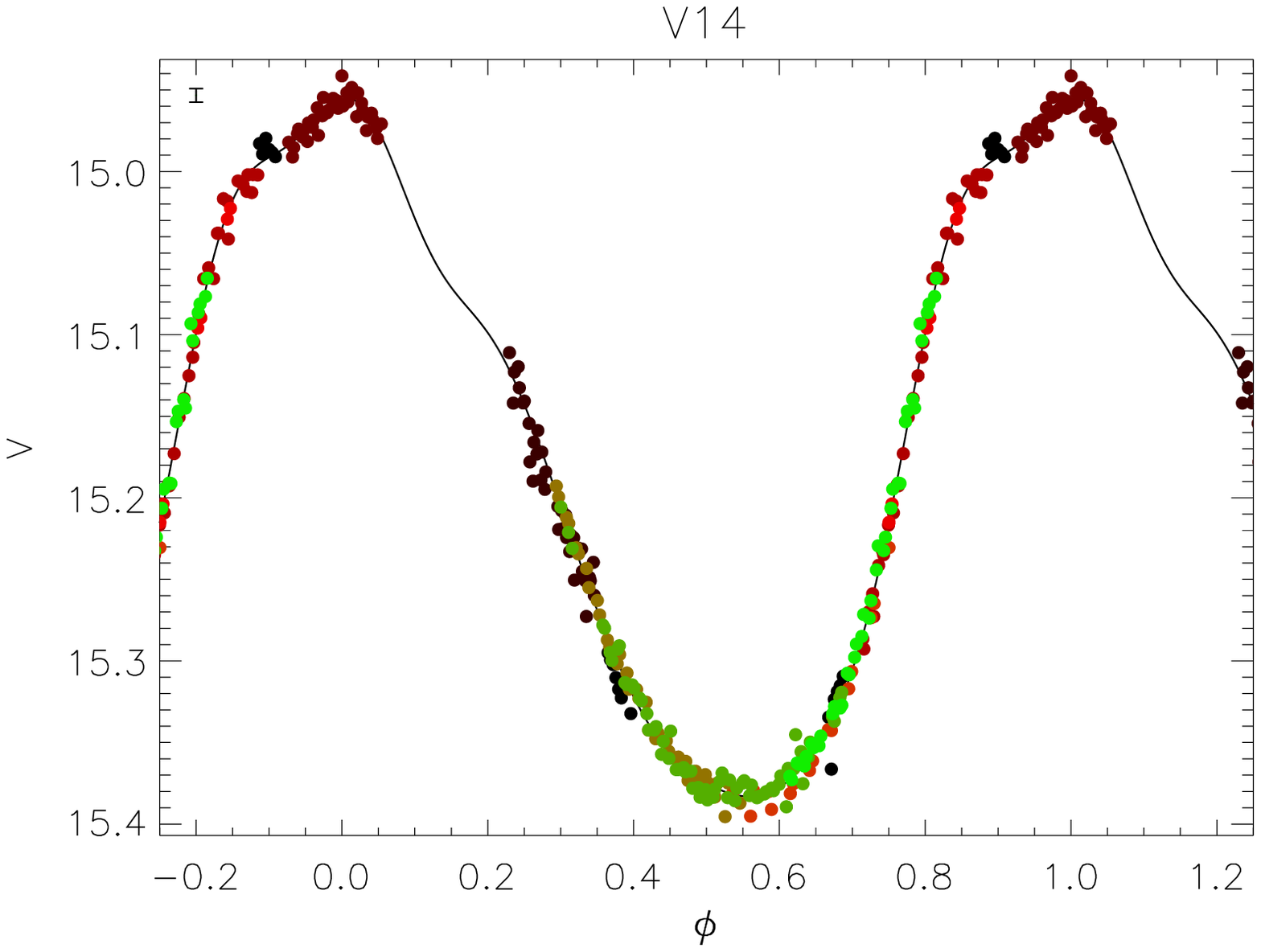}
  \includegraphics[width=6cm, angle=0]{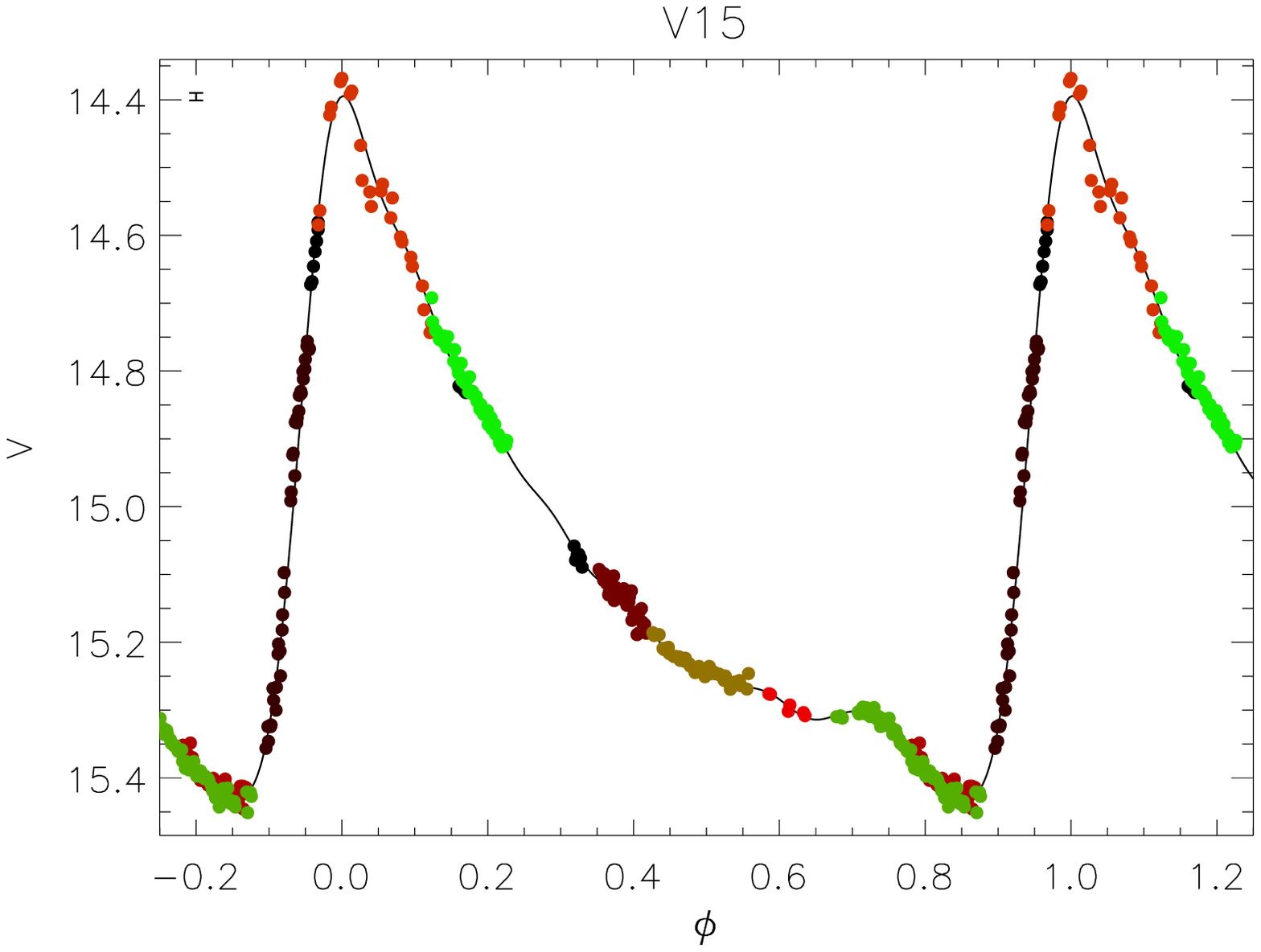}
  \includegraphics[width=6cm, angle=0]{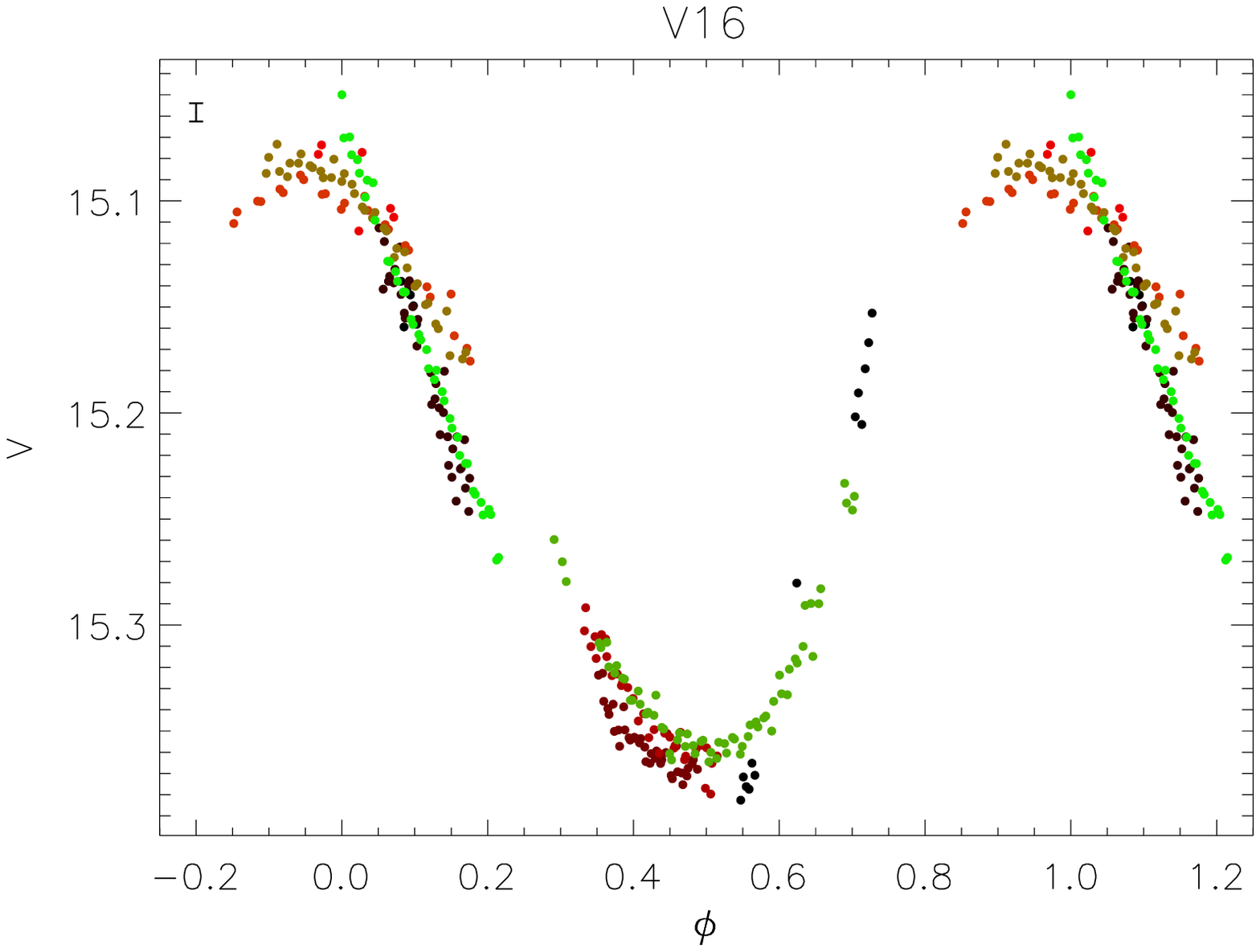}   
  \includegraphics[width=6cm, angle=0]{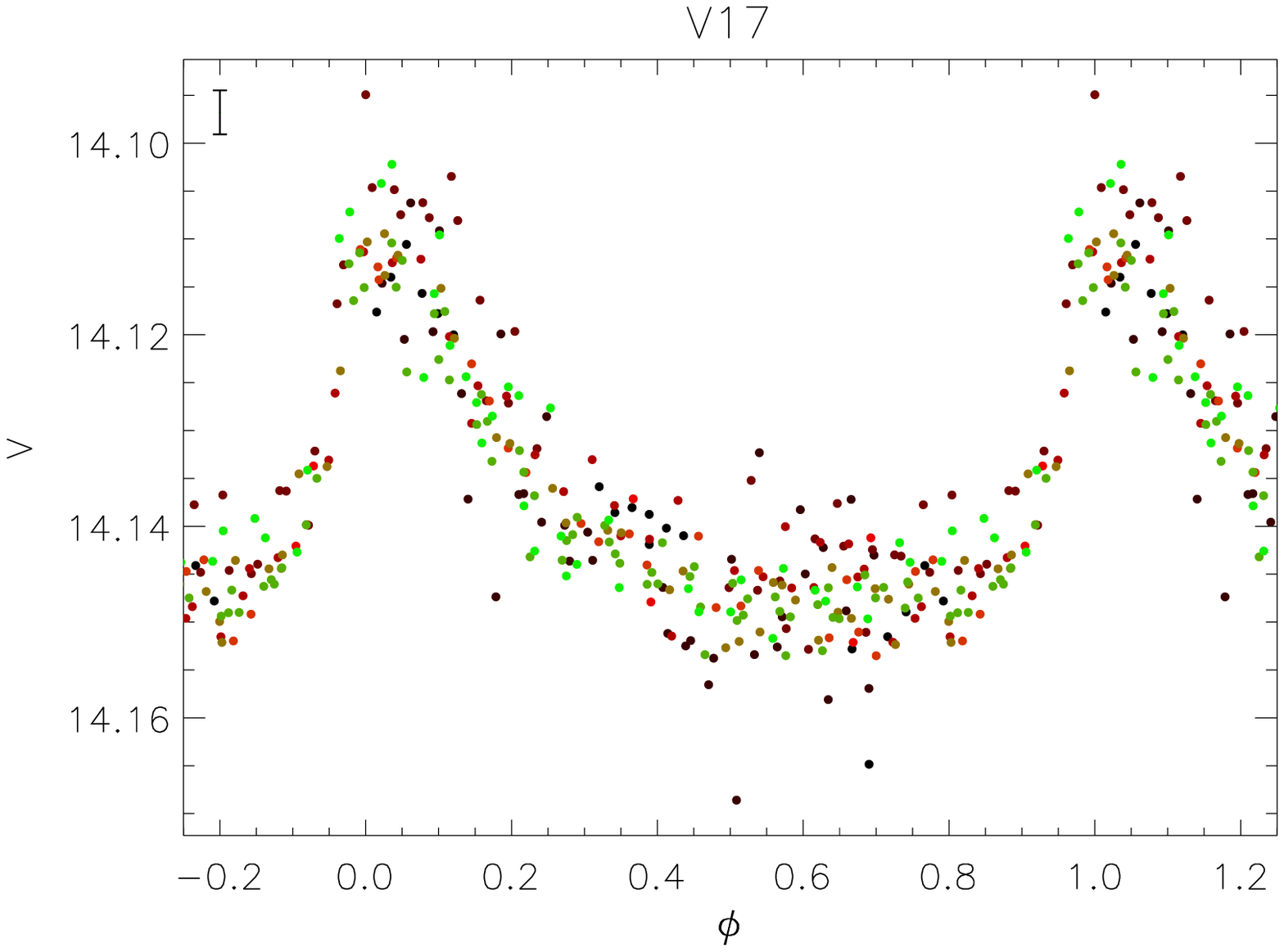}   
  \includegraphics[width=6cm, angle=0]{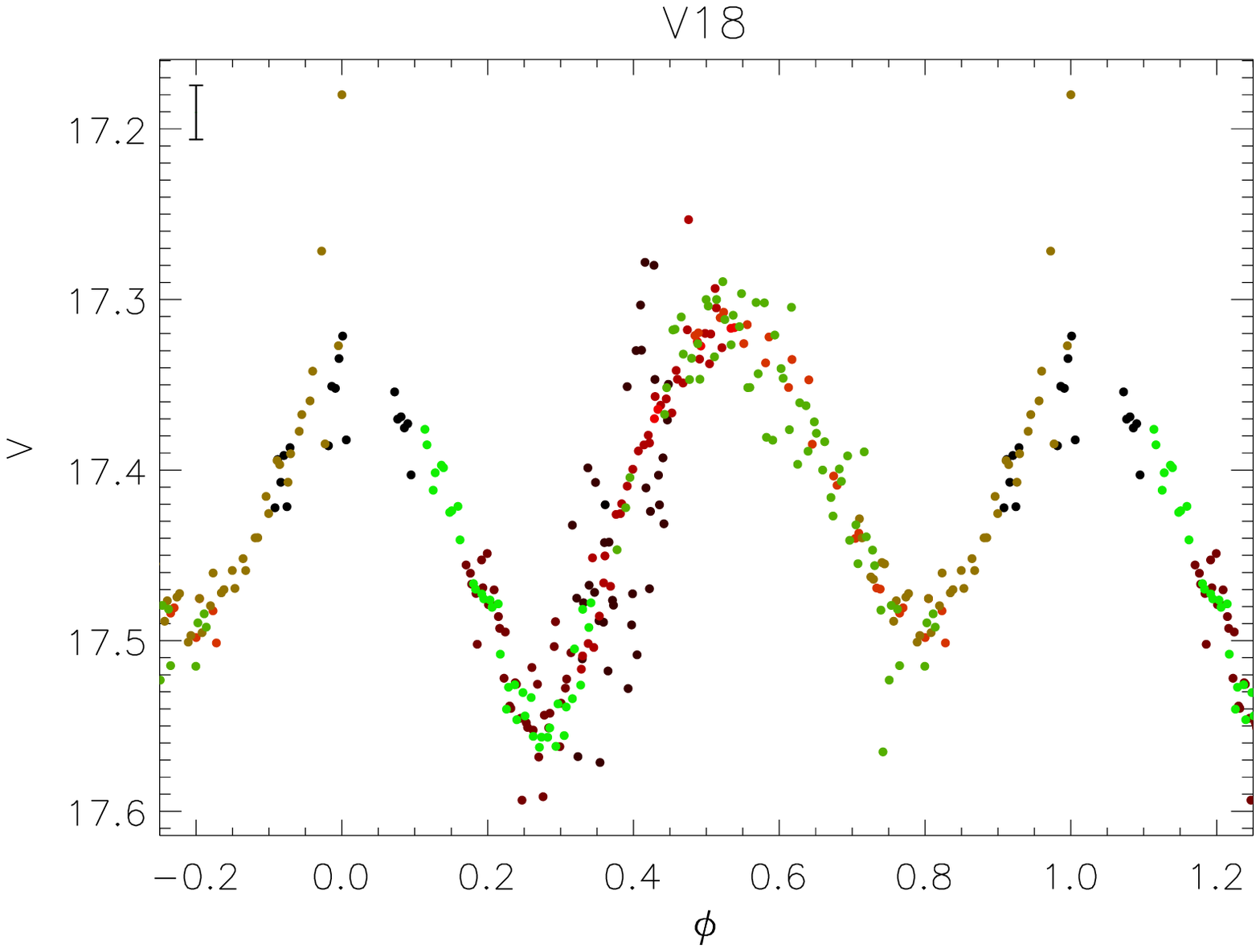}   
  \includegraphics[width=6cm, angle=0]{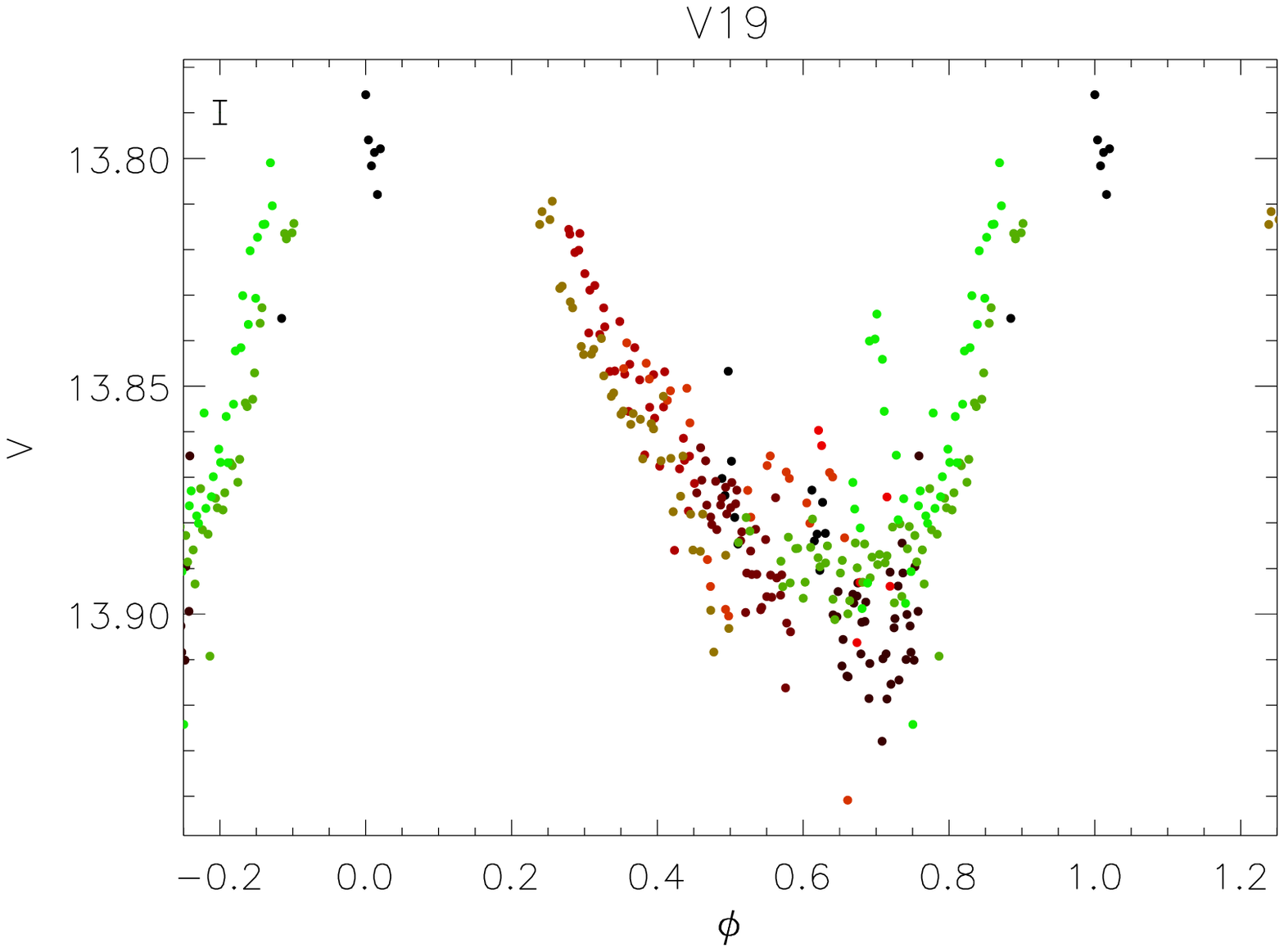}   
  \includegraphics[width=6cm, angle=0]{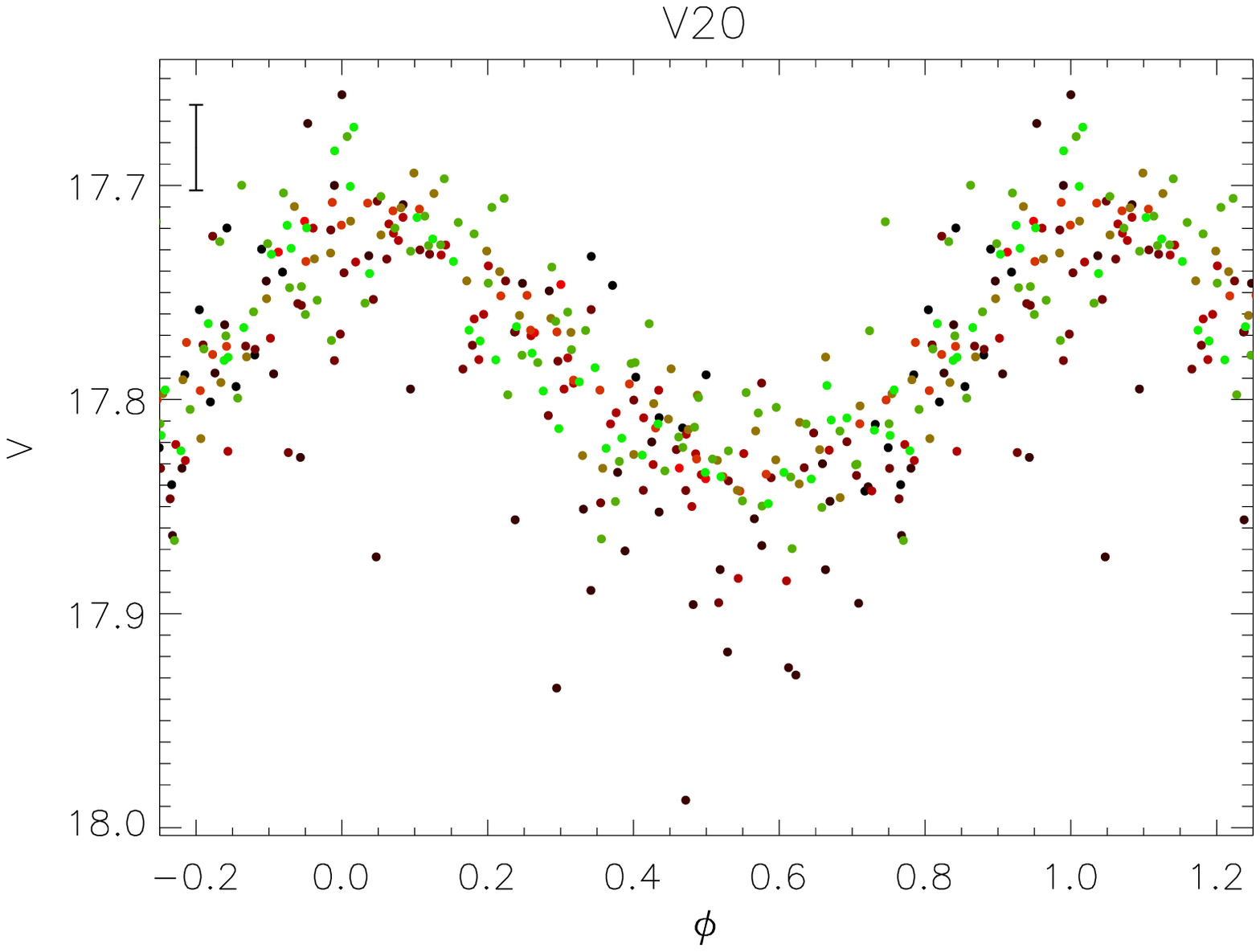}   
  \includegraphics[width=6cm, angle=0]{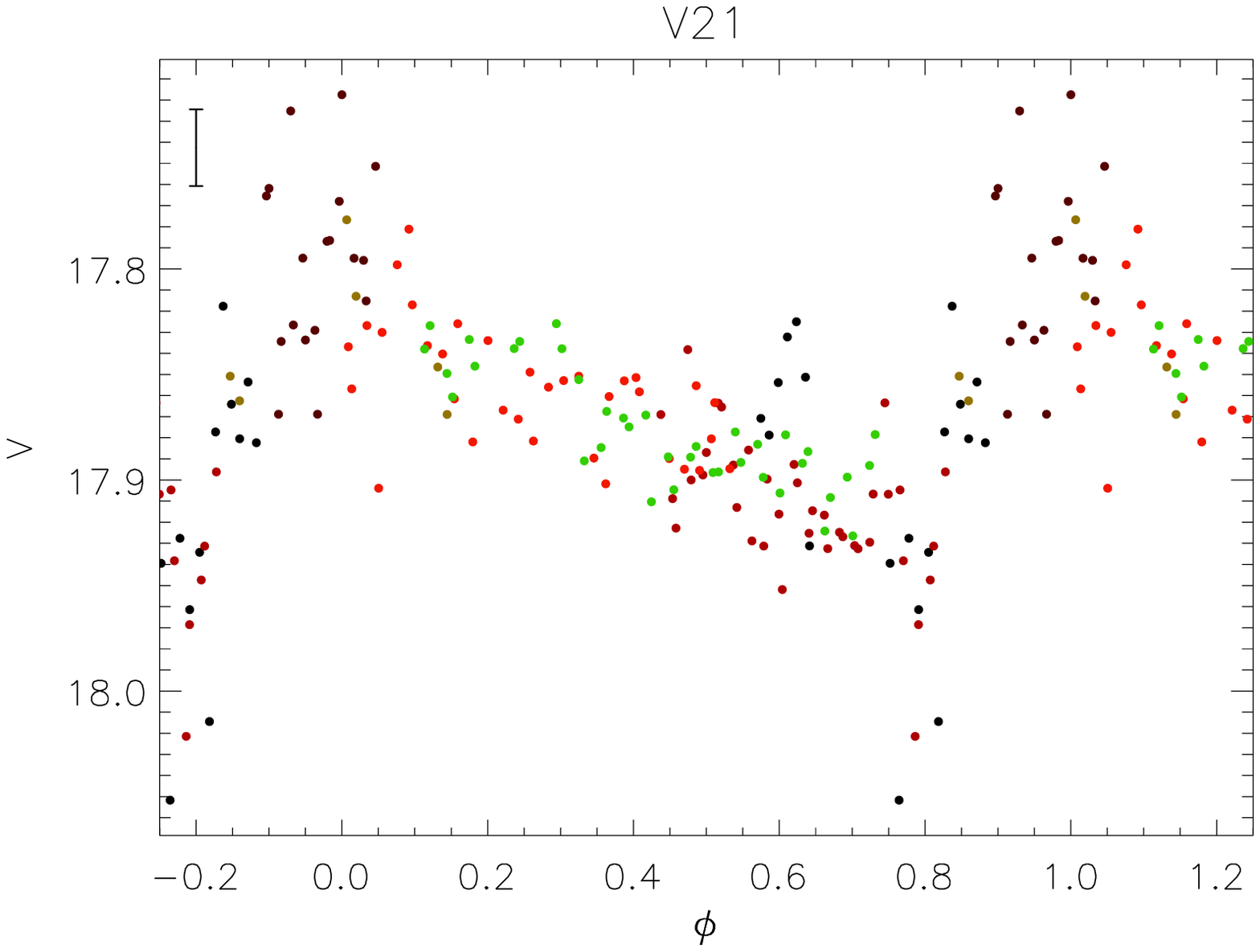}   

  \caption{Phased V-band light curves of the variables with a period estimate. Different colours are used for different nights, with the colour coding given in the form of a colour bar spanning the whole time range of the observations (top panel). Fourier decomposition fits are overplotted for the objects for which those were successful. \label{fig:lc_V}}

\end{figure*}

\begin{figure*}
  \centering
  \includegraphics[width=14cm, angle=0]{fig/colourcode.ps}   \\
  \includegraphics[width=6cm, angle=0]{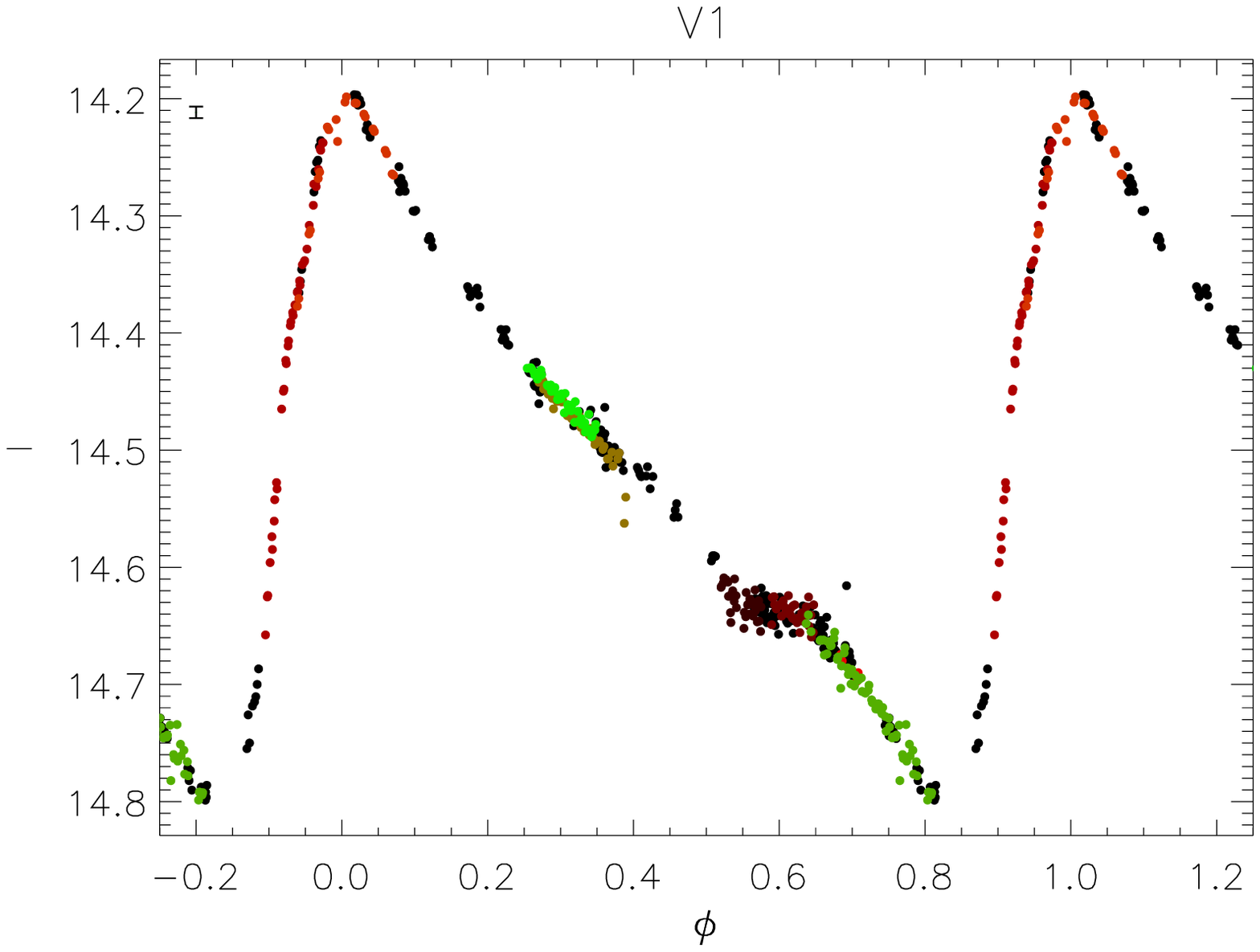}
  \includegraphics[width=6cm, angle=0]{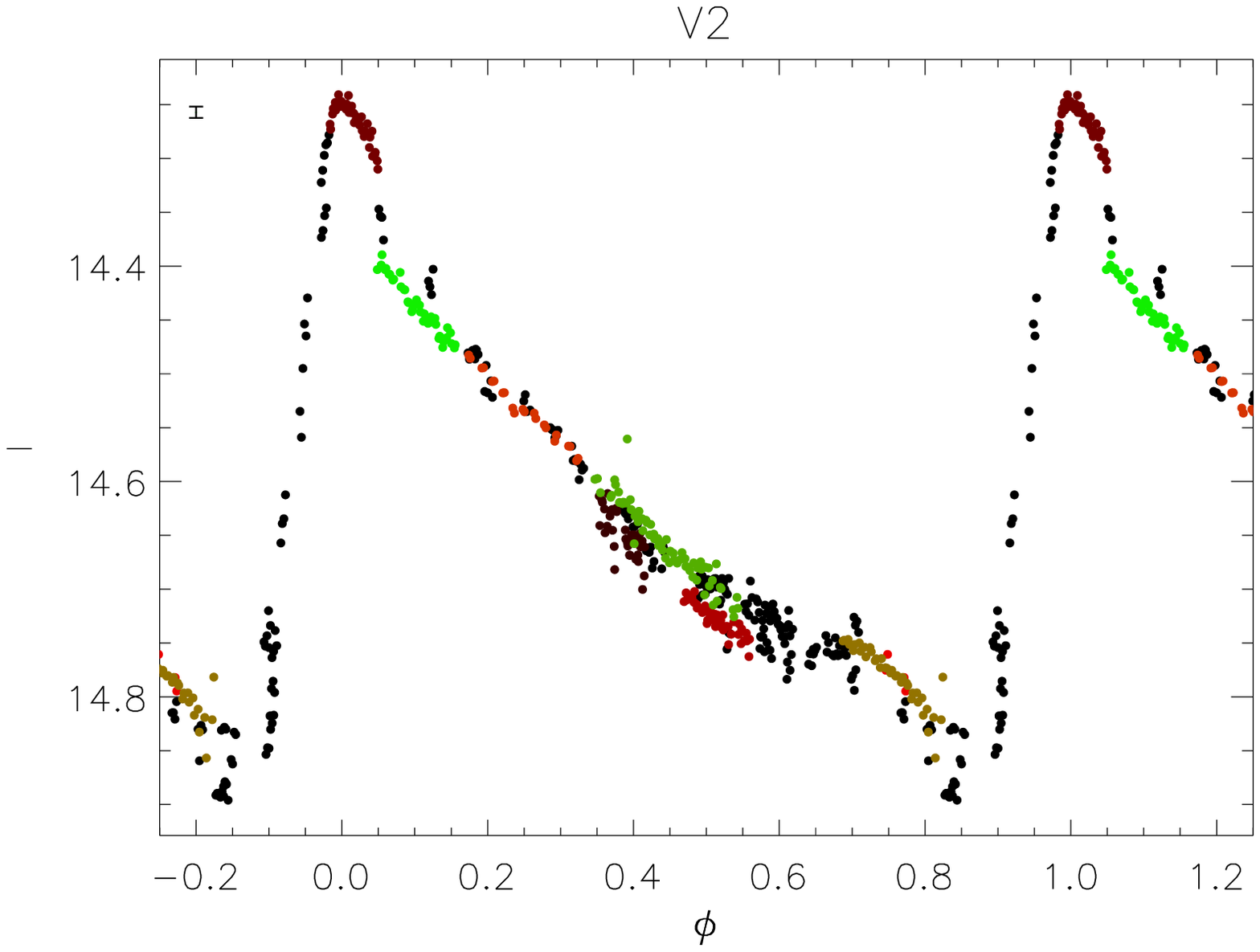}
  \includegraphics[width=6cm, angle=0]{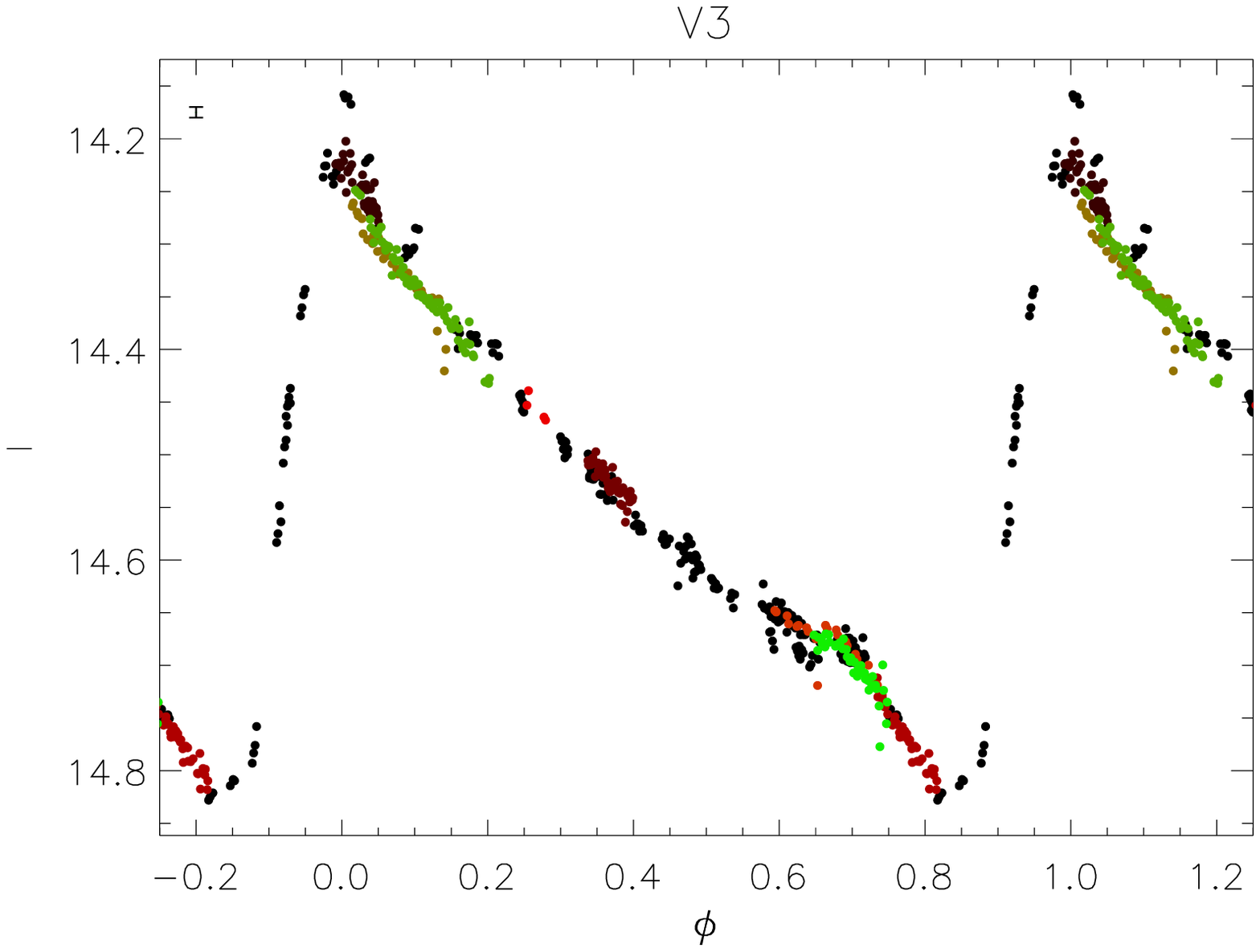}
  \includegraphics[width=6cm, angle=0]{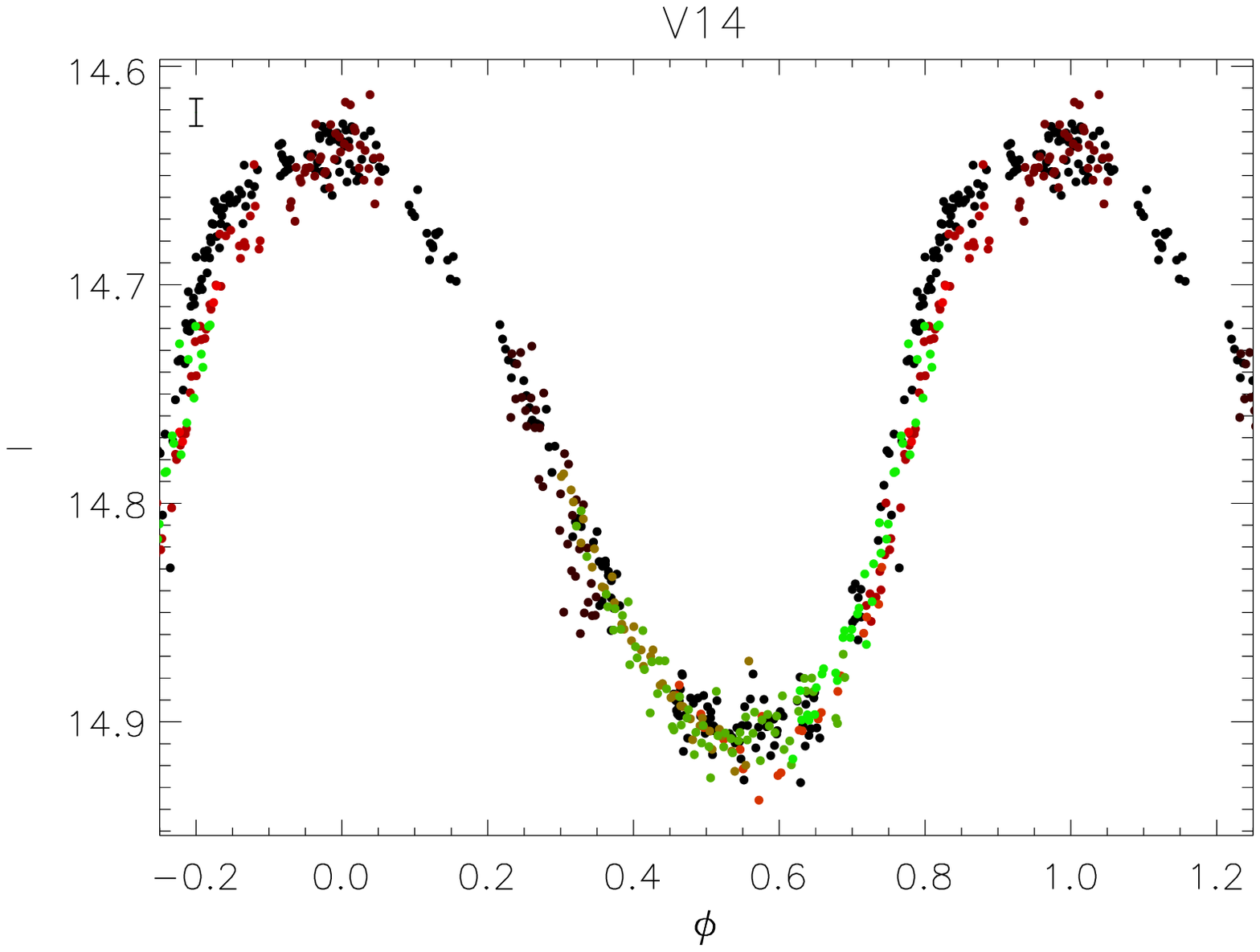}
  \includegraphics[width=6cm, angle=0]{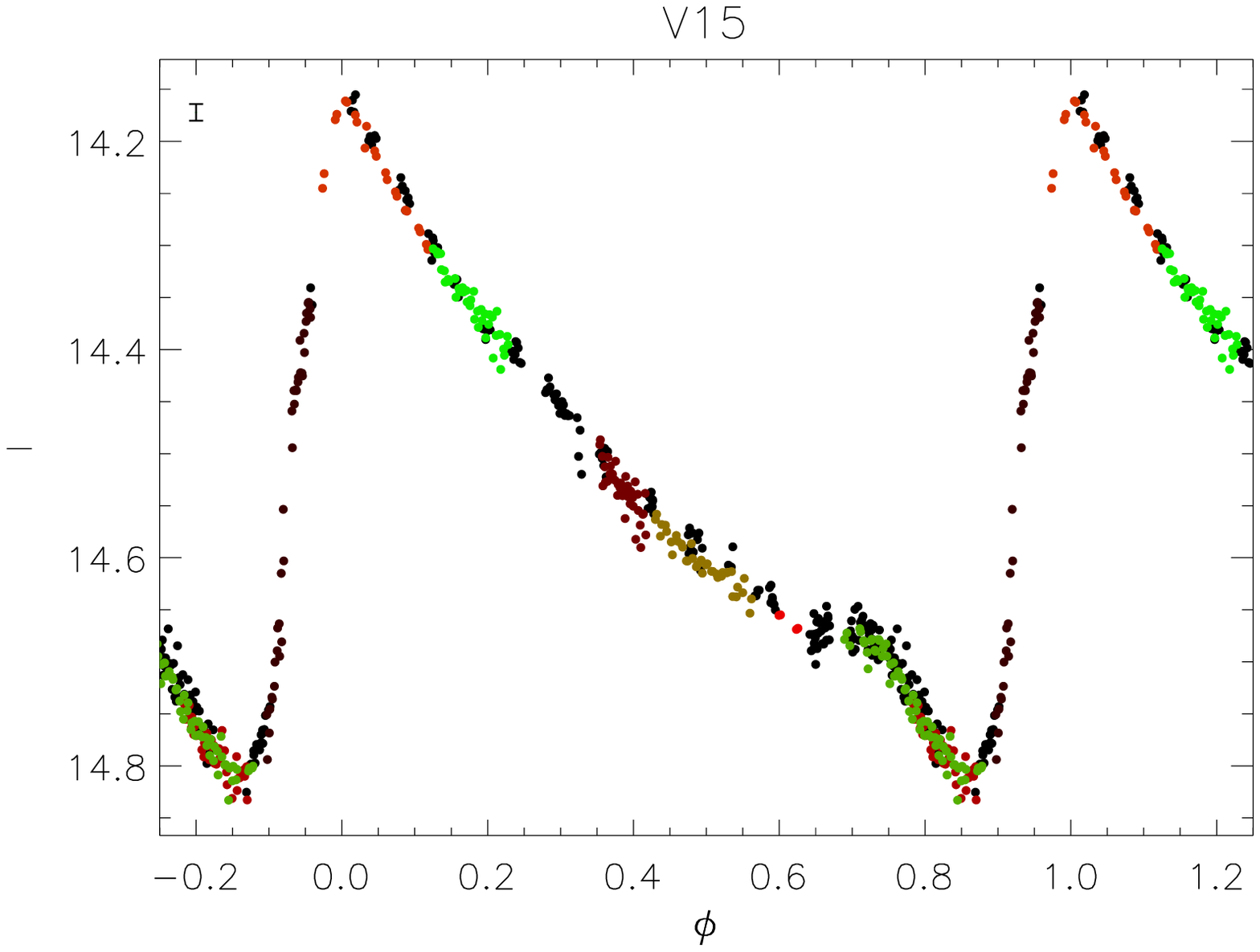}
  \includegraphics[width=6cm, angle=0]{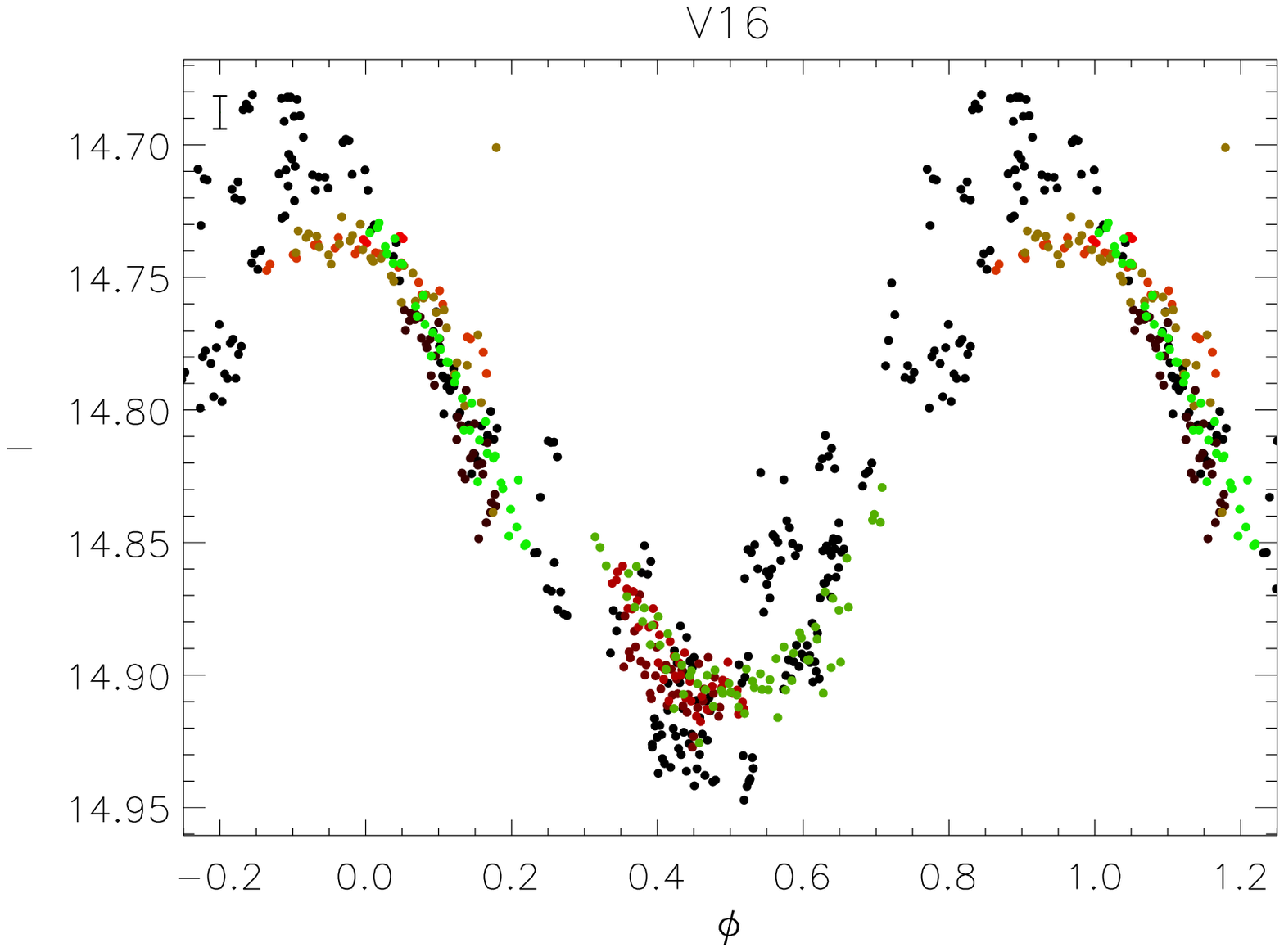}   
  \includegraphics[width=6cm, angle=0]{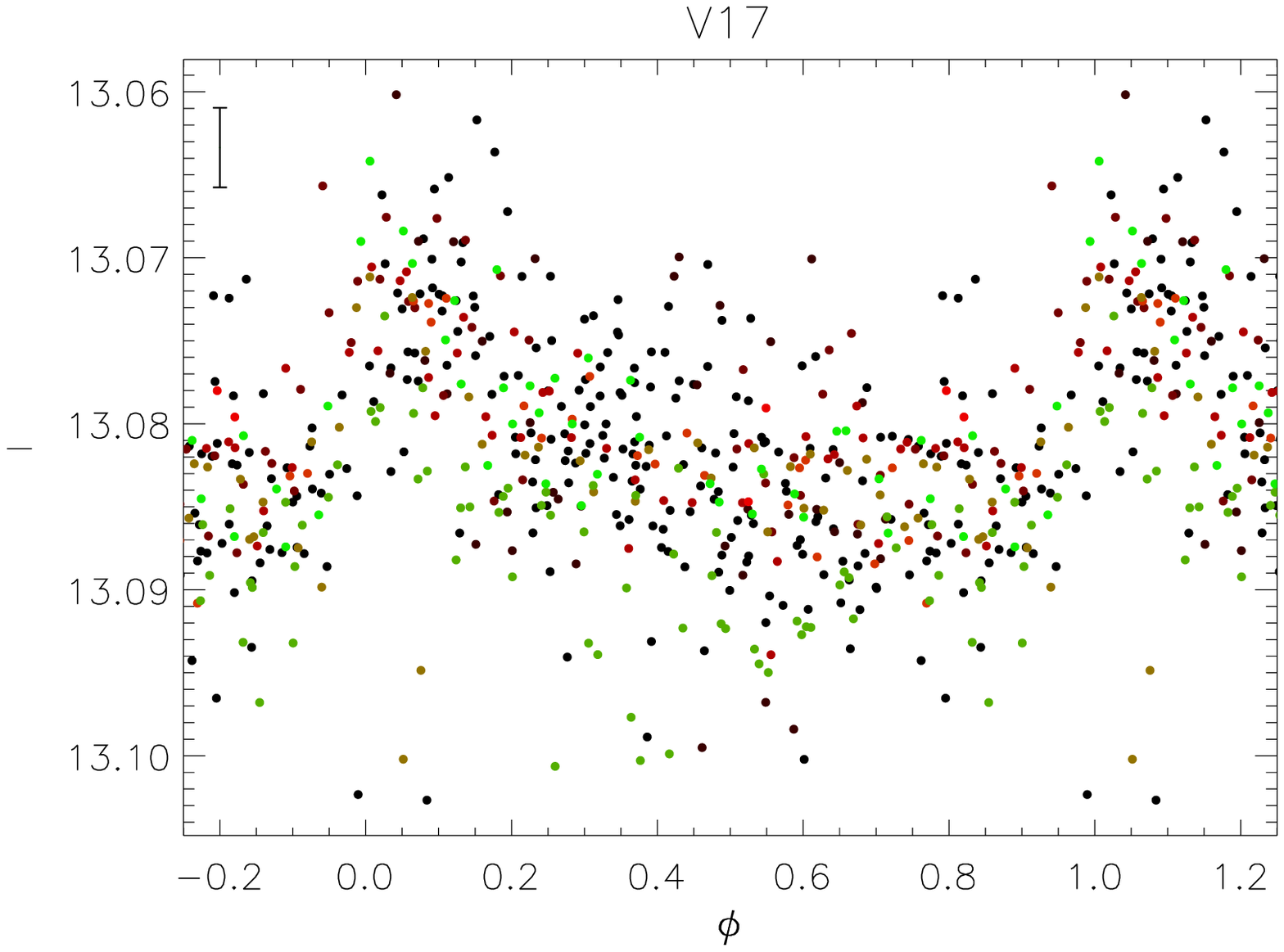}   
   \includegraphics[width=6cm, angle=0]{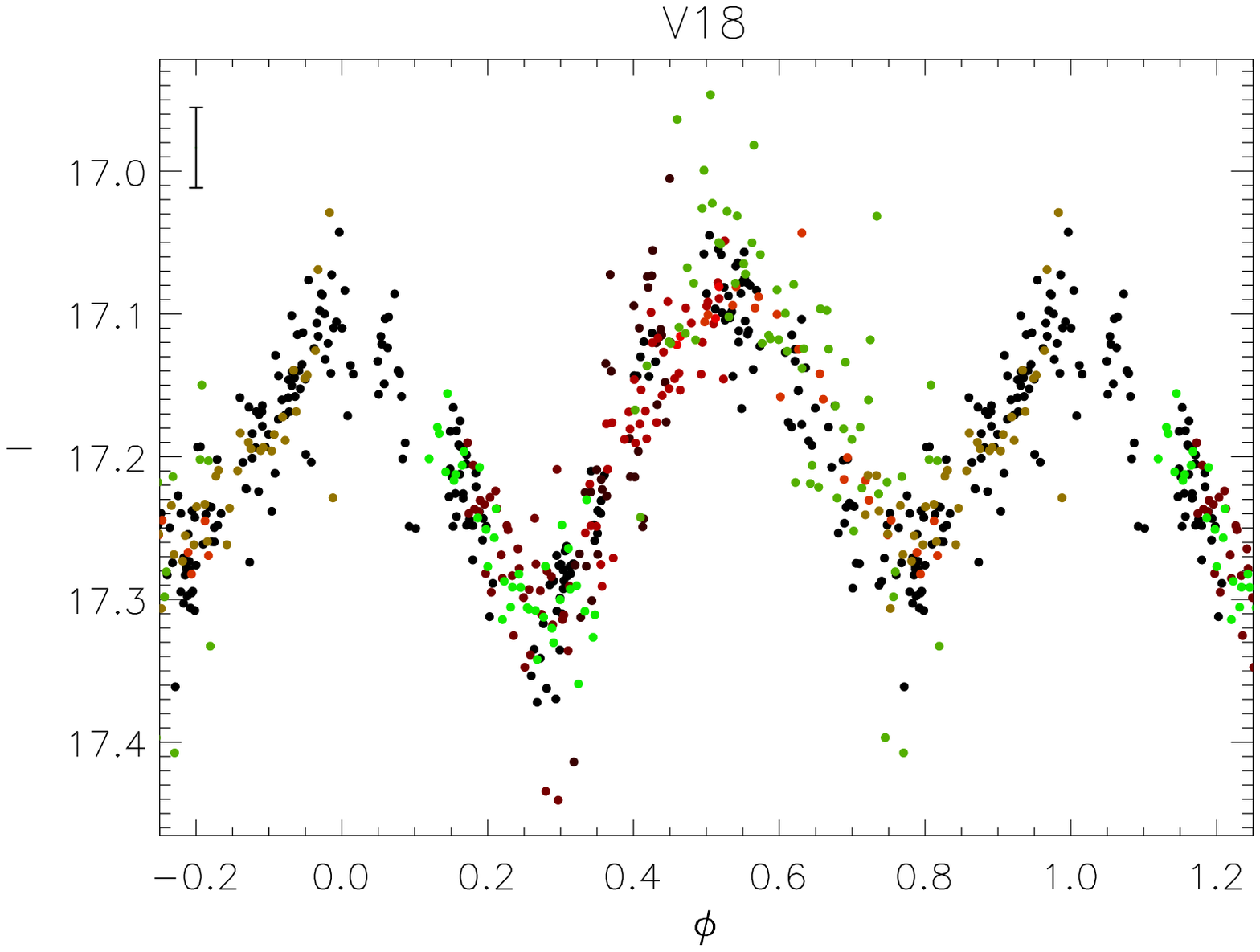}   
 \includegraphics[width=6cm, angle=0]{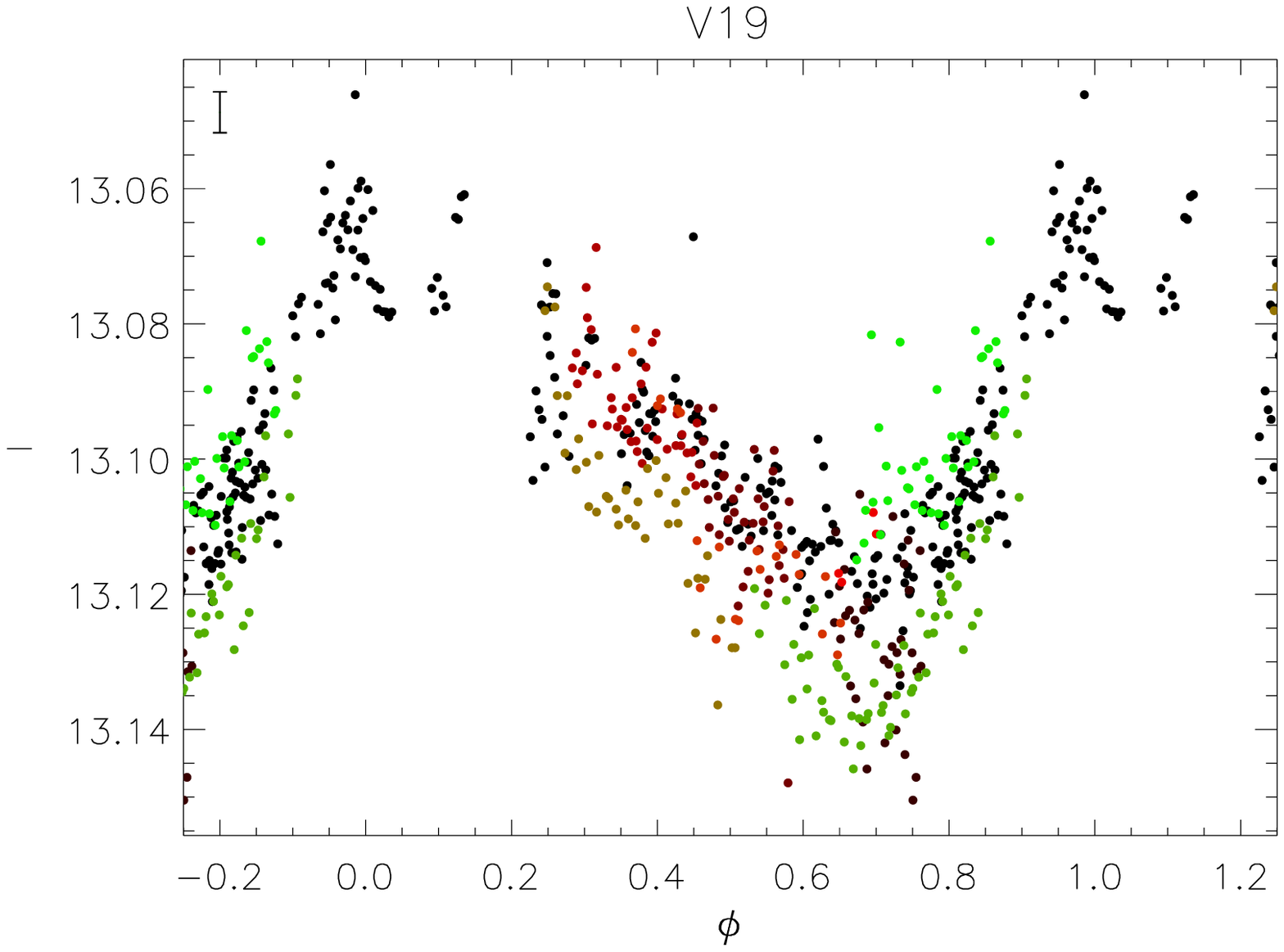}   
  \includegraphics[width=6cm, angle=0]{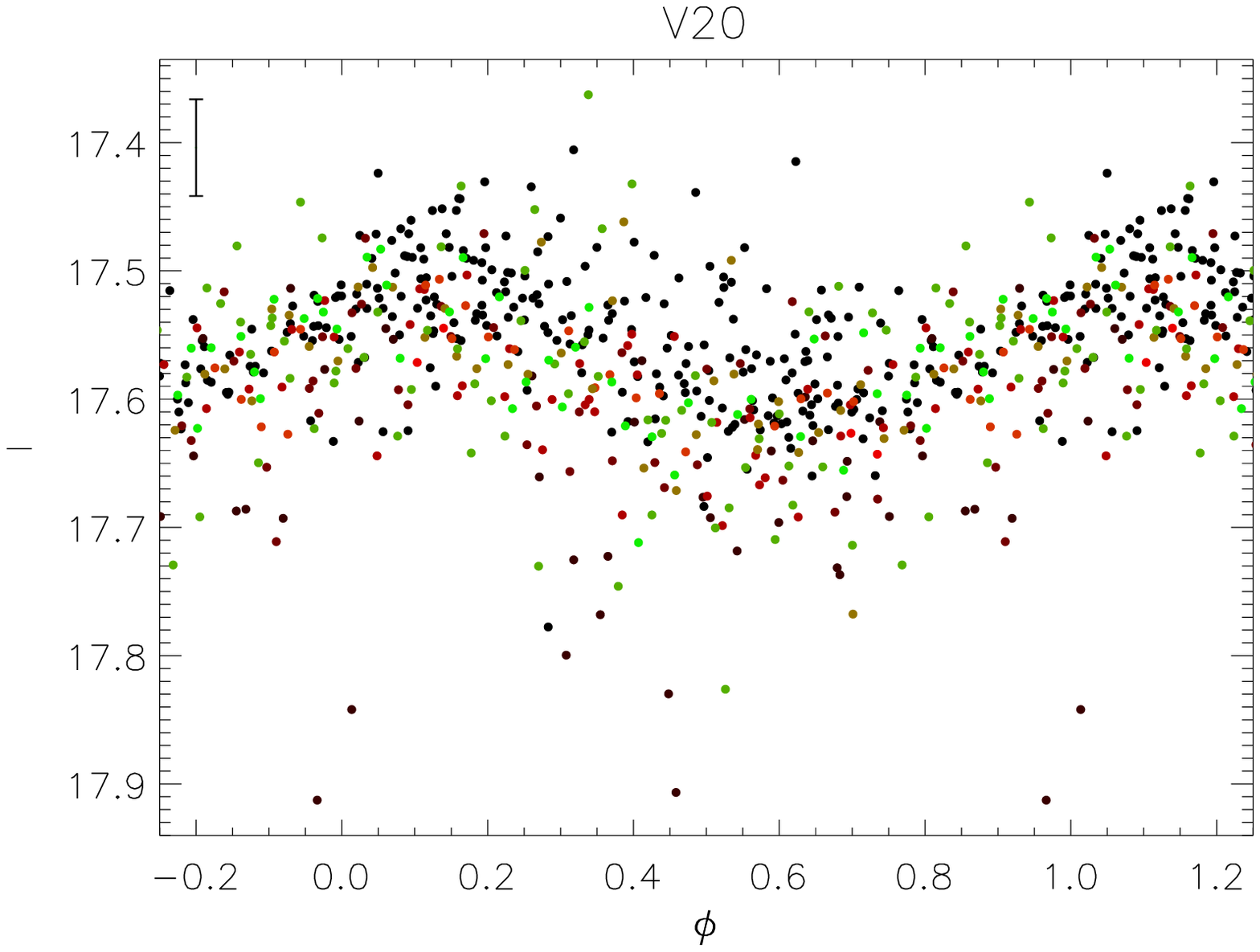}   
  \includegraphics[width=6cm, angle=0]{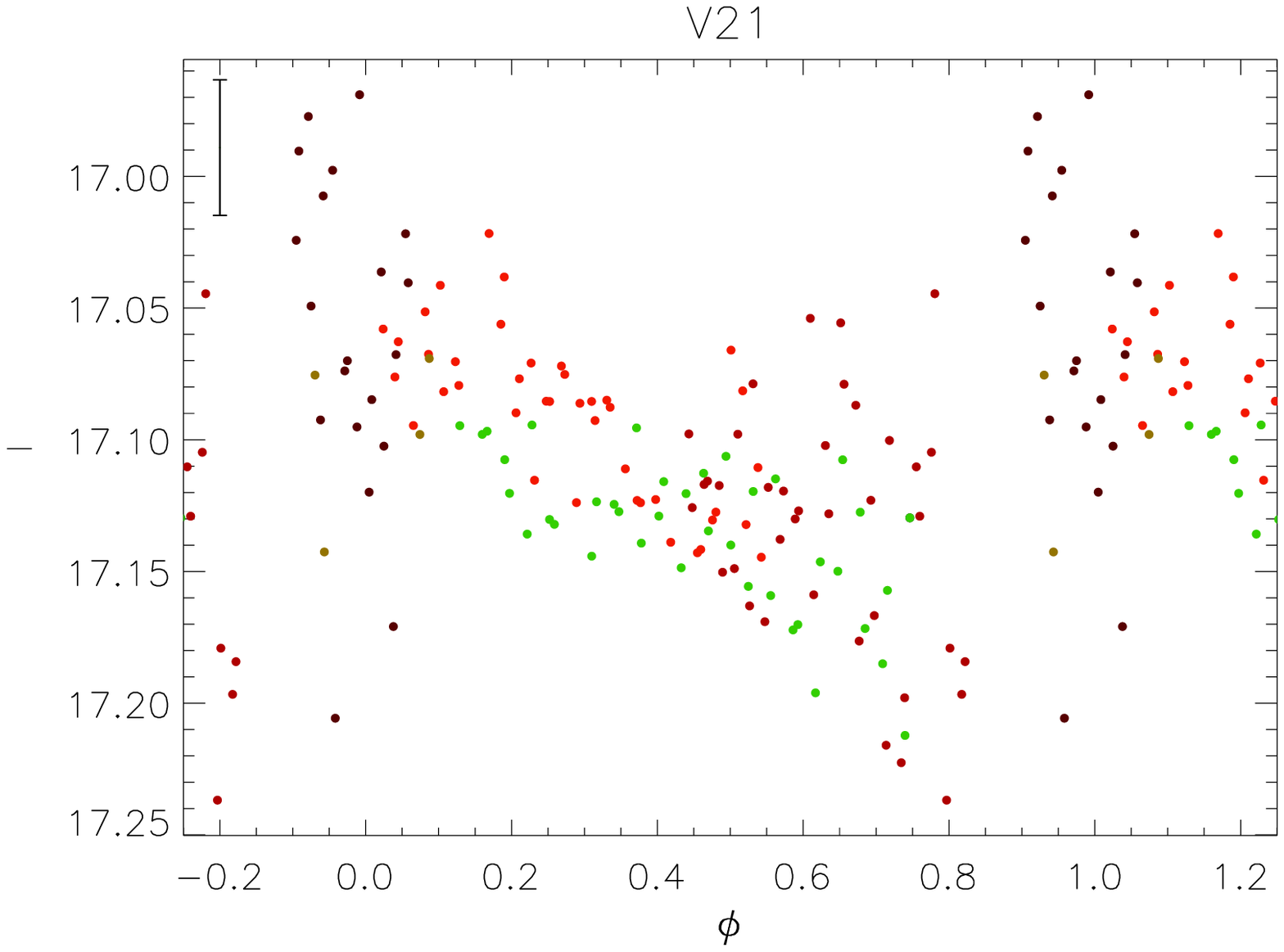}   

  \caption{Same as \Fig{fig:lc_V} but for the $I$ band. For V21, we only plot data points with error $\sigma < 0.3$ mag and do not plot 2011 data due to poor photometry. \label{fig:lc_I}}

\end{figure*}

\begin{figure*}
  \centering
  \includegraphics[width=19cm, angle=0]{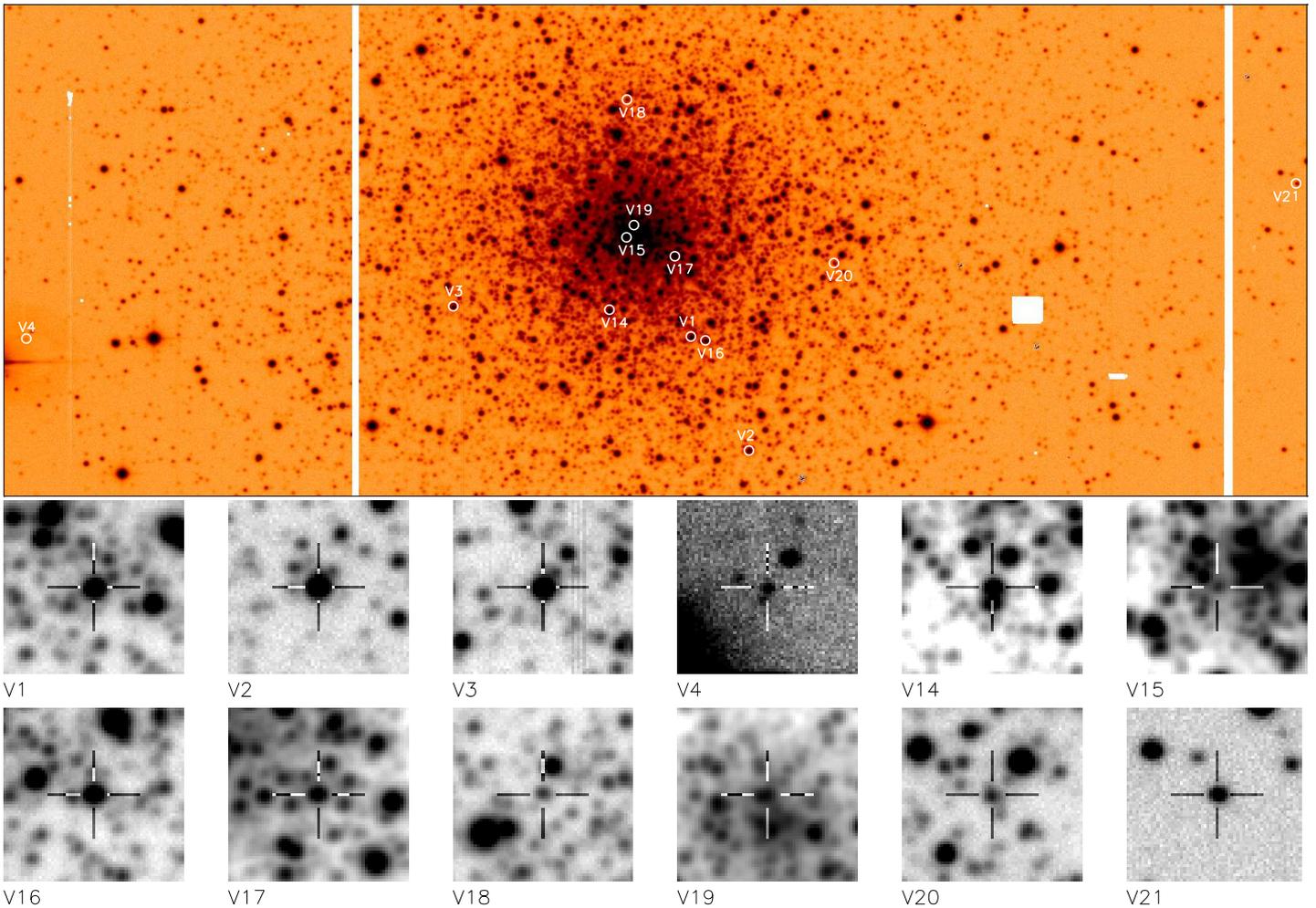}
  \caption{Finding chart for the confirmed variable objects in M 30, using our $V$ reference image. North is up and East is to the right. The image size is 11.57 $\times$ 4.55 arcmin$^2$, while each stamp size is 23.4 $\times$ 23.4 arcsec$^2$. White circles and labels indicate the locations of the variables, and each of the variables we detect in our data is shown with a crosshair at the centre of an individual stamp. Note that the display scale of each stamp is different in order to make the source visible. \label{fig:fchart}}

\end{figure*}

\begin{figure*}
  \centering
  \includegraphics[width=14cm, angle=0]{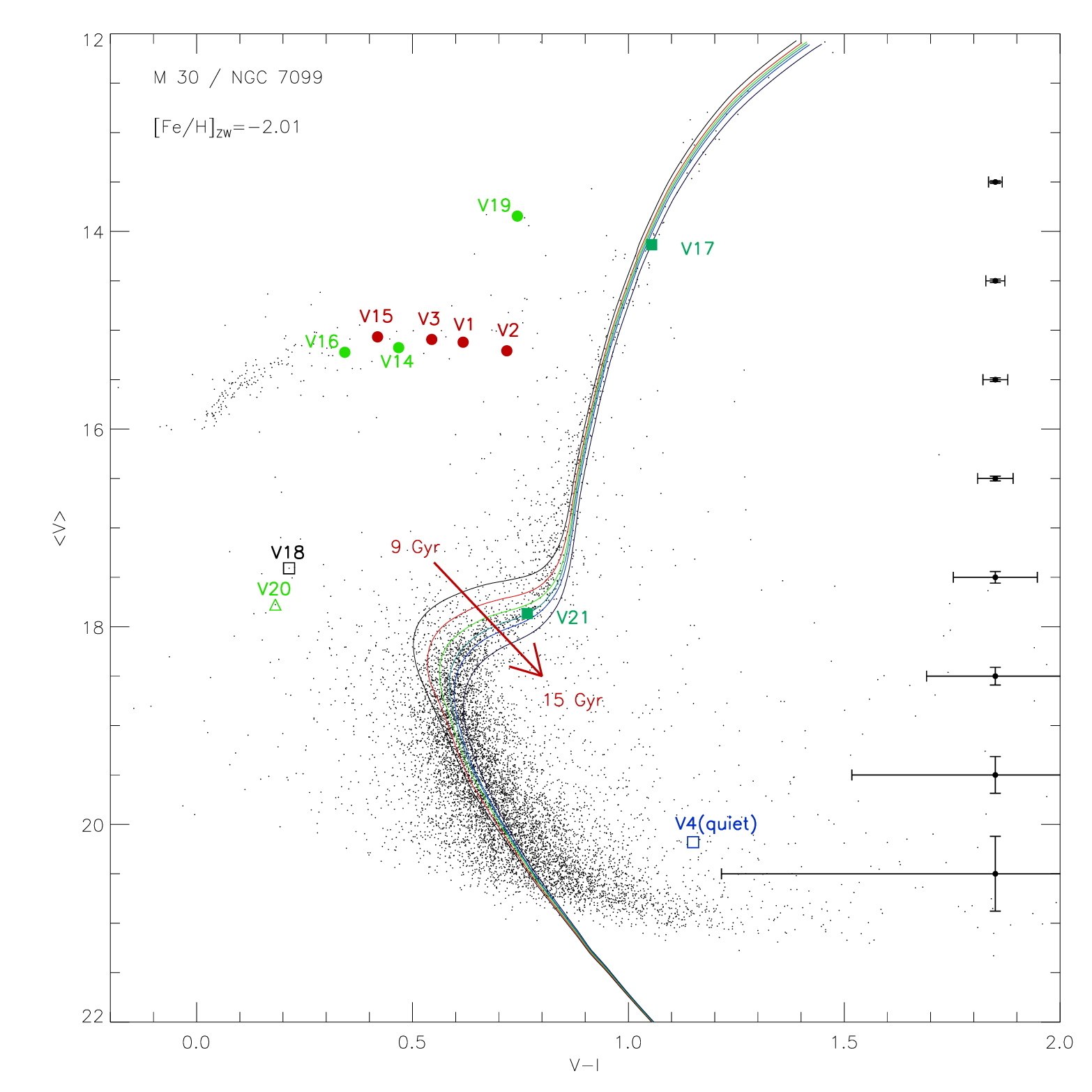}

  \caption{$(V-I), V$ colour-magnitude diagram from our photometry. The location of RR Lyrae (filled circles, RR0 in red, RR1 in green), SX Phoenicis (green triangles), U Geminorum (Blue square), eclipsing blue straggler (black open square) and other variables of uncertain type (green filled squares) are shown. On the right of the plot, typical error bars are plotted for different magnitude levels. Also plotted are isochrones from \cite{vandenberg03} for ages of 9, 10.5, 12, 13.0 (our best-fit isochrone), 13.5 and 15 Gyr. \label{fig:cmd}}

\end{figure*}

All confirmed variables are listed in \Tab{tab:variables}, in which we also give epochs, periods and amplitudes. The corresponding light curves are showin in Figs. \ref{fig:lc_V} and \ref{fig:lc_I}. A finding chart of the cluster with the location of the confirmed variables is shown in \Fig{fig:fchart}, and a CMD showing the locations of all confirmed variables is shown in \Fig{fig:cmd}.

\begin{table*}
\begin{center}
  \begin{tabular}{ccccccccccc}

     \hline
    \#		 	&Epoch	&$P$ 	&$P$ (Rosino	&$P$ (Pietrukowicz		&$<V>$	&$<I>$	& $A_V$	&$A_I$ &Type\\
    		 	&(HJD-2450000)&(d)  	&1949) (d)		&et al. 2004) (d)&(mag)	&(mag)	&(mag) 	&(mag) 	&\\
  \hline  

    V1			&6160.8234	&0.7436296	&0.74365$^a$		&0.751	&15.12	&14.52	&0.93	&0.74	&RR0\\
    V2			&6150.8616	&0.6535119	&0.6535049	&$-$		&15.20	&14.62	&0.92	&0.65	&RR0\\
    V3			&6147.8307	&0.6963265	&0.69632		&$-$		&15.12	&14.54	&0.94	&0.67	&RR0\\
    V4$^b$		&6147.8569	&0.092318	&$-$			&$-$		&20.19	&19.14	&2.74	&1.91	&U Gem\\
    V14		&6150.8758	&0.347953	&$-$		&$-$		&15.18	&14.77	&0.48	&0.32		&RR1\\
    V15		&6160.7960	&0.679015	&$-$		&0.689	&15.07	&14.52	&1.08	&0.68		&RR0\\
    V16		&6190.7559	&0.325366	&$-$		&$-$		&15.22	&14.81	&0.33	&0.27		&RR1\\
    V17		&6150.8805	&0.059954	&$-$		&$-$		&14.14	&13.08	&0.07	&0.04		&SX Phoenicis + blend?\\
    V18		&6161.8536	&0.307099	&$-$		&$-$		&17.43	&17.19	&0.41	&0.49		&Eclipsing blue straggler\\
    V19$^c$	&5779.8451	&0.343379	&$-$		&0.341	&13.85	&13.10	&0.15	&0.11		 &RR1\\
    V20		&6147.8377	&0.040199	&$-$		&$-$		&17.79	&17.57	&0.15	&0.13		&SX Phoenicis\\
    V21		&6147.8589	&0.113151	&$-$		&$-$		&17.87	&17.11	&$\sim$0.2&$-$		&?\\

\hline \hline
  \end{tabular}
  \caption{Epochs, periods, mean magnitudes and amplitudes $A$ in $V$ and $I$ for all confirmed variable stars in M 30. For RR Lyrae stars, $<V>$ and $<I>$ are intensity-weighted mean magnitudes,  while for the other variables, they are inverse-variance-weighted mean magnitudes. $^a$\cite{rosino61} revised the period of V1 to 0.743608 d. $^b$The period given for V4 is that of the sinusoidal variations seen during the quiescent part of the light curve, and the mean magnitudes are also those during quiescent phase; outburst mean magnitudes are given in the text (see \Sec{sec:individual}). $^c$For V19, the mean magnitudes are overestimated and amplitudes are underestimated due to blending, as discussed in the text. \label{tab:variables}}
  \end{center}
\end{table*}


\subsection{Discussion of individual variables}\label{sec:individual}

We are unable to determine the nature of V17 with certainty. The combination of its position on the red giant branch and a short period of $\sim 0.06$d makes it difficult to classify, although the shape of the light curve and the period would both be consistent with an SX Phoenicis variable blended with a red giant star \citep[e.g.][]{darragh12}. This is difficult to quantify, and higher-resolution data would be needed to investigate that possibility. From the position of V18 on the CMD and its light curve showing minima of different depths, we suggest that it is an eclipsing blue straggler binary system, of which only 6 examples are known in globular clusters \citep{kai12}. We classify V20 as an SX Phoenicis variable, from its light curve, period, and position on the CMD; we only identify one pulsation period for this variable. We could not reach any conclusion as to the nature of V21, because of the quality of its light curve; however variability is clear from both the $V-$ and $I$-band light curves, when discarding the poor-quality 2011 data, and data with large error bars, from the $I$ light curve. However we do verify that the variability of V21 is genuine in the difference images, and it is isolated, so there is no reason to believe that the variability is due to contamination from other variables. Below we discuss some of the variables in more detail.

\subsubsection{V2}

Figs. \ref{fig:lc_V} and \ref{fig:lc_I} show that our best period for V2 leads to an unsatisfactory phased light curve in both $V$ and $I$. We suggest that the disjointed light curve may be due to a Blazhko effect \citep{blazhko1907} in this object; more observations are needed to confirm this.

\subsubsection{V3}

Like V2, but to a lesser extent, the light curves V3 seem disjointed, which we suggest may be due to a Blazhko effect. However, and more observations are needed to investigate this further.

\subsubsection{V4}

For the U Geminorum variable V4, we measured a quiet $V$-band median magnitude of $20.21 \pm 0.48$ mag, while during outburst, we find a median magnitude of $17.44 \pm 0.04$ mag, giving an amplitude of 2.74 mag; in the $I$ band we find median values of $19.05 \pm 0.40$ (quiet) and $16.84 \pm 0.48$ (outburst), and an amplitude of 1.91 mag. \cite{machin91} found V4 to be significantly brighter, with $<V>_{\rm quiet} \sim 18.7$ mag, and \cite{pietrukowicz08} also found $<V>_{\rm quiet} < 19$ mag. It is interesting to note that, while our $V$ data only covers one outburst, the $I$ data covers three outburst, which show large variations in magnitude, by as much as $\sim 1$ mag between different outbursts.

\begin{figure}
  \centering
  \includegraphics[width=6cm, angle=0]{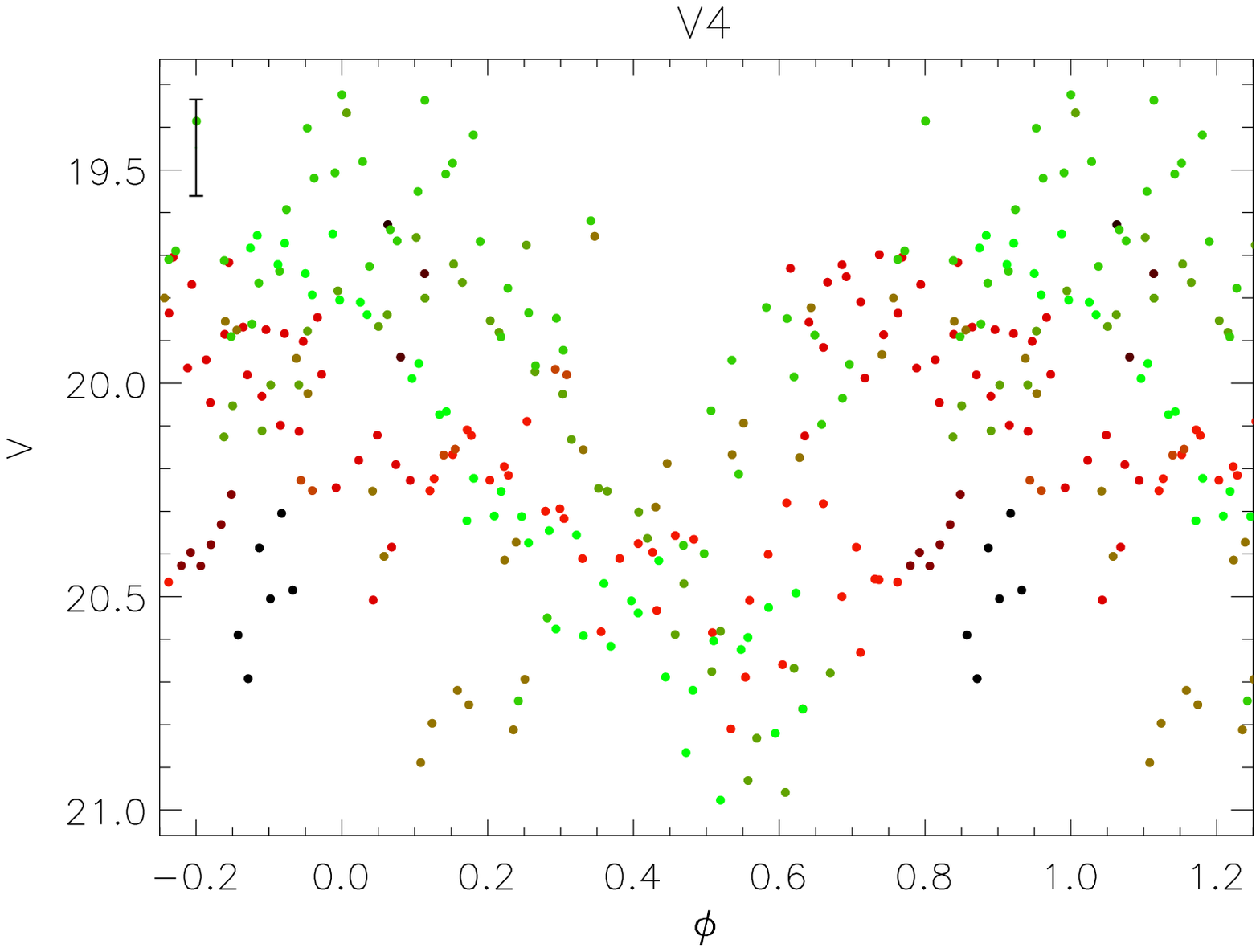}   \\
  \includegraphics[width=6cm, angle=0]{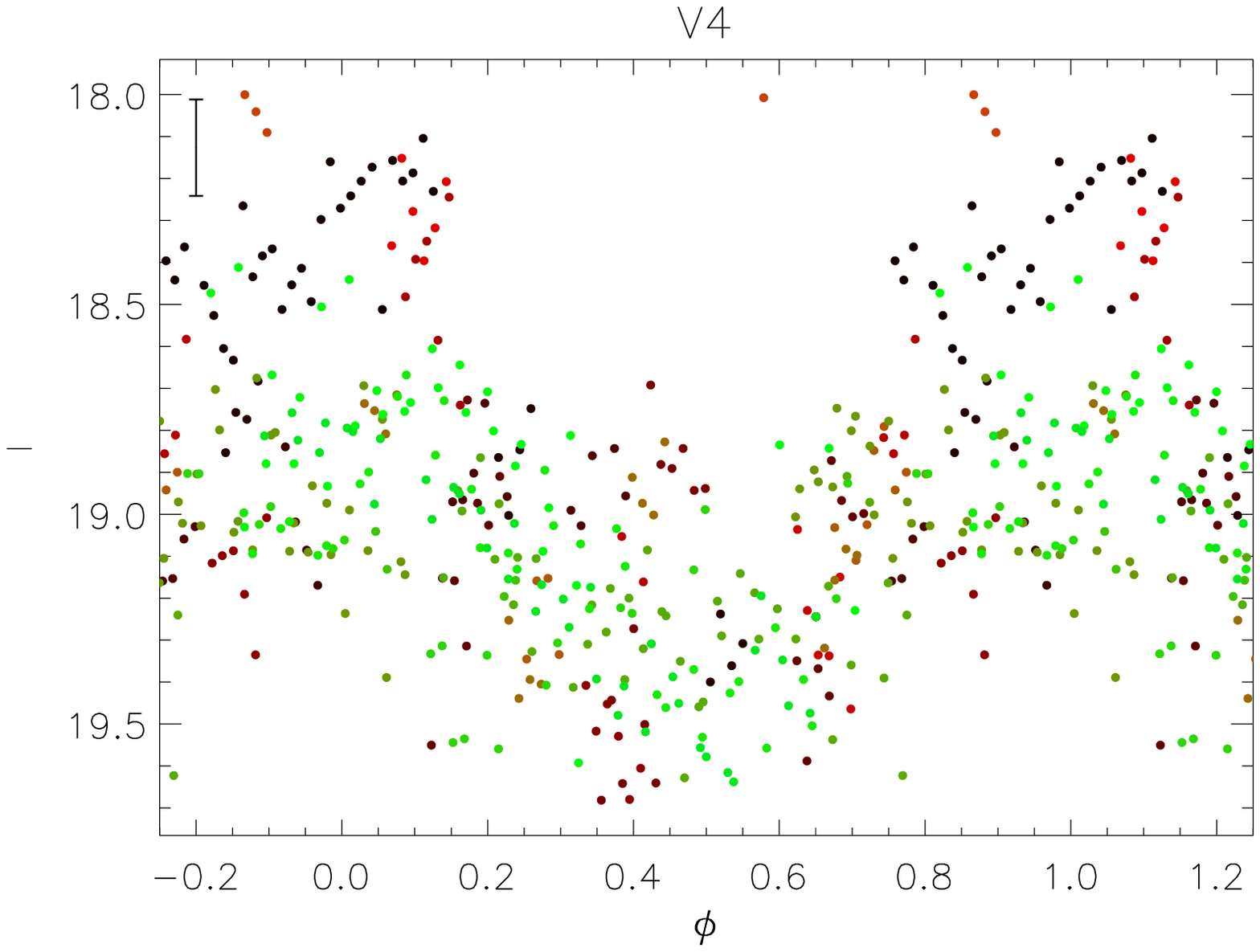}

  \caption{Phased quiescent light curves of V4 in the $V-$ (top) and $I$-band (bottom). Colour coding is the same as for \Fig{fig:lc_V}. For clarity we only plot data points with errors $\sigma < 0.3$ mag. \label{fig:v4_phased}}

\end{figure}

\begin{figure}
  \centering
  \includegraphics[width=7cm, angle=0]{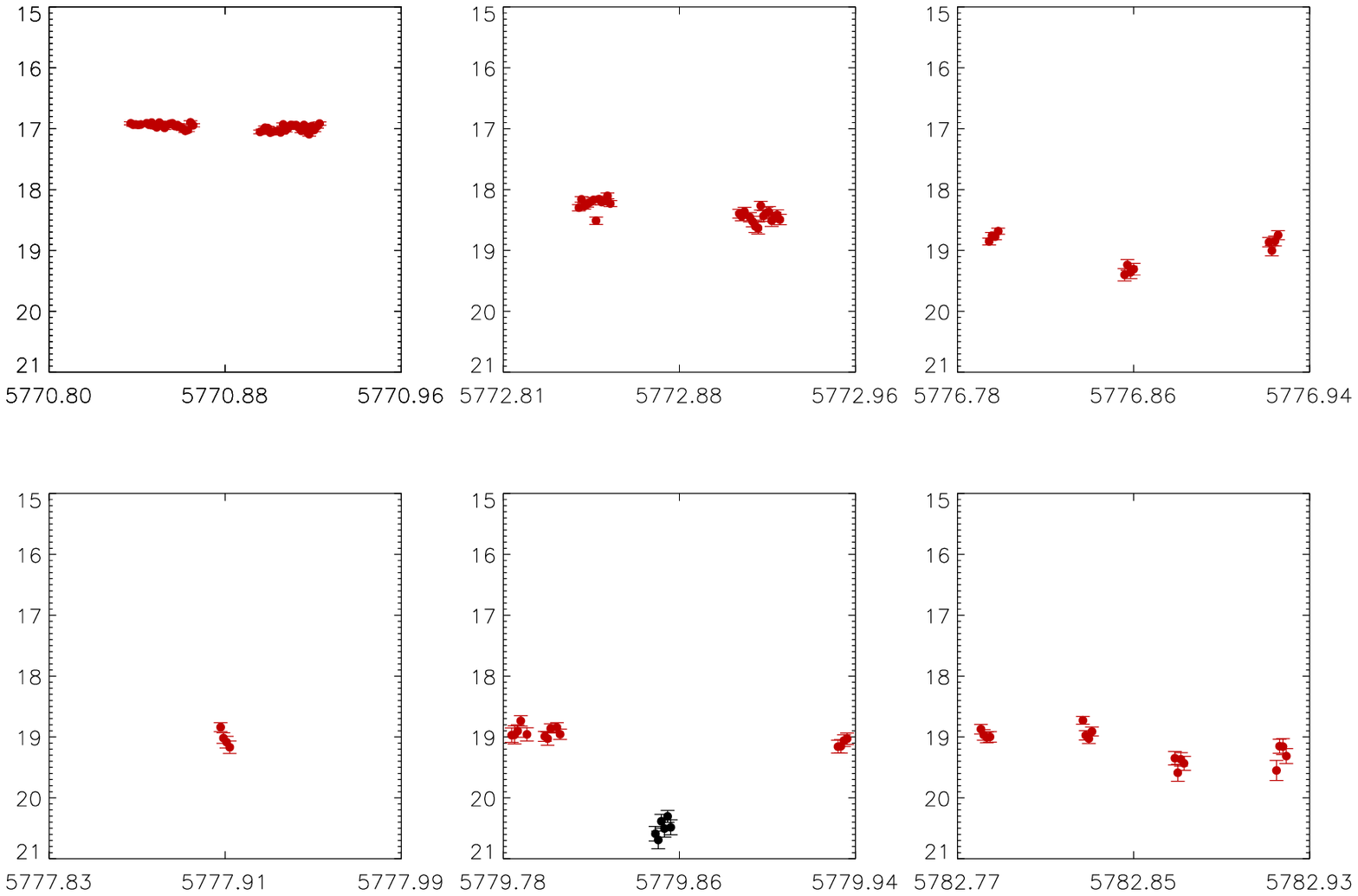} \\ 
  \includegraphics[width=7cm, angle=0]{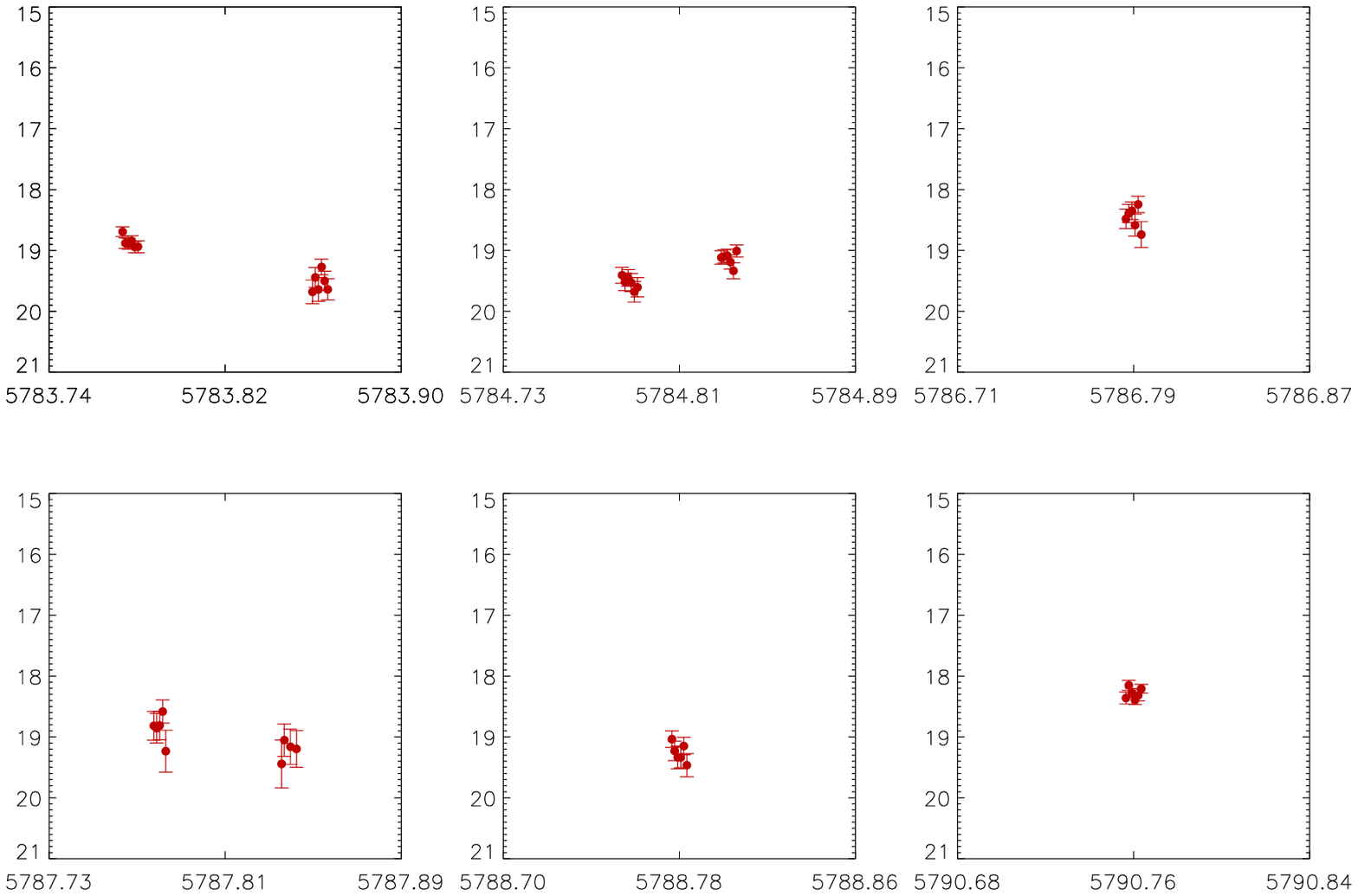}  \\ 
  \includegraphics[width=7cm, angle=0]{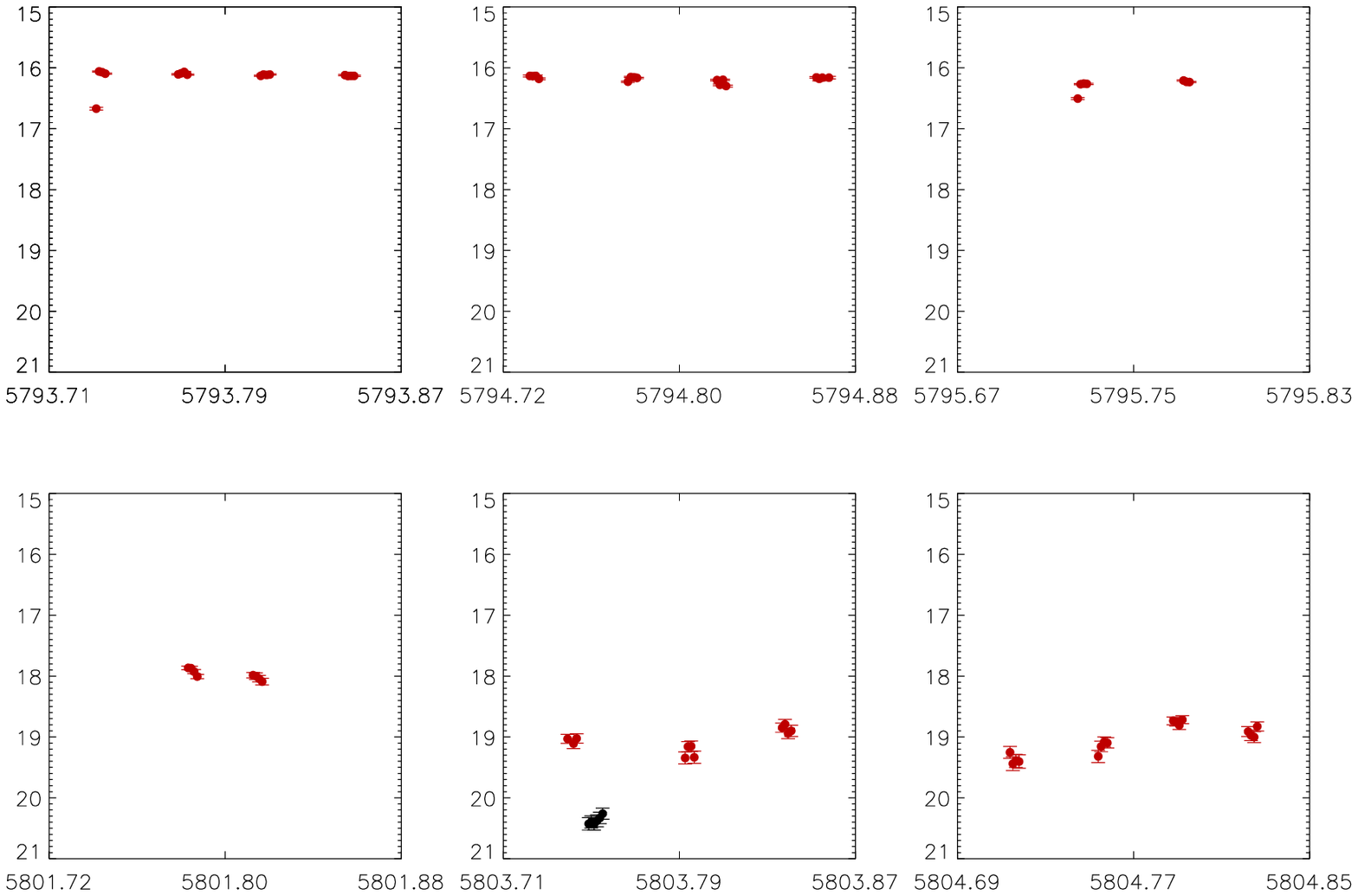}  \\
  \includegraphics[width=7cm, angle=0]{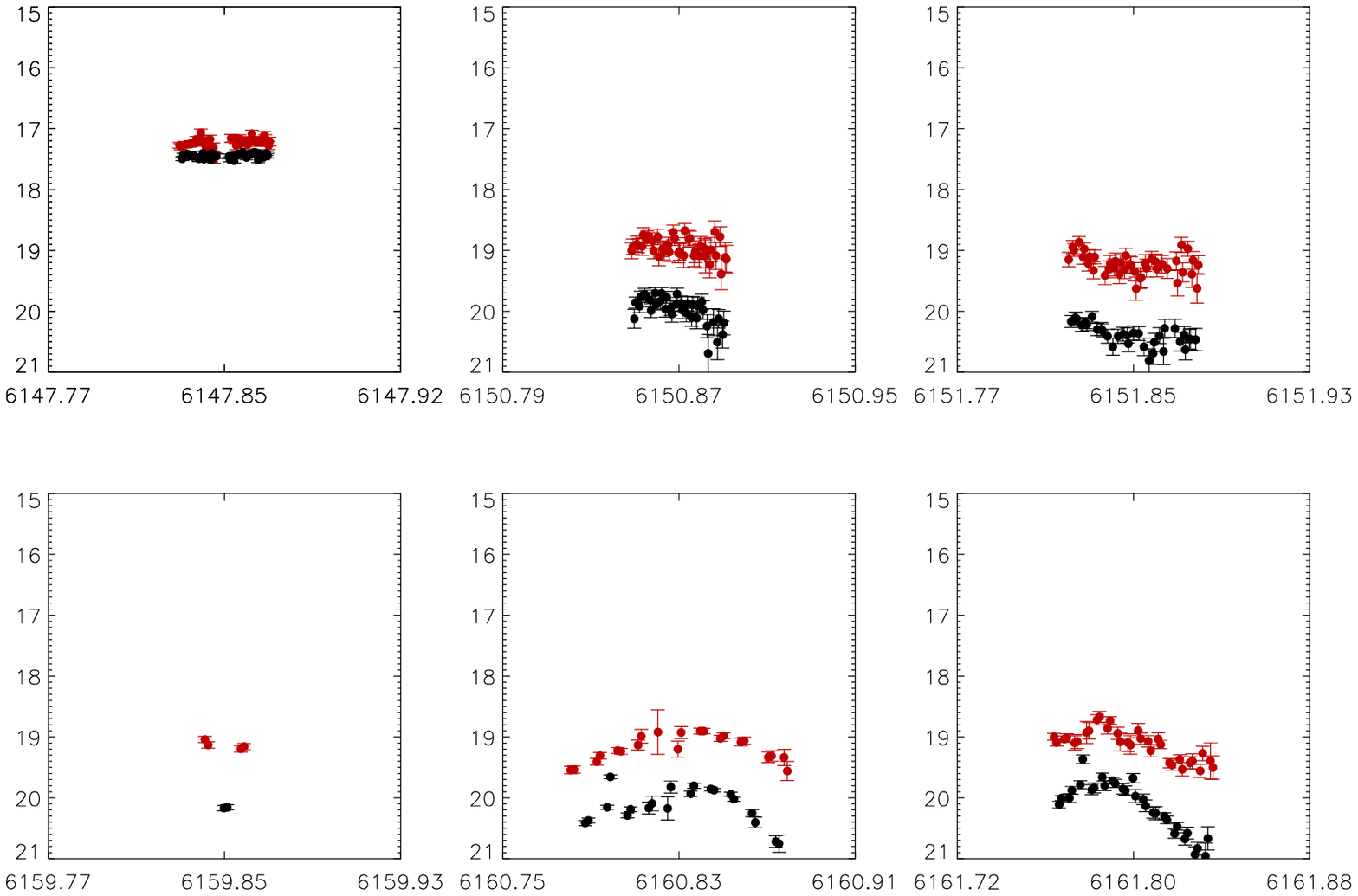}  \\
  \includegraphics[width=7cm, angle=0]{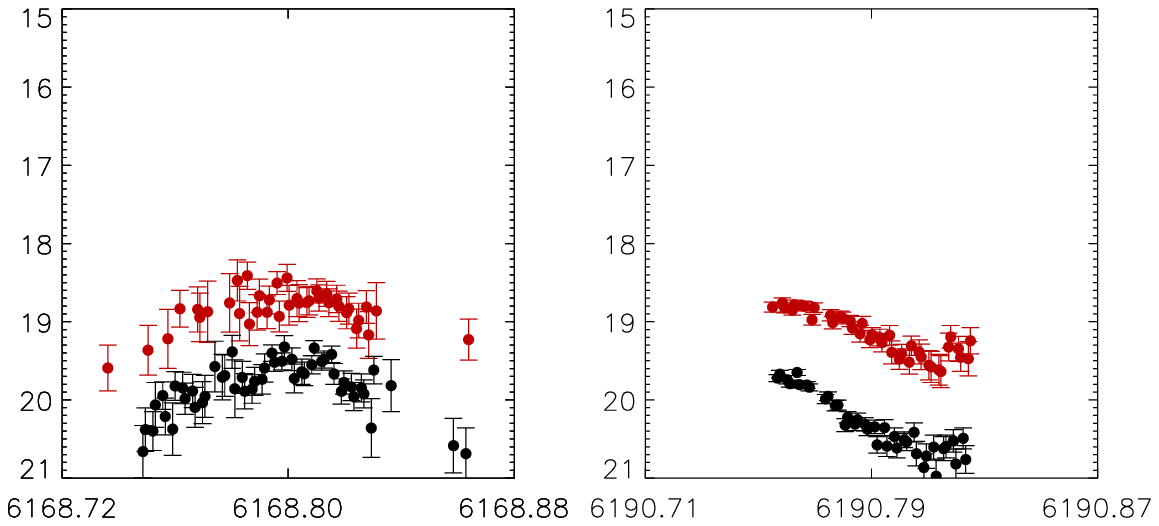}  \\
  
  \caption{2012 $V-$ and $I-$band light curves of V4. $V$ data is plotted in black, and $I$ data in red. For clarity we plot only data points with errors $\sigma < 0.3$ mag. The $x-$axes are given in HJD-2450000. \label{fig:v4_all}}

\end{figure}

We also conducted a period search for the quiet part of the light curve; \cite{machin91} used the quiescent $B-V$ colour they find from their spectra of this object, and the relations of \cite{echevarria84} between $B-V$ and $P$ for dwarf novae, to conclude that the orbital period of V4 must be $<5$ hours. We find a period during quiescence of $P=0.0923$ days, or $P \sim 2.22$ hours, in agreement with that conclusion. In \Fig{fig:v4_phased}, we show the phased quiescent light curves with this period, while the unphased light curve is shown in \Fig{fig:v4_all}.

\subsubsection{V16}

The light curves in both $V$ and $I$ suggest that our single best period does not lead to a satisfactory phased light curve; we discuss possible explanations for this here. First we considered the possibility that V16 might be a double-mode RR Lyrae (RR01) star, which have only been detected in a few Galactic globular clusters \citep[e.g.][and references therein]{nemec85, clement93, clementini04}. To investigate this, a search for the fundamental and first-overtone pulsation periods was conducted using the string-length method. We start by conducting a period search to identify a first period; for this we find $P=0.3254$. Assuming this to be the first-overtone period, and assuming an overtone-to-fundamental period ratio similar to what was found for double-mode RR Lyrae stars in M15, $P_1/P_0 \sim 0.748$ \citep{cox83}, we expect a fundamental period around $P_0 \sim 0.435$. We pre-whiten our data for the first-overtone period to check whether we can recover this period. In the $V$ data, we find no evidence for such a period or any other pulsation period in our resulting power spectrum. We find similar results using the $I$ data, leading us to conclude that the unsatisfactory phasing of the V16 light curves is not due to double-mode pulsation. 

Secondly, we also tried to fit the light curve by including a secular period change. To do this, we minimise the string length, but with a time-dependent period $P(t)$ and phase $\phi(t)$,

\begin{eqnarray}
P(t) &=& P_0 + \beta (t-E) \\
\phi(t) &=& (t-E)/P(t) - \lfloor (t-E)/P(t)\rfloor\, ,
\end{eqnarray}

\noindent
where $\beta$ is the rate of change of the period and $P_0$ is the period at epoch $E$. We varied $P_0$ within a small range near the value we found using the string-length method, and for each value of $P_0$ we explored a grid of values for $\beta$ ranging between $-10^{-7}$ and $10^{-7} dd^{-1}$. From this we found that no such secular period change can explain the scatter in our phased light curve of V16.

This leads us to conclude that V16 either exhibits the Blazhko effect, or shows signs of non-radial pulsation. Amplitude variations are clear when comparing the 2011 (black filled circles on \Fig{fig:lc_V} and \ref{fig:lc_I}) and 2012 data, and Blazhko-like effects cause period modulations as well, which would explain that the light curve is not phased properly when using a single best-fit period. On the other hand, the scatter in the light curves, particularly in the $I-$band, is very similar to that observed in light curves of some of the variables in M 55 detected by \cite{olech99}. They concluded that the scatter in the variations were caused by non-radial oscillations, and we suggest here that non-radial pulsation could also be the source of the scatter in the light curves of V16.

\section{Fourier decomposition of RR Lyrae star light curves}\label{sec:fourdec}

Applying Fourier decomposition to the $V-$band light curves of RR Lyrae variables with sufficient phase coverage allows us to derive several of their properties, which we then use as a proxy for the parameters of their host cluster. Fourier decomposition consists in fitting light curves with the Fourier series

\begin{equation}\label{eq:fourcos}
m(t) = A_0 + \sum_{k=1}^N A_{k} \cos \left[ \frac{2\pi k}{P}(t - E) + \phi_k \right] \, ,
\end{equation}

\noindent
where $m(t)$ is the magnitude at time $t$, $N$ is the number of harmonics used in the fit, $P$ is the period of the variable, $E$ is the epoch, and $A_k$ and $\phi_k$ are the amplitude and phase of the $k^{th}$ harmonic. The Fourier parameters, which are epoch-independent, are then defined as

\begin{eqnarray}
R_{ij} &=& A_i / A_j \\
\phi_{ij} &=& j\phi_i - i\phi_j\, .
\end{eqnarray}

In order to avoid over-fitting light curve features, we used the minimum number of harmonics that provided a good fit. Furthermore, for each variable, we checked the dependence on $N$ of the parameters we derive in the next sections. We found very little variation with $N$, with any changes smaller than the error bars associated with the relevant quantities. In the following analysis, we excluded V16 because we suspect from its light curve that it may be affected by Blazhko effects or by non-radial pulsation (see \Sec{sec:individual}).

We list the coefficients $A_k$ we obtained for the first four harmonics in \Tab{tab:fourier}, as well as the Fourier parameters $\phi_{21}, \phi_{31}$ and $\phi_{41}$ for the variables for which we could obtain a Fourier decomposition. We also list the deviation parameter $D_m$, defined by \cite{jurcsik96} in order to assess whether fit parameters are reliable to derive properties of the RR Lyrae variables. Although \cite{jurcsik96} used a criterion whereby fits should have $D_m < 3$ for their empirical relations to yield reliable estimates of stellar properties, a less stringent criterion of $D_m < 5$ has been used by other authors \citep[e.g.][]{cacciari05}. Here we also adopt $D_m < 5$ as a selection criterion to estimate stellar properties. 

Although some acceptable fits could be obtained for V2, with $D_m < 5$, we exclude it from the analysis as well because we suspect its light curve might be affected by Blazhko effects. Furthermore, the value of the physical parameters is very sensitive to the number of harmonics used in the fit. Although we suggested in \Sec{sec:individual} that V3 might also be affected by Blazhko effects, we find that the Fourier fits and resulting physical parameters are stable for V3, and therefore include it in the following analysis. We note, however, that this may account for the larger value of $D_m$ we find for V3. We also exclude V19 because our phase coverage does not enable us to find a good Fourier fit. This leaves us with 4 RR Lyrae stars with good Fourier decompositions, three RR0 (V1, V3 and V15) and one RR1 (V14).

\begin{table*}
\begin{center}
  \begin{tabular}{ccccccccccc}

     \hline
    \#		&$A_0$	&$A_1$	&$A_2$ 	&$A_3$	&$A_4$	&$\phi_{21}$	&$\phi_{31}$	&$\phi_{41}$	&$N$	&$D_m$	 \\
  \hline
  RR0\\
  \hline  
V1& 15.121(2)&  0.337(2)&  0.160(2)&  0.110(2)&  0.050(2)&  4.045(12)&  8.498(16)&  6.581(22) & 7 &2.6 \\
V2$^a$& 15.196(2)&  0.320(2)&  0.134(2)&  0.107(2)&  0.081(2)&  4.005(11)&  8.511(15)&  6.668(16) & 7 &3.0\\
V3& 15.118(2)&  0.319(2)&  0.152(2)&  0.112(2)&  0.059(2)&  3.973(11)&  8.413(13)&  6.468(19) & 7 &4.9\\
V15& 15.067(2)&  0.344(2)&  0.181(2)&  0.122(1)&  0.081(2)&  3.987(7)&  8.274(12)&  6.431(14) &10 &2.5\\
\hline
RR1\\
\hline
V14& 15.176(2)&  0.214(3)&  0.030(3)&  0.012(2)&  0.010(2)&  4.773(26)&  2.673(137)&  2.338(104) & 6 &$-$\\

\hline \hline
  \end{tabular}
  \caption{Parameters from the Fourier decomposition. A good decomposition could not be obtained for V19, due to poor phase coverage; the Fourier fit to V2 was also poor, and V16 is excluded because it shows signs of amplitude and period modulation due to Blazhko effects or non-radial pulsation. Numbers in parentheses are the 1-$\sigma$ uncertainties on the last decimal place. $^a$We provide our best-fit Fourier parameters but do not use the decomposition of V2 in the derivation of the cluster parameters, due to the sensitivity of the V2 parameters to the number of harmonics used. \label{tab:fourier}}
  \end{center}
\end{table*}

\subsection{Metallicity}\label{sec:rrmet}

We use the empirical relations of \cite{jurcsik96} to derive the metallicity [Fe/H] for each of the variables for which we could obtain a successful Fourier decomposition. The relation is derived from the spectroscopic metallicity measurement of field RR0 variables, and it relates [Fe/H] to the period $P$ and the Fourier parameter $\phi^s_{31}$, where $s$ denotes a parameter obtained by fitting a \textit{sine} series rather than the cosine series we fit with \Eq{eq:fourcos}. [Fe/H] is then expressed as

\begin{equation}\label{eq:metrr0}
\mathrm{[Fe/H]_J} = -5.038 - 5.394\, P + 1.345\, \phi^s_{31}\,
\end{equation}

\noindent
where the subscript J denotes a non-calibrated metallicity, the period P is in days, and $\phi^s_{ij}$ can be calculated via 

\begin{equation}\label{eq:phis}
\phi^s_{ij} = \phi_{ij} - (i-j)\, \frac{\pi}{2}\, .
\end{equation}

We transform these to the metallicity scale of \cite{zinn84} (hereafter ZW) using the relation from \cite{jurcsik95}:

\begin{equation}\label{eq:zw}
\fehzw = \frac{\rm [Fe/H]_J - 0.88}{1.431}\, .
\end{equation}

However \cite{kovacs02} noted that for metal-poor clusters, \Eq{eq:metrr0} yields metallicity values that are too large by $\sim 0.2$ dex. This was also confirmed by \cite{gratton04} and \cite{difabrizio05} by comparing spectroscopic and Fourier-decomposition metallicity values for RR Lyrae in the Large Magellanic Cloud (LMC). Therefore here we include a shift of -0.20 dex (on the $\mathrm{[Fe/H]_J}$ scale) to metallicity values we derive for RR0 stars using \Eq{eq:metrr0}, which corresponds to a shift on the ZW scale of -0.14 dex.

For the RR1 variables, we calculated the metallicity using the empirical relation of \cite{morgan07}, linking [Fe/H], $P$ and $\phi_{31}$:

\begin{eqnarray}\label{eq:metrr1}
\fehzw &=& 2.424 - 30.075\, P +  52.466\, P^2 \\ \nonumber
&&+ 0.982\, \phi_{31} + 0.131 \phi_{31}^2 - 4.198\, \phi_{31}P \,
\end{eqnarray}

Metallicity values calculated using \Eq{eq:metrr0} \& (\ref{eq:zw}) and (\ref{eq:metrr1}) are given in \Tab{tab:starpar}.

\begin{table}
\begin{center}
  \begin{tabular}{ccccc}

     \hline
    \#		&$\fehzw$	&$M_V$	&$\log(L/L_{\bigodot})$ 	&$T_{\rm eff}$	\\
 \hline
RR0 \\
 \hline

V1&  -2.04(2) &  0.351(2) &  1.780(1) & 6187(7) \\
V3&  -1.95(2) &  0.428(2) &  1.746(1) & 6247(6) \\
V15&  -2.01(2) &  0.428(2) &  1.746(1) & 6249(5) \\
\hline
RR1 \\
\hline
V14&  -2.03(4) &  0.540(6) &  1.650(3) & 7180(13) \\
\hline \hline
\end{tabular}

  \caption{Physical parameters for the RR Lyrae variables calculated using the Fourier decomposition parameters and the relations given in the text. Numbers in parentheses are the 1-$\sigma$ uncertainties on the last decimal place. \label{tab:starpar}}
  \end{center}
\end{table}

\subsection{Effective Temperature}

The Fourier parameters can also be used to calculate the effective temperature, using empirical relations derived by \cite{jurcsik98}. Those relations link the $(V-K)_0$ colour to $P$ as well as several of the Fourier coefficients and parameters:

\begin{eqnarray}
(V-K)_0 &=&  1.585 + 1.257\, P - 0.273\, A_1 - 0.234\, \phi^s_{31} \\ \nonumber
&&+ 0.062\, \phi^s_{41} \\\label{eq:teffrr0}
\mathrm{log}\,T_{\rm eff} &=& 3.9291 - 0.1112\,(V-K)_0 \\ \nonumber
&& - 0.0032 \, \mathrm{[Fe/H]_J} \, .
\end{eqnarray}

For RR1 variables, \cite{simon93} used theoretical models to derive an empirical relation,

\begin{equation}\label{eq:teffrr1}
\log\,\teff = 3.7746 - 0.1452 \log\, P + 0.0056\, \phi_{31} \, .
\end{equation}

We list the temperatures we derived for each RR Lyrae star in \Tab{tab:starpar}. As noted in our previous analyses, there are several caveats to deriving temperatures with \Eq{eq:teffrr0} and (\ref{eq:teffrr1}). These relations yield values of $\teff$ for RR0 and RR1 stars on different absolute scales. Furthermore, the effective temperatures we derive here show systematic deviations from the relations predicted by evolutionary models of \cite{castelli99} or the temperature scales of \cite{sekiguchi00}. Bearing these caveats in mind, we use these relations to derive temperature estimates in order to be consistent with our previous studies.

\subsection{Absolute Magnitude}\label{sec:rrmag}

\cite{kovacs01} derived empirical relations to calculate V-band absolute magnitudes for the RR0 variables, linking the magnitude to Fourier coefficients through

\begin{equation}\label{eq:mvrr0}
M_V = -1.876 \log P - 1.158\, A_1 + 0.821 \, A_3 + K_0 \, ,
\end{equation}

\noindent
where $K_0$ is a constant. As in our previous studies, we adopt a value of $K_0=0.41$ mag to be consistent with a true LMC distance modulus of $\mu_0=18.5$ mag \citep{freedman01}. For RR1 variables, we use the relation of \cite{kovacs98},

\begin{equation}\label{eq:mvrr1}
M_V = -0.961\, P - 0.044\, \phi^s_{21} - 4.447 \, A_4 + K_1 \, ,
\end{equation}

\noindent
where $K_1$ is a constant, for which we choose a value of 1.061 with the same justification as for our choice of $K_0$.

We also converted the magnitudes we obtained to luminosities using

\begin{equation}\label{eq:logl}
\log\left( L/L_{\bigodot} \right) = -0.4\, \left[M_V + B_C(\teff) - M_{\rm bol, \bigodot }\right] \, ,
\end{equation}

\noindent
where $M_{\rm bol, \bigodot}$ is the bolometric magnitude of the Sun, $M_{\rm bol, \bigodot }=4.75$, and $B_C(\teff)$ is a bolometric correction which we determine by interpolating from the values of \cite{montegriffo98} for metal-poor stars, and using the value of $\teff$ we derived in the previous section. Values of $M_V$ and $\log(L/L_{\bigodot})$ for the RR0 and RR1 variables are listed in \Tab{tab:starpar}. Using our average values of $M_V$, in conjunction with the average values of $\fehzw$ (\Sec{sec:rrmet}), we find a good agreement with the $M_V - \fehzw$ relation derived in the literature \cite[e.g.][see Fig.9 of that paper]{kains12b}.

\section{Cluster properties}\label{sec:clusterprop}

\subsection{Oosterhoff Type}

We find mean periods $<P_{\rm RR0}>=0.693 \pm 0.038$ d and $<P_{\rm RR1}>=0.346 \pm 0.003$ d, with a proportion of 43\% of the RR Lyrae in this cluster being of RR1 type.

\begin{figure}
  \centering
  \includegraphics[width=8cm, angle=0]{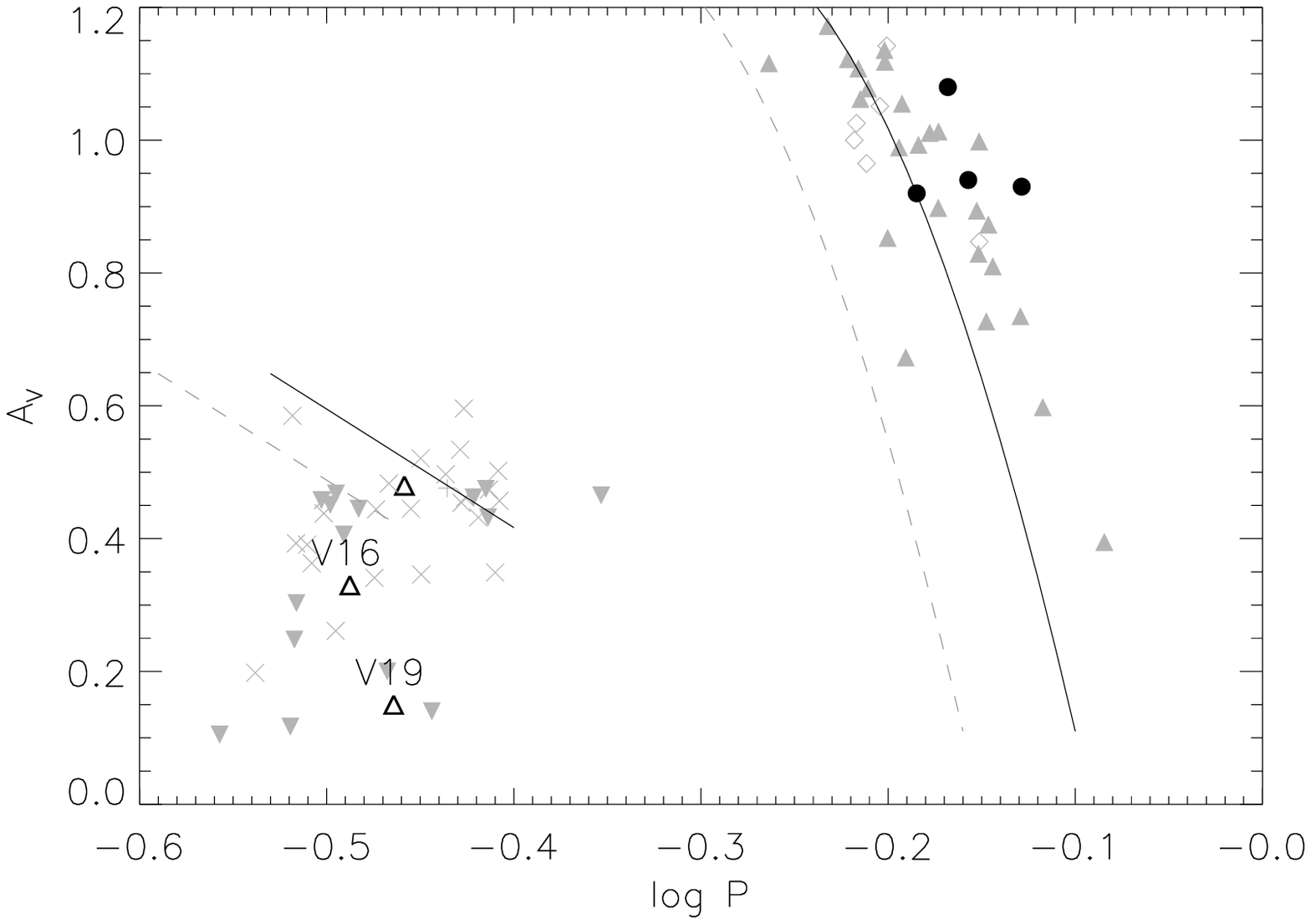}
  \includegraphics[width=8cm, angle=0]{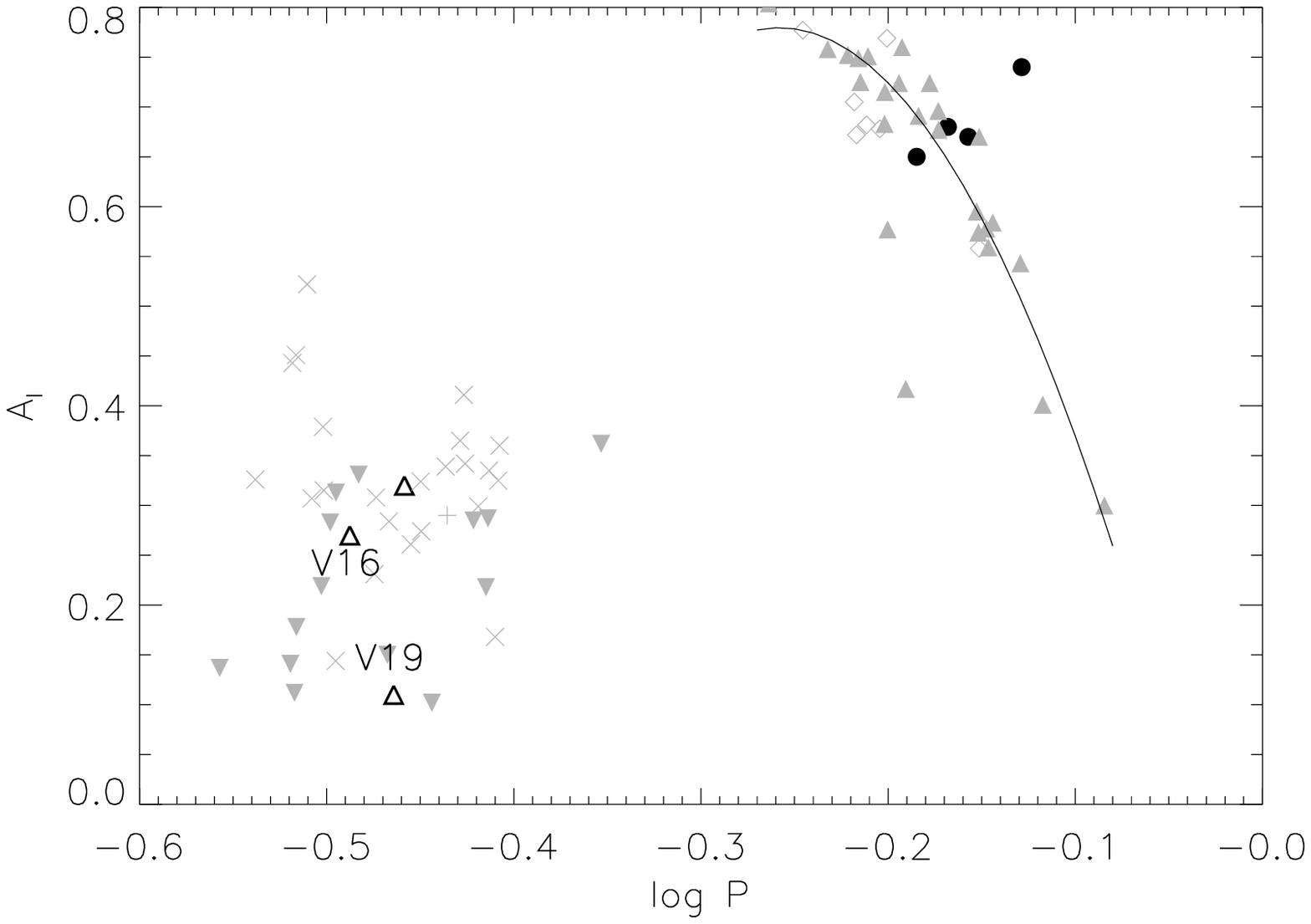}

  \caption{Bailey diagram of the light curve amplitude versus the logarithm of the period for RR Lyrae stars in M 30, in $V$ (top) and $I$ (bottom). RR0 variables are plotted as filled circles, and RR1 variables as open triangles; we mark the locations of V16 which we suggest is affected by Blazhko effects or non-radial pulsation, and of V19, which has an underestimated amplitude due to blending. Also plotted on the $V-$band diagram as solid lines are the relations of \cite{cacciari05} for evolved stars, obtained by applying a shift of +0.06 to $\log\, P$ in the relations derived for the Oosterhoff type I cluster M3; the original relations (without the shift of 0.06 in $\log\, P$) are also plotted in light gray dashed lines. For the $I-$band we plot the relation derived by \cite{arellano11} for RR0 stars. For comparison we also plot the populations of RR Lyrae detected in the Oosterhoff type II clusters M 9 \citep{arellano13} and NGC 5024 \citep{arellano11} in light gray, with filled triangles and inverted triangles for RR0 and RR1 stars respectively. For stars with Blazhko effects, we use open diamonds for RR0 stars and crosses for RR1 stars; the star marked with a + sign is a double-mode RR1 star.\label{fig:bailey}}

\end{figure}

From these values of $<P_{\rm RR0}>$ and the fraction of RR1 stars in M 30, as well as its low metallicity, we confirm previous classification of this cluster in the literature as Oosterhoff type II \citep[e.g.][]{lee99}. This is also confirmed by comparing the locations of our variables on a Bailey diagram (\Fig{fig:bailey}) to the tracks derived by \cite{cacciari05} for evolved stars.

\subsection{Distance}

\subsubsection{Using the RR Lyrae stars}

We can use the $A_0$ parameter from our Fourier decompositions, which corresponds to the mean apparent V magnitude of the RR Lyrae, as well as the absolute magnitudes of the RR Lyrae we derived in \Sec{sec:rrmag} to derive the distance modulus to M 30. The mean value of the $A_0$ for our RR0 variables is 15.09 $\pm$ 0.03 mag, while the mean of the absolute magnitudes is 0.40 $\pm$ 0.04 mag. This yields a distance modulus of $\mu=14.69 \pm 0.05$ mag. Using the parameters for our RR1 variable (V14, see Tables \ref{tab:fourier} \& \ref{tab:starpar}), we find $\mu=14.64$ mag.

The reddening towards this cluster has been estimated in the literature by several authors, with values of $E(B-V)$ ranging from 0.01 to 0.12 mag. \cite{zinn80} derived a value of $E(B-V)=0.01$ mag from integrated light measurements, \cite{bolte87} adopted a value of $E(B-V)=0.02$ mag, while \cite{richer88} found a value of $E(B-V)=0.068 \pm 0.035$ mag; \cite{piotto90} derived $E(B-V)=0.05$ mag from their CMD analysis, and \cite{samus95} found values of $E(B-V)=0.09$ or 0.12 mag, depending on the metallicity of the isochrone they fitted to their CMD. \cite{ferraro99} report a value for this cluster $E(B-V) = 0.03$ mag.

Here we adopt a value of  $E(B-V)=0.03 \pm 0.01$ mag, as well as a value of $R_V=3.1$ for our Galaxy. We use these to derive mean true distance moduli of $\mu_0 = 14.60 \pm 0.05$ mag and $\mu_0 = 14.54$ mag, from our RR0 and RR1 variables respectively. These values correspond to mean physical distances of $8.32 \pm 0.20$ kpc and 8.10 kpc. Since the distance values are sensitive to our chosen value of $E(B-V)$, we also derive distances using the largest value of reddening in the literature $E(B-V)=0.12$ mag, to find a lower limit for the distances of $7.32 \pm 0.17$ and 7.13 kpc. Our values for the distance to the cluster are consistent with previous estimates in the literature, reported in \Tab{tab:lit_dist}.

\subsubsection{Using SX Phoenicis stars}\label{sec:sxphe_dist}

SX Phoenicis stars can be used as standard candles thanks to their period-luminosity ($P-L$) relation \citep[e.g.][]{jeon03}, allowing us to estimate the distance to M 30 using the detected SX Phoenicis star in our sample, V20. We use the $P-L$ relation of \cite{cohen12},

\begin{equation}\label{eq:sxpl}
M_V	= -(1.640 \pm 0.110) - (3.389 \pm 0.090) \log P_f\, ,
\end{equation}

\noindent
where $P_f$ denotes the fundamental-mode pulsation period. Using $P_f = 0.040199$ d for V20, we find $M_V=3.09 \pm 0.11$ mag. Using a mean magnitude of $<V> = 17.79$ mag, $E(B-V)=0.03$ mag, this yields a distance modulus of $14.61 \pm 0.11$ mag, which corresponds to a physical distance of $8.35 \pm 0.42$ kpc, in excellent agreement with the distance calculated using the RR Lyrae stars, and consistent with estimates in the literature.

\begin{table*}
\begin{center}
  \begin{tabular}{cccc}

\hline
    Reference		&$\mu_0$ [mag]		&Distance [kpc]		&Method	\\
\hline

This work			&14.60 $\pm$ 0.05	&8.32 $\pm$ 0.20	&Fourier decomposition of RR0 light curves \\
This work			&14.54 			&8.10			&Fourier decomposition of RR1 light curves \\
This work			&14.61 $\pm$ 0.11	&8.35 $\pm$ 0.42	&SX Phoenicis $P-L$ relation \\
\cite{carretta00}	&14.88 $\pm$ 0.05	&9.46 $\pm$ 0.22	&Parallax of local subdwarfs\\
\cite{ferraro99}		&14.71			&8.75			&Magnitude of the horizontal branch\\
\cite{sandquist99}	&14.65 $\pm$ 0.12	&8.51 $\pm$ 0.47	&Parallax of local subdwarfs\\
\cite{sandquist99}	&14.87 $\pm$ 0.12	&9.42 $\pm$ 0.52	&Parallax of local subdwarfs\\
\cite{gratton97}		&14.94 $\pm$ 0.08 	&9.72 $\pm$ 0.36 	&Parallax of local subdwarfs\\
\cite{bergbusch96}	&14.83			&9.25			&CMD analysis\\
\cite{samus95}		&14.70  $\pm$ 0.10	&8.71 $\pm$ 0.40 			&CMD analysis\\
\cite{piotto90}		&14.65  $\pm$ 0.15	&8.51 $\pm$ 0.59	&CMD analysis\\
\cite{piotto87}		&14.50 $\pm$ 0.50	&7.94 $\pm$ 1.83	&CMD analysis\\
\cite{bolte87}		&14.65			&8.51			&Parallax of local subdwarfs\\

\hline \hline
\end{tabular}

  \caption{Modulus and distance estimates for M 30 in the literature. \label{tab:lit_dist}}
  \end{center}
\end{table*}

\subsection{Metallicity}

Although the relation of \cite{sandage06} relating the mean period of RR0 stars to the cluster metallicity is not applicable for Oosterhoff type II clusters \citep{clement01}, we can use the metallicities we calculated in \Sec{sec:rrmet} for the RR Lyrae stars in the cluster to derive an estimate for the metallicity of M 30. To do this, we simply compute an average of the RR Lyrae metallicities given in \Tab{tab:starpar}, excluding the variables for which the metallicity estimate is unreliable. Assuming that there is no systematic offset between the different types of variables, as in previous studies \citep{kains12b, bramich11}, we find a mean metallicity $\fehzw = -2.01 \pm 0.04$, in good agreement with values found in the literature (see \Tab{tab:lit_met}).

\cite{carretta09} derived a new metallicity scale based on GIRAFFE and UVES spectra of red giant branch (RGB) stars in 19 globular clusters, which is now widely used to quote metallicity values for globular clusters. The transformation from the ZW to the UVES \citep{carretta09} scale is given as

\begin{equation}\label{eq:fehuves}
\fehuves = -0.413 + 0.130\,\fehzw - 0.356\,\fehzw^2\, .
\end{equation}

Using this we find a metallicity for M 30 of $\fehuves=-2.11 \pm 0.06$, significantly higher than the value found for this cluster by \cite{carretta09b} of $\fehuves = -2.34 \pm 0.05$.


\begin{table*}
\begin{center}
  \begin{tabular}{cccc}

     \hline
    Reference		&$\fehzw$	&$\fehuves$	&Method	\\
 \hline
This work			&-2.01 $\pm$ 0.04	&$-2.11 \pm 0.06$	& Fourier decomposition of RR Lyrae light curves\\
\cite{carretta09b}	&-2.04 $\pm$ 0.16	&$-2.34 \pm 0.05$	&UVES spectroscopy of red giants \\
\cite{carretta09a}	&-2.05 $\pm$ 0.16	&$-2.36 \pm 0.05$	&FLAMES/GIRAFFE specta of red giants \\
\cite{sandquist99}	&-2.01 $\pm$ 0.09	&$-2.11 \pm 0.14$	& Simultaneous reddening-metallicity method\\
\cite{bergbusch96}	&-2.03			&-2.14	&CMD isochrone fitting \\
\cite{minniti93}		&-2.11 $\pm$ 0.08	&$-2.27 \pm 0.13$	& Spectroscopy of red giants\\
\cite{claria88}		&-2.4 			&$-2.78 \pm 0.20$	& Spectroscopy of red giants\\
\cite{bolte87}		&-2.03 $\pm$ 0.13	&$-2.14 \pm 0.20$	& CMD isochrone fitting\\
\cite{smith84}		&-2.02 $\pm$ 0.14	&$-2.13 \pm 0.22$	& $Q'_{39}$ spectral index\\
\cite{zinn84}		&-2.13 $\pm$ 0.13	&$-2.31 \pm 0.21$	& $Q_{39}$ spectral index\\
\cite{zinn80}		&-1.96 $\pm$ 0.12	&$-2.04 \pm 0.18$	& $Q_{39}$ spectral index\\

\hline \hline
  \end{tabular}

  \caption{Different metallicity estimates for M 30 in the literature. Values were converted using \Eq{eq:fehuves} where necessary. \label{tab:lit_met}}
  \end{center}
\end{table*}

\subsection{Age}

We use our CMD to derive an estimate for the age of M 30, by fitting to it the isochrones of \cite{vandenberg03}, using our estimate of the cluster metallicity, $\fehzw = -2.01 \pm 0.04$. We also used a value for the $\alpha-$enhancement of $[\alpha/{\rm Fe}]=+0.2$ \citep{dotter10}. From the best-fitting isochrones, we estimate the age of the cluster to be $13.0 \pm 1.0$ Gyr, in good agreement with recent estimates in the literature (see \Tab{tab:lit_age}). A set of isochrones is overplotted on the CMD in \Fig{fig:cmd}.

\begin{table*}
\begin{center}
  \begin{tabular}{cccc}

     \hline
    Reference	&Age [Gyr]		&Method	\\
 \hline
This work			&13 $\pm$ 1 	&CMD isochrone fitting \\
\cite{dotter10}		&13.25 $\pm$ 1.00 	&CMD isochrone fitting  	\\
\cite{carretta00}	&12.3 		&Parallax of local subdwarfs  	\\
\cite{bergbusch96}	&14			&CMD isochrone fitting \\
\cite{samus95}		&17			&CMD isochrone fitting \\
\cite{samus95}		&15			&CMD isochrone fitting \\
\cite{piotto90}		&16 $\pm$ 2	&CMD isochrone fitting\\
\cite{bolte87}		&17 			&CMD isochrone fitting 	\\

	\hline \hline
  \end{tabular}

  \caption{Age estimates for M 30 in the literature. \label{tab:lit_age}}
  \end{center}
\end{table*}

\section{Conclusions}

We have used $V-$ and $I-$band observations spanning a baseline of $\sim 14$ months to survey stellar photometric variability in M 30. This allows us to update the catalogue of variables in this cluster significantly, with detections and refined analysis of three previously known RR Lyrae stars, one known U Gem star, as well as of two candidate RR Lyrae stars, to which we assigned catalogue numbers. Furthermore, we detect previously unknown variables, including two RR1 stars, an SX Phoenicis star, an eclipsing blue straggler system, and two variables that we are unable to classify with certainty. We provide refined period estimates for all variables, and then carry out Fourier decomposition of RR Lyrae stars. Using the Fourier parameters of stars for which this was successful, we derive properties for the RR Lyrae, and use these as proxies for the cluster's properties. 

We find a cluster metallicity of $\fehzw = -2.01 \pm 0.04$, or $\fehuves = -2.11 \pm 0.06$, and distances of $8.32 \pm 0.20$ or 8.10 kpc, using RR0 and RR1 stars respectively. Our light curve of the SX Phoenicis variable V20 also affords another way to estimate the cluster distance; using this, we find a distance of $8.35 \pm 0.42$ kpc. Our CMD also allows us to estimate the age of the cluster by fitting isochrones to it. We find an age of $13 \pm 1$ Gyr. All of these values are in good agreement with estimates in the literature, and are independent estimates calculated for the first time using Fourier decomposition of RR Lyrae in this cluster.

\section*{Acknowledgements}
We thank the referee Christine Clement for constructive comments. NK acknowledges an ESO Fellowship. The research leading to these results has received funding from the European Community's Seventh Framework Programme 
(/FP7/2007-2013/) under grant agreement No 229517. AAF acknowledges the support of DGAPA-UNAM through project IN104612. AAF and SG are 
thankful to the CONACyT (M\'exico) and the Department of Science and Technology (India) for financial support under the Indo-Mexican collaborative project DST/INT/MEXICO/RP001/2008. 
OW (aspirant FRS - FNRS), AE (post-doc PRODEX), YD (post-doc PRODEX) and J. Surdej acknowledge support from the Communaut\'{e} fran\c{c}aise de Belgique -- Actions de recherche concert\'{e}es -- Acad\'{e}mie universitaire Wallonie-Europe. 
TCH gratefully acknowledges financial support from the Korea Research Council for Fundamental Science and Technology (KRCF) through the Young Research Scientist Fellowship Program. TCH acknowledges financial support from KASI (Korea Astronomy and Space Science Institute) grant number 2012-1-410-02.
KA, DB, MD, MH, and CL are supported by NPRP grant NPRP-09-476-1-78 from the Qatar National Research Fund (a member of Qatar Foundation).
The Danish 1.54m telescope is operated based on a grant from the Danish Natural Science Foundation (FNU). 
Funding for the Stellar Astrophysics Centre at Aarhus University is provided by The Danish National Research Foundation. 
This research was supported in part by the ASTERISK project (ASTERoseismic Investigations with SONG and Kepler), funded by the European Research Council (Grant agreement no.: 267864).
CS received funding from the European Union Seventh Framework Programme (FP7/2007-2013) under grant agreement no. 268421. 
CHG and XSF acknowledge support from National Natural Science Foundation of China (grants Nos. 10373023 and 10773027).
HK acknowledges support from a Marie-Curie Intra-European Fellowship.

\bibliographystyle{aa}
\bibliography{../thesisbib}

\begin{thebibliography}{75}
\expandafter\ifx\csname natexlab\endcsname\relax\def\natexlab#1{#1}\fi

\bibitem[{{Allen} {et~al.}(2006){Allen}, {Moreno}, \& {Pichardo}}]{allen06}
{Allen}, C., {Moreno}, E., \& {Pichardo}, B. 2006, \apj, 652, 1150

\bibitem[{{Arellano Ferro} {et~al.}(2013){Arellano Ferro}, {Bramich}, {Figuera
  Jaimes}, {Giridhar}, {Kains}, {Kuppuswamy}, \& {J{\o}rgensen}}]{arellano13}
{Arellano Ferro}, A., {Bramich}, D.~M., {Figuera Jaimes}, R., {et~al.} 2013, in
  prep.

\bibitem[{{Arellano Ferro} {et~al.}(2011){Arellano Ferro}, {Figuera Jaimes},
  {Giridhar}, {Bramich}, {Hern{\'a}ndez Santisteban}, \&
  {Kuppuswamy}}]{arellano11}
{Arellano Ferro}, A., {Figuera Jaimes}, R., {Giridhar}, S., {et~al.} 2011,
  \mnras, 416, 2265

\bibitem[{{Bailey}(1902)}]{bailey1902}
{Bailey}, S.~I. 1902, Annals of Harvard College Observatory, 38, 1

\bibitem[{{Bergbusch}(1996)}]{bergbusch96}
{Bergbusch}, P.~A. 1996, \aj, 112, 1061

\bibitem[{{Bla{\v z}ko}(1907)}]{blazhko1907}
{Bla{\v z}ko}, S. 1907, Astronomische Nachrichten, 175, 325

\bibitem[{{Bolte}(1987)}]{bolte87}
{Bolte}, M. 1987, \apj, 319, 760

\bibitem[{{Bramich}(2008)}]{bramich08}
{Bramich}, D.~M. 2008, \mnras, 386, L77

\bibitem[{{Bramich} {et~al.}(2011){Bramich}, {Figuera Jaimes}, {Giridhar}, \&
  {Arellano Ferro}}]{bramich11}
{Bramich}, D.~M., {Figuera Jaimes}, R., {Giridhar}, S., \& {Arellano Ferro}, A.
  2011, \mnras, 413, 1275

\bibitem[{{Bramich} \& {Freudling}(2012)}]{bramich12b}
{Bramich}, D.~M. \& {Freudling}, W. 2012, \mnras, 424, 1584

\bibitem[{{Bramich} {et~al.}(2013){Bramich}, {Horne}, {Albrow}, {Tsapras},
  {Snodgrass}, {Street}, {Hundertmark}, {Kains}, {Arellano}, {Figuera}, \&
  {Giridhar}}]{bramich13}
{Bramich}, D.~M., {Horne}, K., {Albrow}, M.~D., {et~al.} 2013, \mnras, 428,
  2275

\bibitem[{{Cacciari} {et~al.}(2005){Cacciari}, {Corwin}, \&
  {Carney}}]{cacciari05}
{Cacciari}, C., {Corwin}, T.~M., \& {Carney}, B.~W. 2005, \aj, 129, 267

\bibitem[{{Carretta} {et~al.}(2009{\natexlab{a}}){Carretta}, {Bragaglia},
  {Gratton}, {D'Orazi}, \& {Lucatello}}]{carretta09}
{Carretta}, E., {Bragaglia}, A., {Gratton}, R., {D'Orazi}, V., \& {Lucatello},
  S. 2009{\natexlab{a}}, \aap, 508, 695

\bibitem[{{Carretta} {et~al.}(2009{\natexlab{b}}){Carretta}, {Bragaglia},
  {Gratton}, \& {Lucatello}}]{carretta09b}
{Carretta}, E., {Bragaglia}, A., {Gratton}, R., \& {Lucatello}, S.
  2009{\natexlab{b}}, \aap, 505, 139

\bibitem[{{Carretta} {et~al.}(2009{\natexlab{c}}){Carretta}, {Bragaglia},
  {Gratton}, {Lucatello}, {Catanzaro}, {Leone}, {Bellazzini}, {Claudi},
  {D'Orazi}, {Momany}, {Ortolani}, {Pancino}, {Piotto}, {Recio-Blanco}, \&
  {Sabbi}}]{carretta09a}
{Carretta}, E., {Bragaglia}, A., {Gratton}, R.~G., {et~al.} 2009{\natexlab{c}},
  \aap, 505, 117

\bibitem[{{Carretta} {et~al.}(2000){Carretta}, {Gratton}, {Clementini}, \&
  {Fusi Pecci}}]{carretta00}
{Carretta}, E., {Gratton}, R.~G., {Clementini}, G., \& {Fusi Pecci}, F. 2000,
  \apj, 533, 215

\bibitem[{{Castelli}(1999)}]{castelli99}
{Castelli}, F. 1999, \aap, 346, 564

\bibitem[{{Claria} {et~al.}(1988){Claria}, {Minniti}, \& {Gomez}}]{claria88}
{Claria}, J.~J., {Minniti}, D., \& {Gomez}, M.~N. 1988, \aap, 199, L1

\bibitem[{{Clement} {et~al.}(1993){Clement}, {Ferance}, \& {Simon}}]{clement93}
{Clement}, C.~M., {Ferance}, S., \& {Simon}, N.~R. 1993, \apj, 412, 183

\bibitem[{{Clement} {et~al.}(2001){Clement}, {Muzzin}, {Dufton}, {Ponnampalam},
  {Wang}, {Burford}, {Richardson}, {Rosebery}, {Rowe}, \& {Hogg}}]{clement01}
{Clement}, C.~M., {Muzzin}, A., {Dufton}, Q., {et~al.} 2001, \aj, 122, 2587

\bibitem[{{Clementini} {et~al.}(2004){Clementini}, {Corwin}, {Carney}, \&
  {Sumerel}}]{clementini04}
{Clementini}, G., {Corwin}, T.~M., {Carney}, B.~W., \& {Sumerel}, A.~N. 2004,
  \aj, 127, 938

\bibitem[{{Cohen} \& {Sarajedini}(2012)}]{cohen12}
{Cohen}, R.~E. \& {Sarajedini}, A. 2012, \mnras, 419, 342

\bibitem[{{Cox} {et~al.}(1983){Cox}, {Hodson}, \& {Clancy}}]{cox83}
{Cox}, A.~N., {Hodson}, S.~W., \& {Clancy}, S.~P. 1983, \apj, 266, 94

\bibitem[{{Darragh} \& {Murphy}(2012)}]{darragh12}
{Darragh}, A.~N. \& {Murphy}, B.~W. 2012, Journal of the Southeastern
  Association for Research in Astronomy, 6, 72

\bibitem[{{Di Fabrizio} {et~al.}(2005){Di Fabrizio}, {Clementini}, {Maio},
  {Bragaglia}, {Carretta}, {Gratton}, {Montegriffo}, \&
  {Zoccali}}]{difabrizio05}
{Di Fabrizio}, L., {Clementini}, G., {Maio}, M., {et~al.} 2005, \aap, 430, 603

\bibitem[{{Djorgovski} \& {King}(1986)}]{djorgovski86}
{Djorgovski}, S. \& {King}, I.~R. 1986, \apjl, 305, L61

\bibitem[{{Dotter} {et~al.}(2010){Dotter}, {Sarajedini}, {Anderson},
  {Aparicio}, {Bedin}, {Chaboyer}, {Majewski}, {Mar{\'{\i}}n-Franch}, {Milone},
  {Paust}, {Piotto}, {Reid}, {Rosenberg}, \& {Siegel}}]{dotter10}
{Dotter}, A., {Sarajedini}, A., {Anderson}, J., {et~al.} 2010, \apj, 708, 698

\bibitem[{{Echevarria} \& {Jones}(1984)}]{echevarria84}
{Echevarria}, J. \& {Jones}, D.~H.~P. 1984, \mnras, 206, 919

\bibitem[{{Ferraro} {et~al.}(1999){Ferraro}, {Messineo}, {Fusi Pecci}, {de
  Palo}, {Straniero}, {Chieffi}, \& {Limongi}}]{ferraro99}
{Ferraro}, F.~R., {Messineo}, M., {Fusi Pecci}, F., {et~al.} 1999, \aj, 118,
  1738

\bibitem[{{Forbes} \& {Bridges}(2010)}]{forbes10}
{Forbes}, D.~A. \& {Bridges}, T. 2010, \mnras, 404, 1203

\bibitem[{{Freedman} {et~al.}(2001){Freedman}, {Madore}, {Gibson}, {Ferrarese},
  {Kelson}, {Sakai}, {Mould}, {Kennicutt}, {Ford}, {Graham}, {Huchra},
  {Hughes}, {Illingworth}, {Macri}, \& {Stetson}}]{freedman01}
{Freedman}, W.~L., {Madore}, B.~F., {Gibson}, B.~K., {et~al.} 2001, \apj, 553,
  47

\bibitem[{{Gratton} {et~al.}(2004){Gratton}, {Bragaglia}, {Clementini},
  {Carretta}, {Di Fabrizio}, {Maio}, \& {Taribello}}]{gratton04}
{Gratton}, R.~G., {Bragaglia}, A., {Clementini}, G., {et~al.} 2004, \aap, 421,
  937

\bibitem[{{Gratton} {et~al.}(1997){Gratton}, {Fusi Pecci}, {Carretta},
  {Clementini}, {Corsi}, \& {Lattanzi}}]{gratton97}
{Gratton}, R.~G., {Fusi Pecci}, F., {Carretta}, E., {et~al.} 1997, \apj, 491,
  749

\bibitem[{{Jeon} {et~al.}(2003){Jeon}, {Lee}, {Kim}, \& {Lee}}]{jeon03}
{Jeon}, Y.-B., {Lee}, M.~G., {Kim}, S.-L., \& {Lee}, H. 2003, \aj, 125, 3165

\bibitem[{{Jurcsik}(1995)}]{jurcsik95}
{Jurcsik}, J. 1995, \actaa, 45, 653

\bibitem[{{Jurcsik}(1998)}]{jurcsik98}
{Jurcsik}, J. 1998, \aap, 333, 571

\bibitem[{{Jurcsik} \& {Kov\'{a}cs}(1996)}]{jurcsik96}
{Jurcsik}, J. \& {Kov\'{a}cs}, G. 1996, \aap, 312, 111

\bibitem[{{Kai} \& {Sheng-Bang}(2012)}]{kai12}
{Kai}, L. \& {Sheng-Bang}, Q. 2012, \aj, 144, 161

\bibitem[{{Kains} {et~al.}(2012){Kains}, {Bramich}, {Figuera Jaimes}, {Arellano
  Ferro}, {Giridhar}, \& {Kuppuswamy}}]{kains12b}
{Kains}, N., {Bramich}, D.~M., {Figuera Jaimes}, R., {et~al.} 2012, \aap, 548,
  A92

\bibitem[{{Kov{\'a}cs}(1998)}]{kovacs98}
{Kov{\'a}cs}, G. 1998, \memsai, 69, 49

\bibitem[{{Kov{\'a}cs}(2002)}]{kovacs02}
{Kov{\'a}cs}, G. 2002, in Astronomical Society of the Pacific Conference
  Series, Vol. 265, Omega Centauri, A Unique Window into Astrophysics, ed.
  F.~{van Leeuwen}, J.~D. {Hughes}, \& G.~{Piotto}, 163

\bibitem[{{Kov{\'a}cs} \& {Walker}(2001)}]{kovacs01}
{Kov{\'a}cs}, G. \& {Walker}, A.~R. 2001, \aap, 371, 579

\bibitem[{{Lafler} \& {Kinman}(1965)}]{lafler65}
{Lafler}, J. \& {Kinman}, T.~D. 1965, \apjs, 11, 216

\bibitem[{{Lee} \& {Carney}(1999)}]{lee99}
{Lee}, J.-W. \& {Carney}, B.~W. 1999, \aj, 118, 1373

\bibitem[{{Machin} {et~al.}(1991){Machin}, {Callanan}, {Allington-Smith},
  {Charles}, \& {Hassall}}]{machin91}
{Machin}, G., {Callanan}, P.~J., {Allington-Smith}, J., {Charles}, P.~A., \&
  {Hassall}, B.~J.~M. 1991, \mnras, 250, 602

\bibitem[{{Margon} \& {Downes}(1983)}]{margon83}
{Margon}, B. \& {Downes}, R.~A. 1983, \apjl, 274, L31

\bibitem[{{Minniti} {et~al.}(1993){Minniti}, {Geisler}, {Peterson}, \&
  {Claria}}]{minniti93}
{Minniti}, D., {Geisler}, D., {Peterson}, R.~C., \& {Claria}, J.~J. 1993, \apj,
  413, 548

\bibitem[{{Monet} {et~al.}(2003){Monet}, {Levine}, {Canzian}, {Ables}, {Bird},
  {Dahn}, {Guetter}, {Harris}, {Henden}, {Leggett}, {Levison}, {Luginbuhl},
  {Martini}, {Monet}, {Munn}, {Pier}, {Rhodes}, {Riepe}, {Sell}, {Stone},
  {Vrba}, {Walker}, {Westerhout}, {Brucato}, {Reid}, {Schoening}, {Hartley},
  {Read}, \& {Tritton}}]{monet03}
{Monet}, D.~G., {Levine}, S.~E., {Canzian}, B., {et~al.} 2003, \aj, 125, 984

\bibitem[{{Montegriffo} {et~al.}(1998){Montegriffo}, {Ferraro}, {Origlia}, \&
  {Fusi Pecci}}]{montegriffo98}
{Montegriffo}, P., {Ferraro}, F.~R., {Origlia}, L., \& {Fusi Pecci}, F. 1998,
  \mnras, 297, 872

\bibitem[{{Morgan} {et~al.}(2007){Morgan}, {Wahl}, \& {Wieckhorst}}]{morgan07}
{Morgan}, S.~M., {Wahl}, J.~N., \& {Wieckhorst}, R.~M. 2007, \mnras, 374, 1421

\bibitem[{{Nemec}(1985)}]{nemec85}
{Nemec}, J.~M. 1985, \aj, 90, 240

\bibitem[{{Olech} {et~al.}(1999){Olech}, {Kaluzny}, {Thompson}, {Pych},
  {Krzeminski}, \& {Shwarzenberg-Czerny}}]{olech99}
{Olech}, A., {Kaluzny}, J., {Thompson}, I.~B., {et~al.} 1999, \aj, 118, 442

\bibitem[{{Padmanabhan} {et~al.}(2008){Padmanabhan}, {Schlegel}, {Finkbeiner},
  {Barentine}, {Blanton}, {Brewington}, {Gunn}, {Harvanek}, {Hogg},
  {Ivezi{\'c}}, {Johnston}, {Kent}, {Kleinman}, {Knapp}, {Krzesinski}, {Long},
  {Neilsen}, {Nitta}, {Loomis}, {Lupton}, {Roweis}, {Snedden}, {Strauss}, \&
  {Tucker}}]{padmanabhan08}
{Padmanabhan}, N., {Schlegel}, D.~J., {Finkbeiner}, D.~P., {et~al.} 2008, \apj,
  674, 1217

\bibitem[{{Pietrukowicz} \& {Kaluzny}(2004)}]{pietrukowicz04}
{Pietrukowicz}, P. \& {Kaluzny}, J. 2004, \actaa, 54, 19

\bibitem[{{Pietrukowicz} {et~al.}(2008){Pietrukowicz}, {Kaluzny},
  {Schwarzenberg-Czerny}, {Thompson}, {Pych}, {Krzeminski}, \&
  {Mazur}}]{pietrukowicz08}
{Pietrukowicz}, P., {Kaluzny}, J., {Schwarzenberg-Czerny}, A., {et~al.} 2008,
  \mnras, 388, 1111

\bibitem[{{Piotto} {et~al.}(1987){Piotto}, {Capaccioli}, {Ortolani}, {Rosino},
  {Alcaino}, \& {Liller}}]{piotto87}
{Piotto}, G., {Capaccioli}, M., {Ortolani}, S., {et~al.} 1987, \aj, 94, 360

\bibitem[{{Piotto} {et~al.}(1990){Piotto}, {King}, {Capaccioli}, {Ortolani}, \&
  {Djorgovski}}]{piotto90}
{Piotto}, G., {King}, I.~R., {Capaccioli}, M., {Ortolani}, S., \& {Djorgovski},
  S. 1990, \apj, 350, 662

\bibitem[{{Richer} {et~al.}(1988){Richer}, {Fahlman}, \&
  {Vandenberg}}]{richer88}
{Richer}, H.~B., {Fahlman}, G.~G., \& {Vandenberg}, D.~A. 1988, \apj, 329, 187

\bibitem[{{Rosino}(1949)}]{rosino49}
{Rosino}, L. 1949, \memsai, 20, 63

\bibitem[{{Rosino}(1961)}]{rosino61}
{Rosino}, L. 1961, Asiago-Padova Cont., 117

\bibitem[{{Samus} {et~al.}(1995){Samus}, {Ipatov}, {Smirnov}, {Kravtsov},
  {Alcaino}, {Liller}, \& {Alvarado}}]{samus95}
{Samus}, N.~N., {Ipatov}, A.~P., {Smirnov}, O.~M., {et~al.} 1995, Astronomy
  Letters, 21, 810

\bibitem[{{Sandage}(2006)}]{sandage06}
{Sandage}, A. 2006, \aj, 131, 1750

\bibitem[{{Sandquist} {et~al.}(1999){Sandquist}, {Bolte}, {Langer}, {Hesser},
  \& {Mendes de Oliveira}}]{sandquist99}
{Sandquist}, E.~L., {Bolte}, M., {Langer}, G.~E., {Hesser}, J.~E., \& {Mendes
  de Oliveira}, C. 1999, \apj, 518, 262

\bibitem[{{Sekiguchi} \& {Fukugita}(2000)}]{sekiguchi00}
{Sekiguchi}, M. \& {Fukugita}, M. 2000, \aj, 120, 1072

\bibitem[{{Simon} \& {Clement}(1993)}]{simon93}
{Simon}, N.~R. \& {Clement}, C.~M. 1993, \apj, 410, 526

\bibitem[{{Skottfelt} {et~al.}(2013){Skottfelt}, {Bramich}, {Figuera Jaimes},
  {J{\o}rgensen}, {Kains}, {Harpsoe}, {Liebig}, {Penny}, {Alsubai}, {Andersen},
  {Bozza}, {Browne}, {Calchi Novati}, {Damerdji}, {Diehl}, {Dominik}, {Elyiv},
  {Giannini}, {Hessman}, {Hinse}, {Hundertmark}, {Juncher}, {Kerins},
  {Korhonen}, {Mancini}, {Martin}, {Rabus}, {Rahvar}, {Southworth},
  {Snodgrass}, {Street}, {Surdej}, {Tregloan-Reed}, {Vilela}, \&
  {Williams}}]{skottfelt13}
{Skottfelt}, J., {Bramich}, D.~M., {Figuera Jaimes}, R., {et~al.} 2013, ArXiv
  e-prints

\bibitem[{{Smith}(1984)}]{smith84}
{Smith}, H.~A. 1984, \apj, 281, 148

\bibitem[{{Smitka} \& {Layden}(2010)}]{smitka07}
{Smitka}, M.~T. \& {Layden}, A.~C. 2010, in Bulletin of the American
  Astronomical Society, Vol.~36, American Astronomical Society Meeting
  Abstracts 215, 606.18

\bibitem[{{Stellingwerf}(1978)}]{stellingwerf78}
{Stellingwerf}, R.~F. 1978, \apj, 224, 953

\bibitem[{{Stetson}(2000)}]{stetson00}
{Stetson}, P.~B. 2000, \pasp, 112, 925

\bibitem[{{Terzan}(1968)}]{terzan68}
{Terzan}, A. 1968, Publications of the Observatoire Haute-Provence, 9, 24

\bibitem[{{Terzan} \& {Rutily}(1975)}]{terzan75}
{Terzan}, A. \& {Rutily}, B. 1975, \aap, 38, 307

\bibitem[{{VandenBerg} \& {Clem}(2003)}]{vandenberg03}
{VandenBerg}, D.~A. \& {Clem}, J.~L. 2003, \aj, 126, 778

\bibitem[{{Zinn}(1980)}]{zinn80}
{Zinn}, R. 1980, \apjs, 42, 19

\bibitem[{{Zinn} \& {West}(1984)}]{zinn84}
{Zinn}, R. \& {West}, M.~J. 1984, \apjs, 55, 45

\end{thebibliography}

\label{lastpage}

\end{document}